\documentclass{aa}
\usepackage{graphicx}
\usepackage{natbib}
\usepackage{amssymb}
\usepackage{amstext}
\usepackage{latexsym} 
\usepackage{threeparttable} 
\usepackage[sumlimits]{amsmath}
\newcommand{\Ha}{H$\alpha$}			
\newcommand{\HII}{H{\sc ii}}			
\newcommand{\NII}{[N{\sc ii}]}			

\begin{document}
   \title{The H$\alpha$ Galaxy Survey \thanks{
Based on observations made with the Jacobus Kapteyn Telescope operated 
on the island of La Palma by the Isaac Newton Group in the Spanish 
Observatorio del Roque de los Muchachos of the Instituto de Astrofisica 
de Canarias
   }}
   \subtitle{VII. The spatial distribution of star formation within 
disks and bulges.}
   \author{P. A. James
          \inst{1},
          C. F. Bretherton
          \inst{1,2} and
	  J. H. Knapen
	  \inst{3}
          } \offprints{P.A. James} 
          \institute{Astrophysics Research
	  Institute, Liverpool John Moores University, Twelve Quays
	  House, Egerton Wharf, Birkenhead CH41 1LD, UK \\
	  \email{paj,cft@astro.livjm.ac.uk}
          \and Royal Observatory, Greenwich, London, SE10 9LF, UK\\
	  \and Instituto de Astrof\'\i sica de Canarias, 
	  E-38200 La Laguna, Spain\\
	  \email{jhk@.iac.es}
	  }
          \date{Received ; accepted }

\abstract
{}
{We analyse the current build-up of stellar mass within the disks
and bulges of nearby galaxies through a comparison of the spatial
distributions of forming and old stellar populations.}
{\Ha\ and $R$-band imaging are used to determine the distributions
of young and old stellar populations in 313 S0a - Im field galaxies
out to $\sim$40~Mpc.  Concentration indices
and mean normalised light profiles are calculated as a function
of galaxy type and bar classification.}
{The mean profiles and concentration indices show a strong and smooth
dependence on galaxy type.  Apart from a central deficit due to bulge/bar
light in some galaxy types, mean \Ha\ and $R$-band profiles are 
very similar.  Mean profiles within a given type are remarkably
constant even given wide ranges in galaxy luminosity and size.
SBc, SBbc and particularly
SBb galaxies have profiles that are markedly different from those of 
unbarred galaxies.  The \Ha\ emission from individual SBb galaxies
is studied in detail; virtually all show resolved central components 
and concentrations of star formation at or just outside the
bar-end radius.}
{Galaxy type is an excellent predictor of $R$-band light profile.
In field galaxies, star formation has the same radial distribution
as $R$-band light, i.e. stellar mass is building at approximately
constant morphology, with no strong evidence for outer truncation
or inside-out disk formation.  Bars have a strong impact on the 
radial distribution of star formation, particularly in SBb galaxies.}

   
\authorrunning{James et al.}
\titlerunning{H$\alpha$ Galaxy Survey. VII.}
\maketitle
%
\section{Introduction}
\label{sec:intro}

Disks of spiral galaxies host the majority of the star formation
activity in the local Universe \citep[][henceforth Paper
  IV]{tins80,some01,hani06,paper4}.  However, there are many remaining
questions regarding the formation, stability and growth of stellar
disks, and the interrelation of these with other galaxy components,
particularly bars and central bulge components.  Star formation (SF)
is shown by a wide range of indicators to be virtually ubiquitous in
disks, unlike elliptical galaxies where most of the SF was completed
early in their evolution.  Typical current SF rates within the disks
of bright, nearby spiral galaxies are 1--2 M$_{\odot}$~yr$^{-1}$
\citep[][Paper I]{kenn83,paper1}, sufficient to accumulate the mass of
a substantial disk if continued over a Hubble time.  Thus it is
reasonable to ask whether the spatial distribution of new stars is
consistent with that of the overall stellar mass in disks of the same
type, thus motivating a picture in which disks are constructed largely
through the types of SF we see at the current epoch.  Alternatives are
that disks could have grown from the inside outwards
\citep{truj05,muno07}, resulting in the youngest stars lying on
average at greater radial distances than the old stellar population;
or indeed outside-in models of galaxy formation have been suggested
\citep{gall08}.  It is also possible that spatial distributions of
young and old stars could differ because of radial drifts of stellar
orbits, driven by tidal torques from spiral arms, bars or external
tidal interactions.

The origin of the characteristic exponential profile, found to describe
the radial light distribution of many disks, has also been studied through
theoretical studies \citep{fall80,free02,elme05}, most of which start
from the observation by \citet{mest63} that the angular momentum
distribution of an exponential disk resembles that of a sphere 
undergoing solid-body rotation.

Observational determinations of the radial distribution of SF in
galaxy disks have been performed in the past, using several approaches
and observational techniques.  The approaches include detailed studies
of individual galaxies, e.g., \citet{comt82} who looked at the Sbc
spiral NGC~1566, to statistical studies of typically several tens of
galaxies.  An early study of the latter type was carried out by
\citet{hodg83}, who considered the distributions of \HII\ regions
revealed by narrow-band H$\alpha$ imaging, finding them to have
characteristic ring or `doughnut' shaped distributions in early-type
spirals, with more extended distributions being found in late-type
spirals, and `oscillating' (non-monotonic) distributions occurring in
many barred galaxies.  \citet{ryde94} studied the relative scale
lengths of H$\alpha$, $V$- and $I$-band emission from 34 S0--Sm
galaxies, finding the line emission to have larger scale lengths than
those of the continuum emission.  \citet{garc96} investigated
H$\alpha$ imaging of 52 barred spiral galaxies, finding nuclear rings
in 10, and emission indicating SF along the bar in 18.

\citet{dale01} analysed H$\alpha$ extents of galaxies normalised by 
their $I$-band sizes, finding no strong type dependences; similar
results for H$\alpha$ to $R$-band scalelength ratios were found by
\citet{koop04} and \citet{koop06}.
\citet{hatt04} studied the effects of galaxy-galaxy interactions on the
distributions of H$\alpha$ emission within disks, finding that extended 
starbursts are common in such galaxies.
Finally, \citet{bend07} looked in detail at the SF distribution in 
galaxies from the SINGS sample, using 24~$\mu$m emission as the primary
SF tracer. They found this emission to be typically compact and symmetric 
for early type galaxies, but more extended and asymmetric for late type 
galaxies.

The aim of this paper is to address these questions through an
analysis of the spatial distribution of SF as traced by the
\Ha\ emission line in a large sample of local disk and irregular
galaxies, and to compare it with the distribution of older stars.
\Ha\ provides essentially a snapshot view of SF activity, since it is
powered by massive stars with main sequence lifetimes of $\sim$10$^7$
years.  This can cause problems with the interpretation of the
\Ha\ properties of individual galaxies, since the emission is driven
by a small number of \HII\ regions with short lifetimes, and so the
stochastic uncertainties are substantial.  Thus the approach adopted
here is to look at the mean properties of at least several galaxies of
the same type, giving a statistical basis for studies of the
distribution of SF, the growth of disks, and the effects of bars on
the SF process.  The galaxies used are a field sample, and hence the
effects of environment should be small, though some of the galaxies do
have fairly close companions.  These will be studied in a subsequent
paper (J. H. Knapen \& P. A. James ApJ submitted, Paper VIII).

All data are taken from the \Ha\ Galaxy Survey (\Ha GS), a survey of
327 nearby galaxies (plus a further 7 serendipitously observed
objects) which have been imaged in both the \Ha\ line and the $R$-band
continuum.  The narrow-band filters used encompassed the \Ha\ and
neighbouring \NII\ lines, which should be borne in mind when
interpreting the narrow-band flux distributions in the present paper.
For convenience, this combined emission is referred to as
\Ha\ throughout.  The \Ha GS sample contains all Hubble types from
S0/a to Im and galaxies are selected to have heliocentric
recession velocities less than 3000~km~s$^{-1}$.  All galaxies were
observed with the 1.0 metre Jacobus Kapteyn Telescope (JKT), part of
the Isaac Newton Group of Telescopes (ING) situated on La Palma in the
Canary Islands.  The selection and the observation of the sample are
discussed in Paper I.  The overall aim of the survey is to quantify as
fully as possible the star formation properties in field galaxies at
the current epoch. Earlier \Ha GS papers have looked at total SF rates
and H$\alpha$ equivalent widths in galaxies (Paper I), and the
contributions of galaxies of different types to the integrated SF rate
(SFR) per unit volume of the local Universe (Paper IV).  Whereas
previous papers derived from the \Ha GS have mainly focussed on
integrated SF properties for each of the galaxies studied, the present
paper will focus on the spatial resolution of SF provided by the
\Ha\ imaging technique.  This will be done first through an analysis
of three concentration indices applied to both the $R$-band and
\Ha\ light distributions, and then through more detailed study of mean
normalised radial light distributions for galaxies of each
morphological type, and for barred and unbarred galaxies.

Some results from the earlier papers that are relevant for the
present study will first be summarised.  Paper I contained a brief
analysis of the effect of bars on total SFRs of galaxies, and on the
equivalent width of \Ha\ emission.  Bars give modest, barely
significant increases in both quantities overall; however, barred
galaxies of types Sab -- Sc inclusive were found to have larger SFRs by
factors 1.5--2.0 compared with unbarred galaxies of the same types.
The SFR per galaxy is highest in intermediate disk types with
classifications of Sbc and Sc.  In Paper IV it was found that
these same types also dominate the SFR density, i.e., SFR per unit
volume of the local Universe, summed over all Hubble types.  Disk
regions, defined as those lying more than 1~kpc from the centres of
galaxies, were found to contribute more than 80\% of the total SF
currently occurring in the local Universe.

\begin{figure*}
\centering
\rotatebox{-90}{
\includegraphics[height=5.9cm]{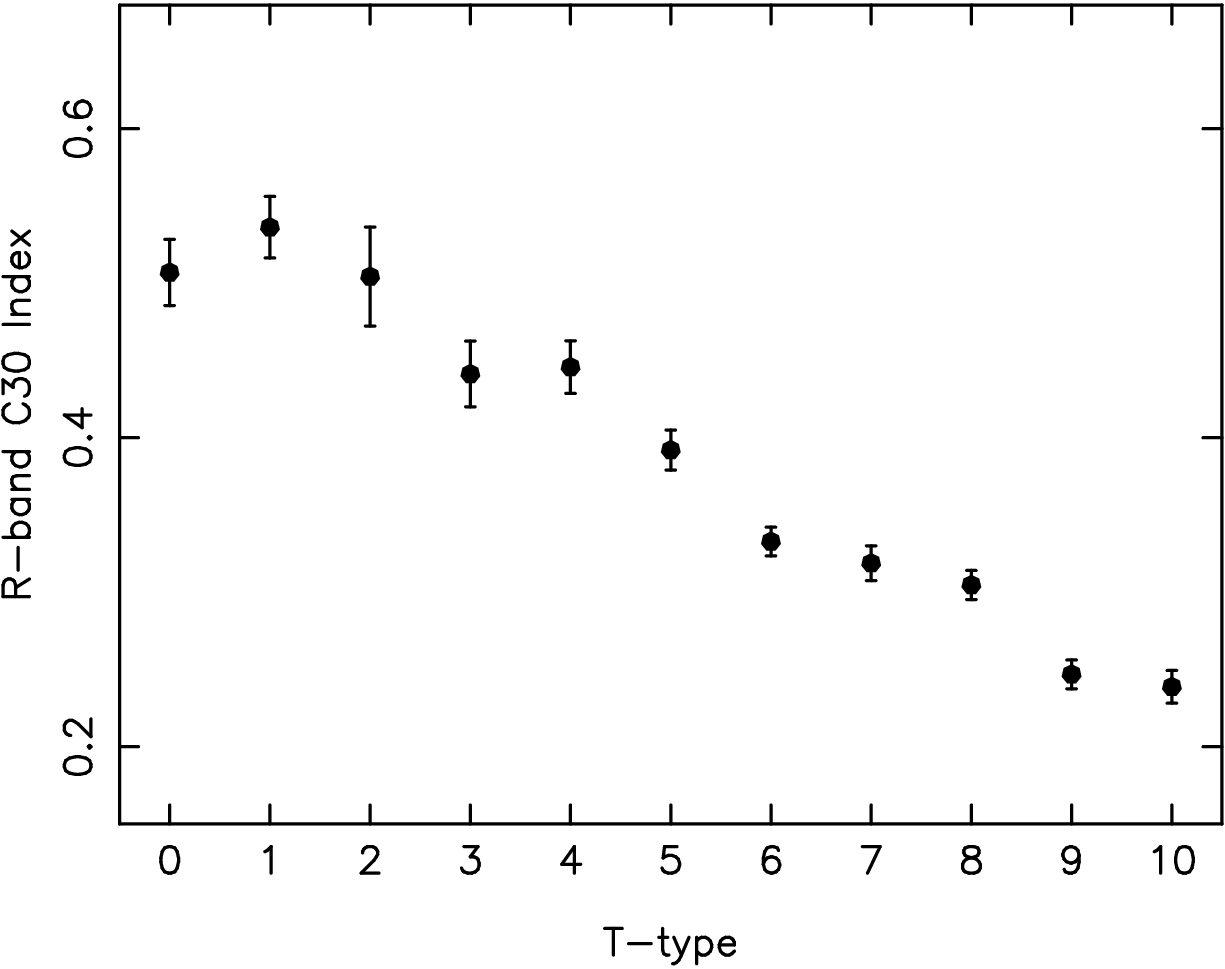}
}
\rotatebox{-90}{
\includegraphics[height=5.9cm]{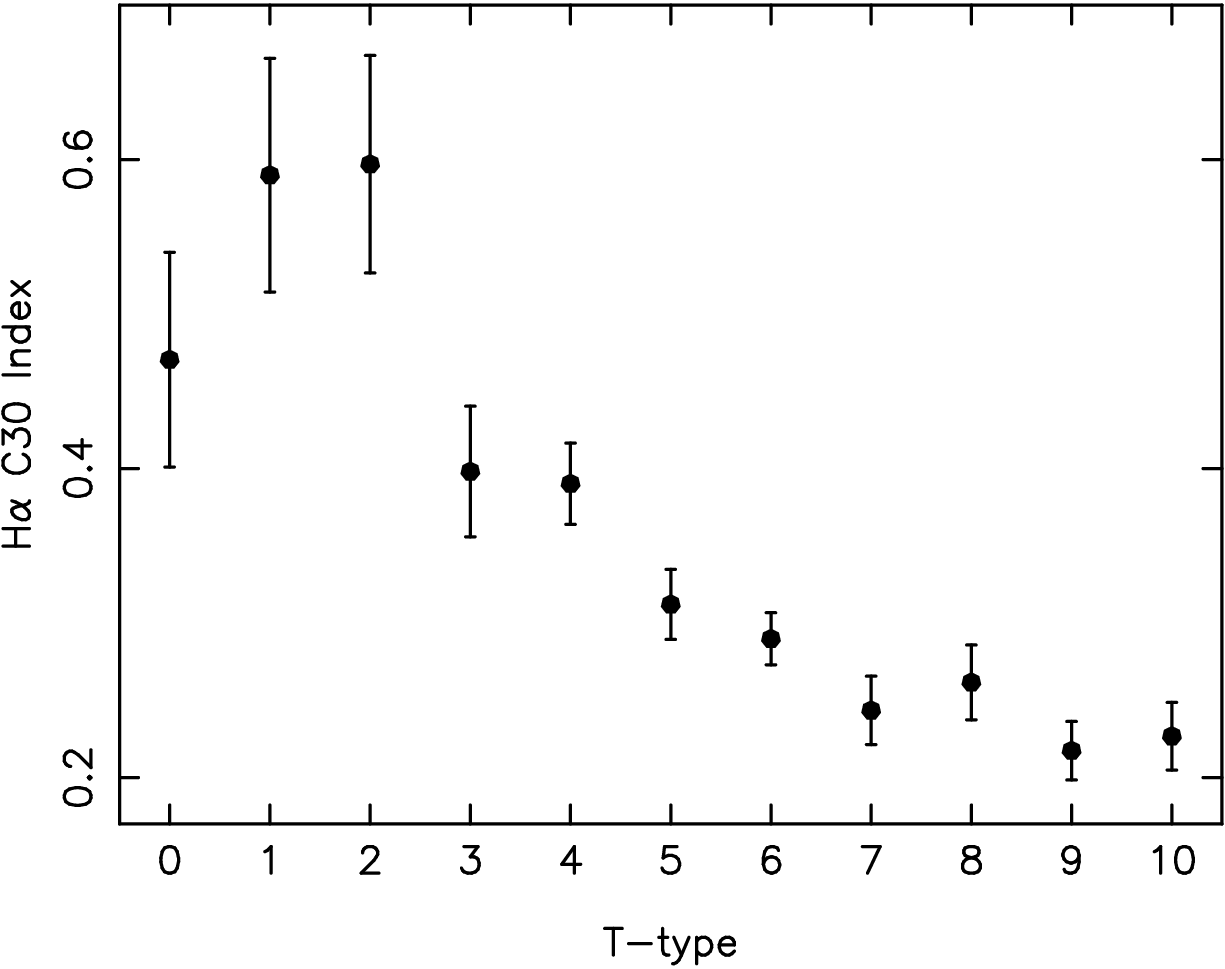}
}
\rotatebox{-90}{
\includegraphics[height=5.9cm]{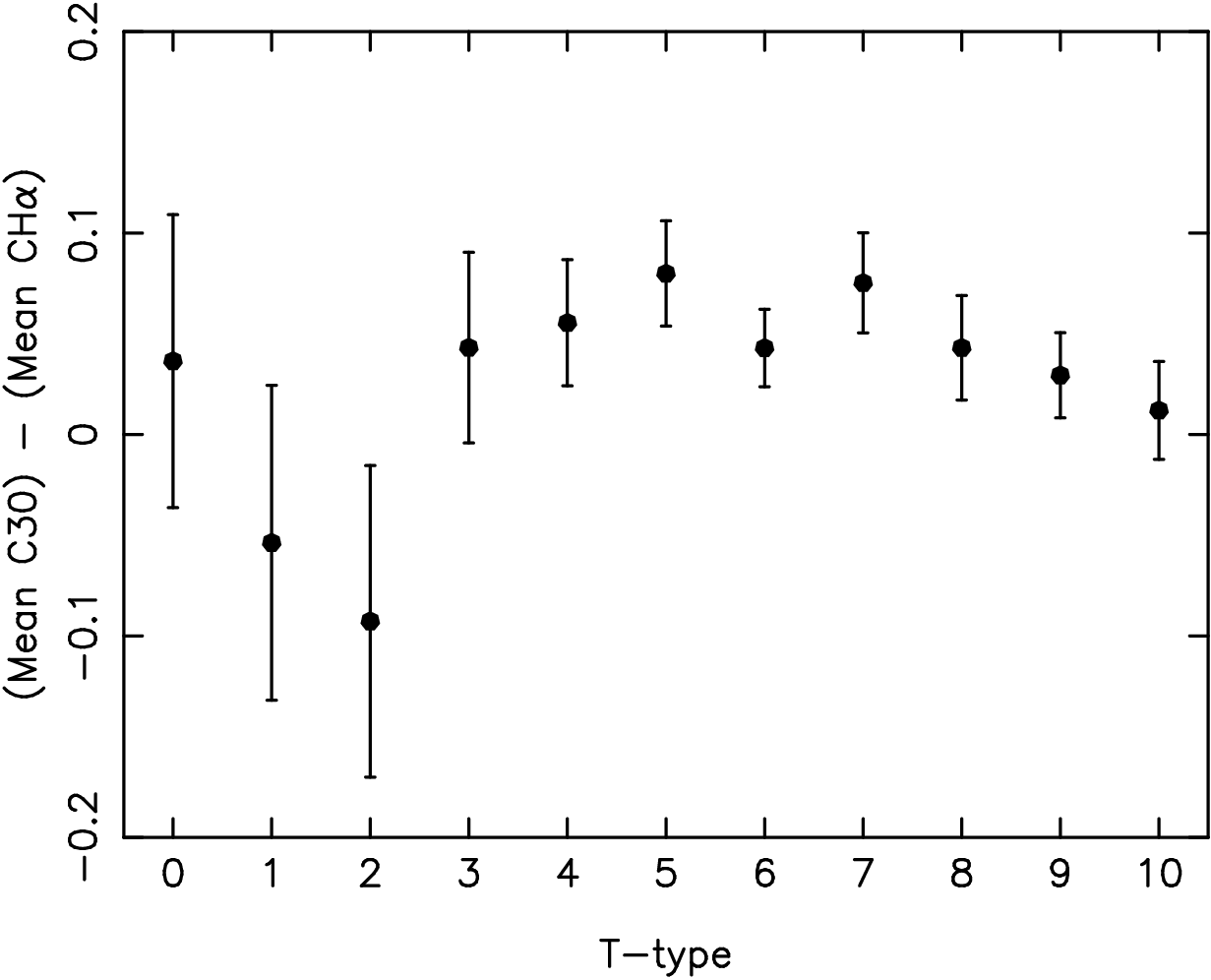}
}
\caption
{The concentration index C30 plotted against
Hubble type T.  The plot on the left shows the mean value of the
$R$-band C30 index for each type, with the error bars representing the
standard error on the mean for all 
galaxies of that type.  The plot in the centre shows the same mean
concentration indices for \Ha\ emission, and the plot on the right is
the difference, in the sense ($R$-band $-$ \Ha\ ).  }
\label{fig:c30_v_t}
\end{figure*}

Sample \Ha\ and $R$-band images of several galaxies from the \Ha GS
database are presented in Paper I, and a large number of \Ha\ images
of those galaxies that have hosted supernovae are presented in the
online version of another paper in this series 
\citep[][Paper III]{paper3}.  These
provide a good indication of the overall quality of the data used in
the present analysis, and in the interests of brevity no images are
shown in the present paper.

The structure of this paper is as follows.  Section 2 contains an
investigation of three concentration indices, which are applied to
both \Ha\ and $R$-band light distributions. The strength of
correlation between indices and Hubble type is studied, leading to
preliminary conclusions on the relative distributions of young and old
stellar populations as a function of galaxy type. The effect of strong
bars on the radial distribution of SF as traced by these indices is
also analysed.  Section 3 looks at mean radial profiles as a more
detailed tracer of stellar distributions.  \Ha\ and $R$-band mean
profiles, binned by type, are presented for the full sequence of
spiral types, and compared for barred and unbarred types, leading to
the identification of an effect on profiles for bars in SBb types that
is particularly marked in \Ha\ profiles but also clear in the $R$-band
light distributions.  The effect of continuum-subtraction errors
  on profiles is also studied in this section. Section 4 contains a
detailed investigation of the central and bar-end \Ha\ emission in SBb
galaxies.  Section 5 contains a discussion of some of the main
results, and the conclusions are presented in Sect. 6.


\section{Concentration indices}
\label{sec:concind}

\subsection{The concentration indices studied here}

Concentration indices provide a simple measure of the observed
radial distribution of the luminosity of a galaxy in a
specified bandpass.  They are of use since they contain much of the
information contained in a galaxy classification (as will be
demonstrated below) but in a quantified and less subjective form.  The
three indices investigated here are all measures of the `cuspiness' of
the light distribution, and are sensitive to the difference between,
for example, an $r^{1/4}$ law profile characteristic of an elliptical
galaxy or a luminous classical bulge, where a large fraction of the
light is concentrated in a central spike, and a more extended
distribution such as an exponential, as would be expected for a galaxy
disk. All are scale-independent, and can be applied without distance
information.

The first index is 
the C30 index of \citet{abra94} and \citet{koop98},
which is the ratio of the flux within 0.3 times the $R =$ 24.0 mag per
square arcsec isophotal radius (henceforth $r_{24}$) and the total
flux within that same radius.

Secondly, the C$_{31}$ index \citep{deva77} is
defined as the ratio of the radii of the apertures containing 75\% and
25\% of the total light, although for convenience here we plot the log
to the base 10 of this quantity.  Other variants make use of different
fractions of light, with 80\% and 20\% also being widely used. One
drawback of C$_{31}$ is that it requires an estimate of the total
luminosity, which is generally a poorly defined quantity.

Finally, we study 
the Petrosian index $CI_p$ \citep{shim01} which is based on the Petrosian
radius \citep{petr76} of the galaxy under study.  This is the radius
at which the local surface brightness (SB) is fainter by a given
factor $\eta$ than the average SB within that radius.  This is a
useful quantity since this radius is independent of galaxy distance,
$K$-corrections and extinction (if uniform).  Here we define the
Petrosian radius for an index $\eta =$ 0.2; the concentration index is
then the ratio of the radii containing 50\% and 90\% of the flux
within the Petrosian radius.

In this section we investigate the correlation between these different
indices and galaxy Hubble type for the \Ha GS sample; the sample size
is sufficient for us to separate galaxies by Hubble $T$-type \citep{deva59}, 
with the
latter running from $T=0$ (S0a) to $T=$ 10 (Im).  The data used are
multi-aperture photometry from the CCD $R$-band and narrow-band
continuum-subtracted \Ha\ images from the \Ha GS survey.  For the
spiral types S0a-Sm ($T$-types 0-9), elliptical apertures are
used, with constant ellipticity and position angle at all values of
the aperture semi-major axis; these parameters are defined by galaxy
isophotal shapes in the outer regions of their disks.  For Im
Magellanic irregulars of type $T=10$ (and face-on spirals), circular
apertures are used.

\subsection{Concentration indices for $R$-band and \Ha\ light as a function
of galaxy type}

\subsubsection{The C30 index vs type}

\begin{figure*}
\centering
\rotatebox{-90}{
\includegraphics[height=8cm]{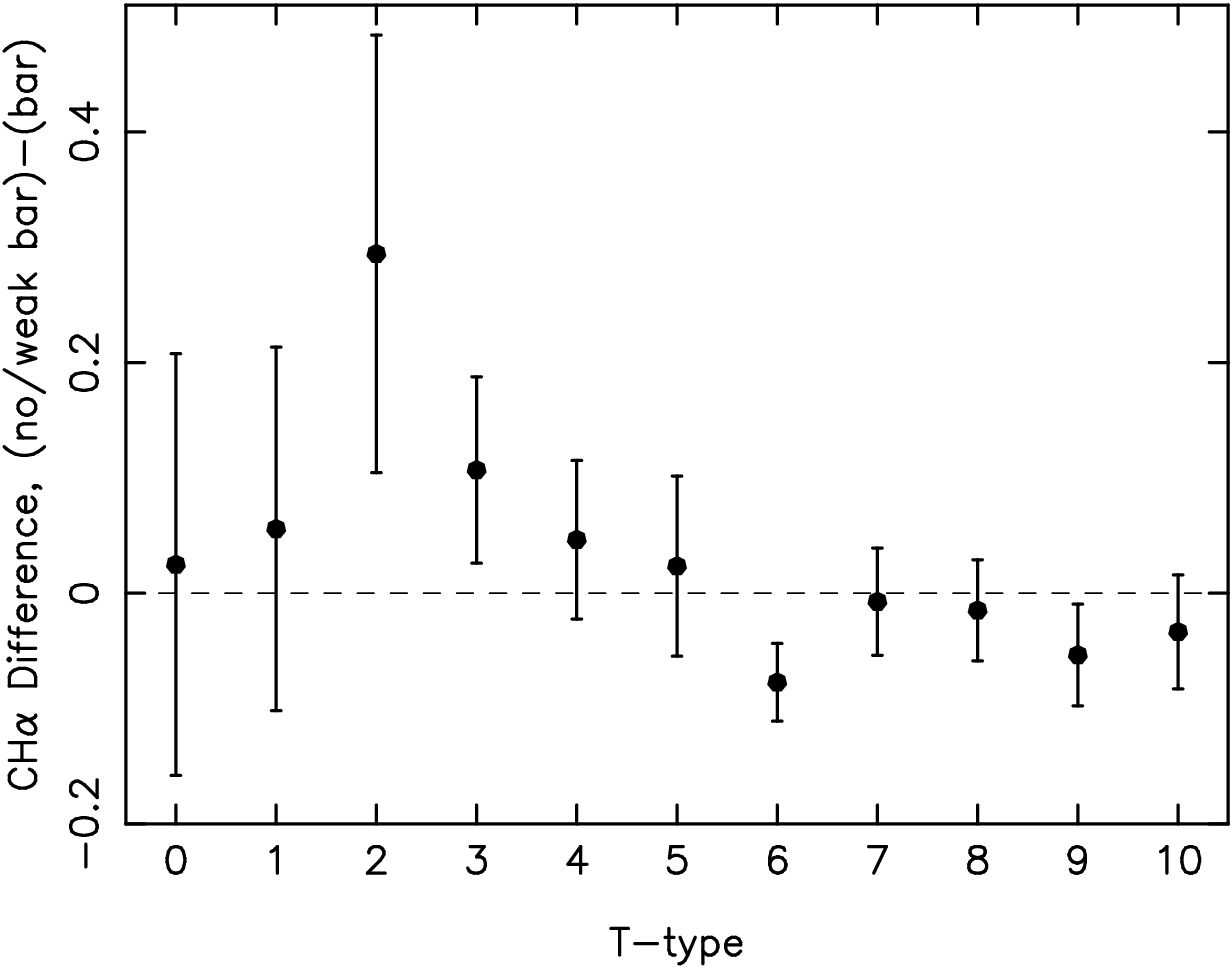}
}
\rotatebox{-90}{
\includegraphics[height=8cm]{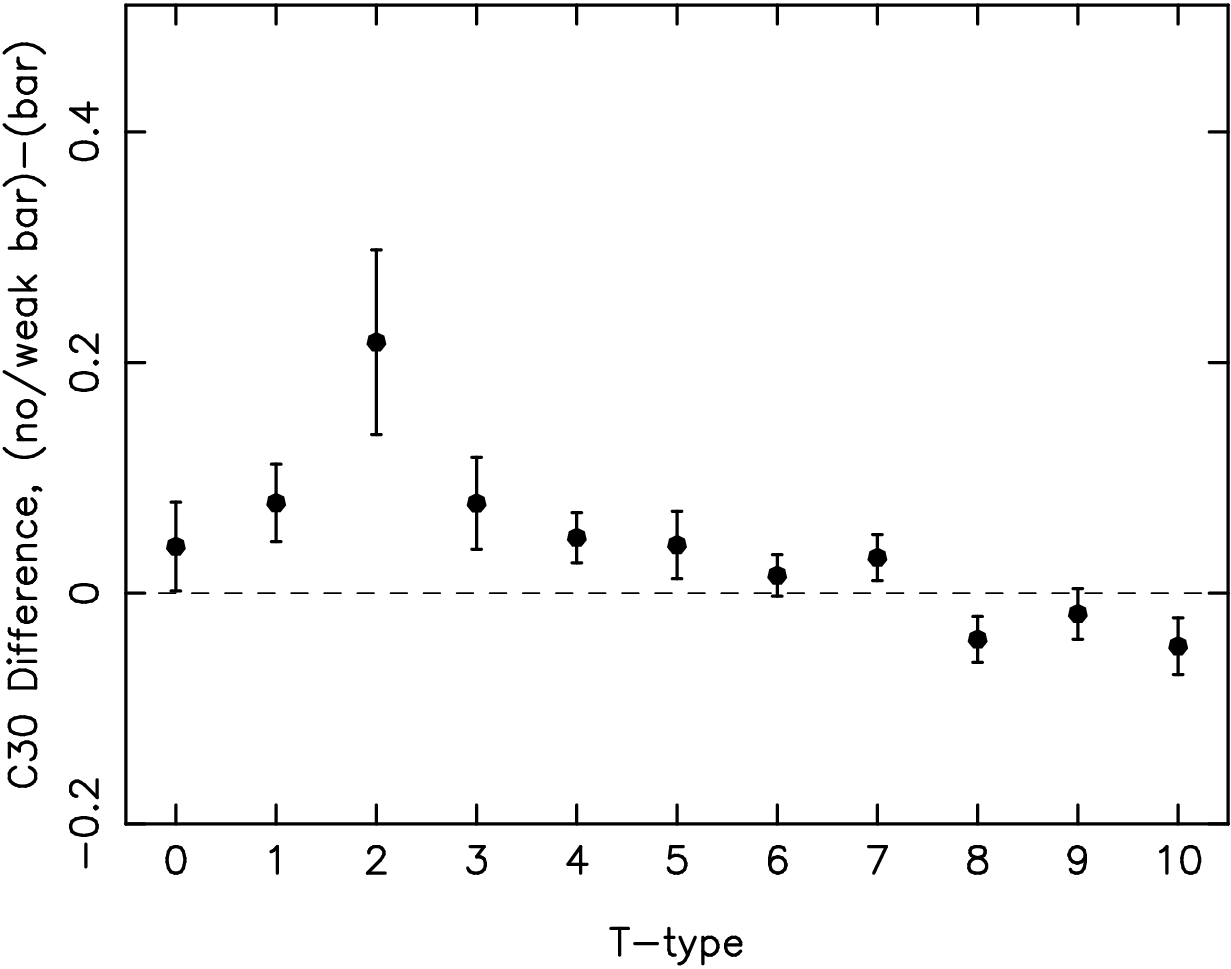}
}
\caption{The difference in mean concentration index between
  unbarred/weakly barred galaxies and strongly barred galaxies,
  plotted against Hubble type T.  For
  points lying above the line, the unbarred/weakly barred galaxies
  have more centrally concentrated emission than do their strongly
  barred counterparts.  The plot on the left shows the mean
  difference in this index for the \Ha\ light distribution, and the 
equivalent plot for $R$-band continuum light
is on the right.}
\label{fig:dc30_v_t}
\end{figure*}

Figure \ref{fig:c30_v_t} shows the mean values of the C30
concentration index for $R$-band and \Ha\ light (left and central
plots), with the right hand plot showing the difference between the
$R$-band and \Ha\ mean indices.  The $R$-band C30 index plot shows
small error bars, and hence a small scatter in the index, for each
Hubble $T$-type, both in an absolute sense and relative to the
variations between $T$-types.  This correlation of C30 with Hubble
type was investigated using two methods.  Firstly, a non-parametric
Spearman rank analysis yields a $\rho$ value of 0.98, well above the 
99\% confidence value which requires a $\rho$ of 0.79 or greater.
At the referee's suggestion, here and elsewhere we calculated an 
error-weighted fit assuming a linear model for the relation between these
parameters.  The significance of the deviations from this fit is 96\%
(4\% probability of these residuals arising by chance).
 A similar correlation was also found by \citet{koop98,koop04}
for a sample of 29 isolated galaxies of types S0 - Sc (they found
interestingly different results for Virgo cluster galaxies but that
sample is less relevant here).  \citet{koop04} report C30 values
decreasing smoothly from 0.61 - 0.72 for S0 galaxies (slightly early
than any galaxies in the present sample) to 0.24 - 0.37 for Sc types.
Thus the range they find is very similar to that of the present study,
although their Sc indices correspond more closely to those of our Sd -
Im types.

The index applied to the \Ha\ light 
distribution shows a substantially larger scatter, but an overall trend
that is still significant according the Spearman test ($\rho =$ 0.94). 
The significance of deviations from the best-fit linear trend is
again 96\%.
For most types, the $R$-band
light is more centrally concentrated in the mean than the \Ha\ light.
This is as expected given that
most star formation takes place in the disks of galaxies, whereas
bulges contain predominantly old stars.  
Types Sa and Sab are exceptions to this trend, showing marginally the
opposite result.  In this index, the latest types (Im; $T=$ 10) show
no significant difference in the mean $R$-band and \Ha\ indices.

\subsubsection{The C$_{31}$ index vs type }

Secondly we investigated the mean values of C$_{31}$ binned by galaxy
type, for both $R$-band and \Ha\ emission.  An exponential profile has
a concentration index of 0.447, whereas an $r^{1/4}$ law profile has
one of 0.845.  The mean concentration index variation as a function of
galaxy type shows a very similar pattern to the C30 index so we omit
the equivalent figure.  The bulge-dominated S0/a and Sa galaxies have
the highest values of $R$-band index, 0.64$\pm$0.04, intermediate
between the expectations for pure disk and pure $r^{1/4}$ bulge, as
would be expected.  The later type, disk-dominated, spiral galaxies
and irregulars have indices $\sim$0.44$\pm$0.02, similar to those
expected for pure exponential profiles, in accordance with the
findings of \citet{kent85}.  
The correlation between $R$-band
C$_{31}$ index and $T$-type is significant, with a Spearman $\rho$ of 0.99. 
The significance of deviations from the best-fit linear trend is 99\%.



The mean \Ha\ C$_{31}$ concentration index also shows a clear
relationship with galaxy classification, with the late-type spirals
and irregulars possessing values lower than the early types.  Again,
there is a larger scatter in \Ha\ indices than for the $R$-band.  For
the later spiral types, Sc -- Im, the \Ha\ concentration index is
lower than that of a pure exponential.  
The overall relation
between \Ha\ C$_{31}$ index and $T$-type yields a Spearman $\rho$ of 0.91
indicating high significance.  
The significance of deviations from the best-fit linear trend is 71\%.

We also investigated the difference
between $R$-band and \Ha\ C$_{31}$ indices.  In most cases this
difference is positive, implying that the continuum light is more
centrally concentrated than the \Ha.  
However, for this index, the higher concentration
of old stars compared with star formation occurs even for the latest
Hubble types, including the Magellanic irregulars, which are not
thought to contain significant bulge components.

\begin{table*}
\begin{center}
\begin{tabular}{cccccccccccc}
\hline 
\hline 
T-type & 0 & 1 & 2 & 3 & 4 & 5 & 6 & 7 & 8 & 9 & 10 \cr
\hline A & 8 & 6 & 5 & 5 & 10 & 16 & 10 & 9 & 8 & 24 & 45 \cr 
AB & 2 & 0 & 2 & 5 & 10 & 13 & 9 & 8 & 14 & 5 & 4 \cr 
B & 4 & 6 & 1 & 13 & 4 & 7 & 11 & 11 & 7 & 13 & 18 \cr 
\hline 
Total & 14 & 12 & 8 & 23 & 24 & 36 & 30 & 28 & 29 & 42 & 67 
\cr 
\hline
\end{tabular}
\caption[]{Number of galaxies of each Hubble T-type
and bar classification contributing to the mean profiles presented in
this paper}
\label{tbl:ngal}
\end{center}
\end{table*}

\subsubsection{The Petrosian index vs type }

The Petrosian index is an inverse concentration index, so a more
centrally-concentrated galaxy will have a lower value of $CI_p$.  An
exponential surface brightness profile has an index of 0.501 and
an $r^{1/4}$ law profile results in a $CI_p$ value of 0.371.
Again we find the bulge-dominated S0/a and Sa
galaxies to have light profiles that are close to the value expected for
a pure $r^{1/4}$ law,
whereas the irregulars and disk-dominated spirals have concentrations
that are closer to that predicted for a pure exponential profile.
The relation between $R$-band $CI_p$ index and $T$-type 
is found to give a Spearman $\rho$ of 0.98, again indicating strong
correlation.  
The significance of deviations from the best-fit linear trend is 99.7\%



A similar analysis is presented Fig. 10 in \citet{shim01} which shows
the Petrosian concentration indices for 426 SDSS galaxies against
their T-type classification.  The values for the \Ha GS galaxies appear to
be systematically higher than those for the SDSS galaxies.
The majority of SDSS Sa-Im galaxies have indices between 0.35 and
0.50, whereas the equivalent range for \Ha GS galaxies is between 0.44
and 0.57.  The reasons for this offset are not clear, but it should be 
noted that the \Ha GS sample contains a large proportion of low luminosity
and low surface brightness galaxies, particularly amongst the late types, 
which may be less fully represented in other samples.

The \Ha\ Petrosian concentration index again has a less clean trend as
a function of galaxy type than the $R$-band index, but the
  Spearman rank test still indicates a strong correlation,
  $\rho =$ 0.90. 
The significance of deviations from the best-fit linear trend is 92\%.
For most types, the continuum emission
appears more centrally concentrated than the \Ha\ emission.  Sab and
Sb galaxies are the exception and appear to have statistically similar
light distributions in $R$-band and \Ha\ emission.

Overall we conclude that 
the C30 index seems marginally the best proxy for morphological type
of the three investigated here.

\subsubsection{The C30 index as an indicator of bar presence}

We also carried out an analysis of the effect of bars on the mean
concentration indices.  Of the three indices discussed above, the C30
index was preferred for this analysis as a result of the small scatter
about mean values found above for this index. Figure
\ref{fig:dc30_v_t} shows the difference in the mean value of this
index between unbarred or weakly barred and strongly barred galaxies
(i.e., A and AB vs B classifications).  The left-hand plot shows the
effect of bars on the \Ha -derived index, whereas the right-hand plot
is for $R$-band emission.  The error bars show the standard error on
the mean value of the difference for each type.  For points lying
above the line, the unbarred/weakly barred galaxies have more
centrally concentrated emission than do their strongly barred
counterparts.  For the $R$-band plot, this appears to be the case for
most spiral types: all the points for $T=$ 0 - 7 have positive values,
indicating that strong bars are associated with less centrally
concentrated $R$-band luminosity.  For the later types ($T=$ 8 - 10)
any effect of bars seems to be in the opposite sense, with the
$R$-band light being somewhat more centrally concentrated than for
those galaxies with no or weak bars.  The relation between bar
  effect and galaxy $T$-type was investigated using the two
  statistical tests introduced above.  The Spearman test indicates a
  significant correlation, with the $\rho$ value being 0.82.  
The significance of deviations from the best-fit linear trend is 71\%. 
Similar trends are seen in the left-hand plot of
Fig. \ref{fig:dc30_v_t} for the distributions of \Ha\ light.  For all
$T$-types, the scatter is larger, but again the earlier type spirals
show less central concentration in the barred galaxies, with the
reverse being true for later types.  In this case the transition takes
place between types 5 and 6 (Sc and Scd), although given the size of
the error bars the differences for any individual type are not
significant. The Spearman test indicates that this correlation
  is significant ($\rho=$0.83). 
The significance of deviations from the best-fit linear trend is 38\%.

The main result from this initial analysis of barred galaxies is the
lower central concentration of both old and forming stars in
earlier-type barred spiral galaxies than non-barred.  This effect will
be studied in more detail in Sect. \ref{sec:bars}.

\section{Mean light profiles in $R$ and \Ha\ light}
\label{sec:profiles}

\begin{figure*}
\centering
\rotatebox{-90}{
\includegraphics[height=6.1cm]{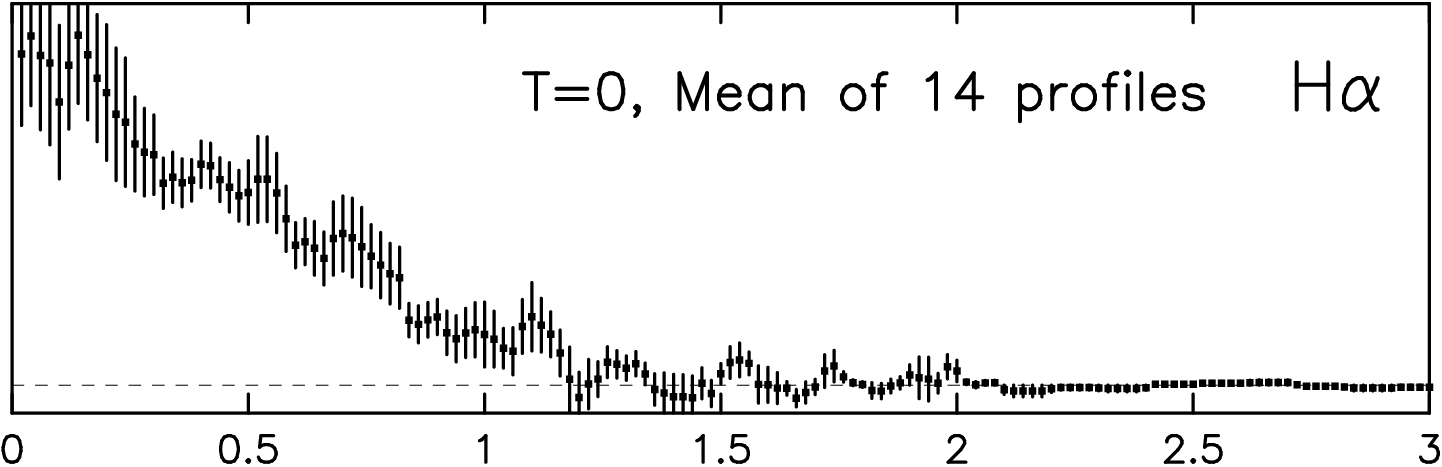}
}
\rotatebox{-90}{
\includegraphics[height=6.1cm]{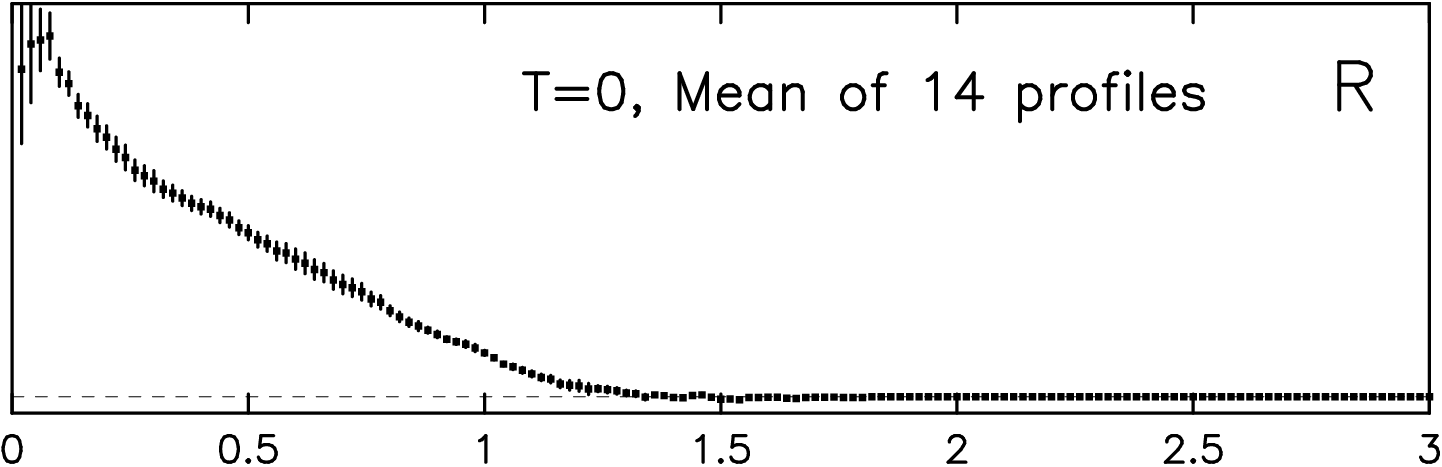}
}
\rotatebox{-90}{
\includegraphics[height=6.1cm]{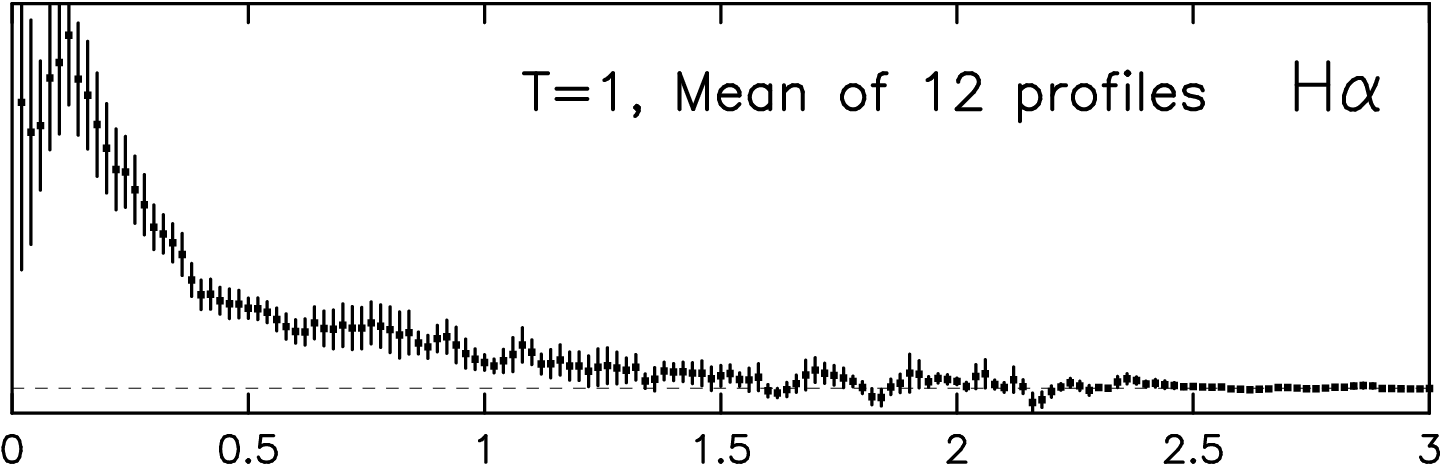}
}
\rotatebox{-90}{
\includegraphics[height=6.1cm]{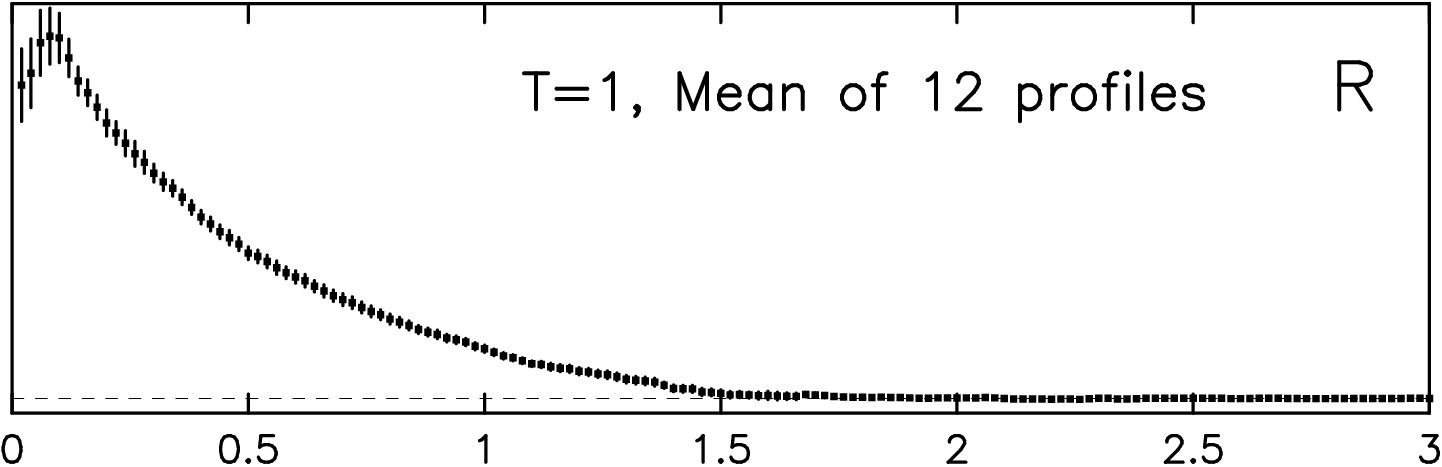}
}
\rotatebox{-90}{
\includegraphics[height=6.1cm]{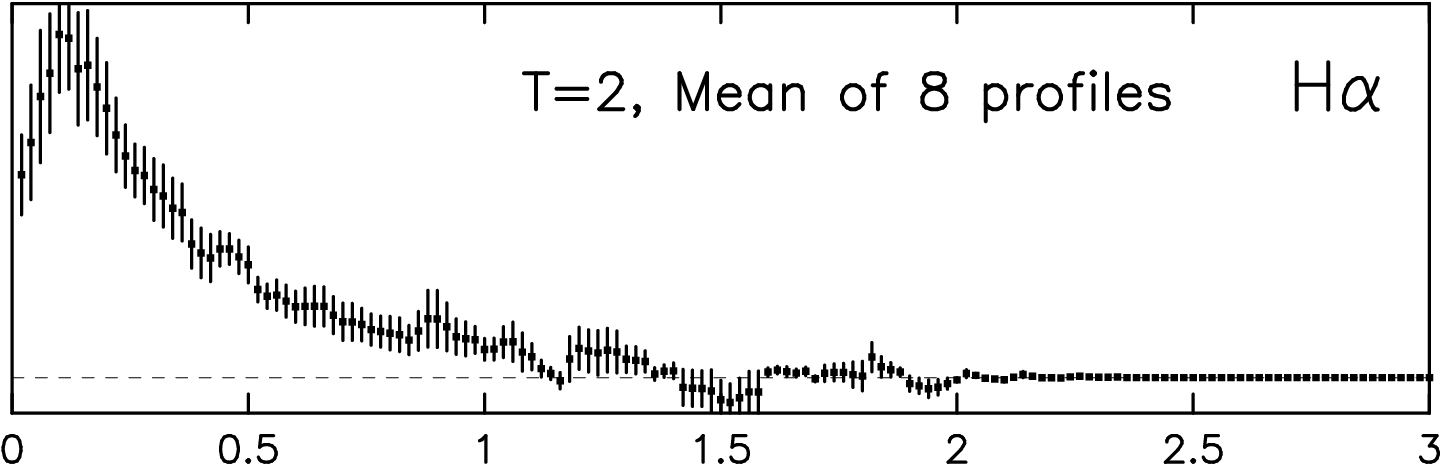}
}
\rotatebox{-90}{
\includegraphics[height=6.1cm]{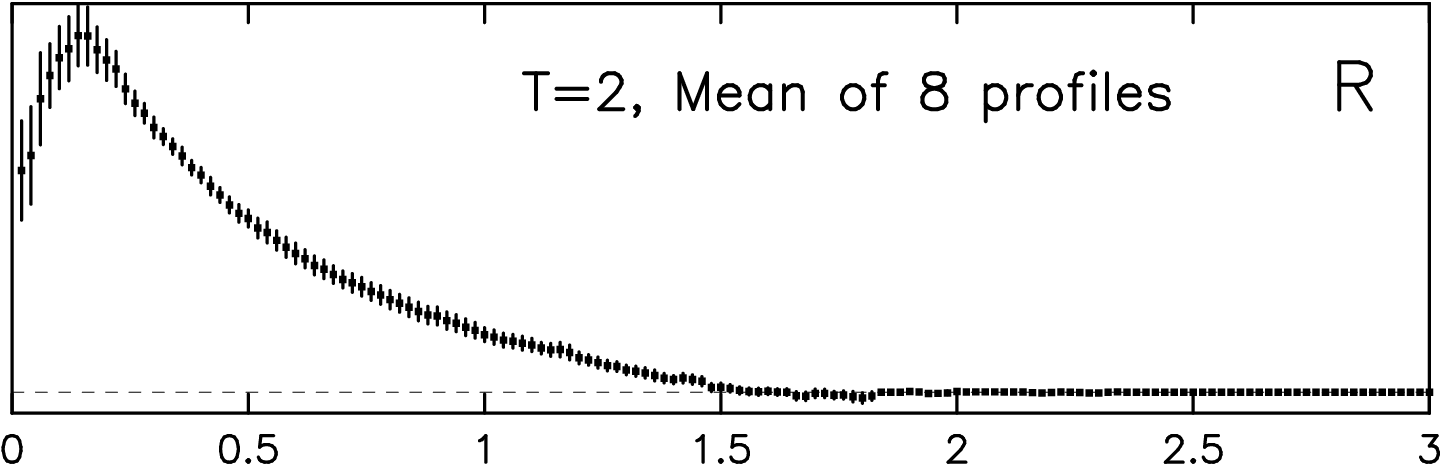}
}
\rotatebox{-90}{
\includegraphics[height=6.1cm]{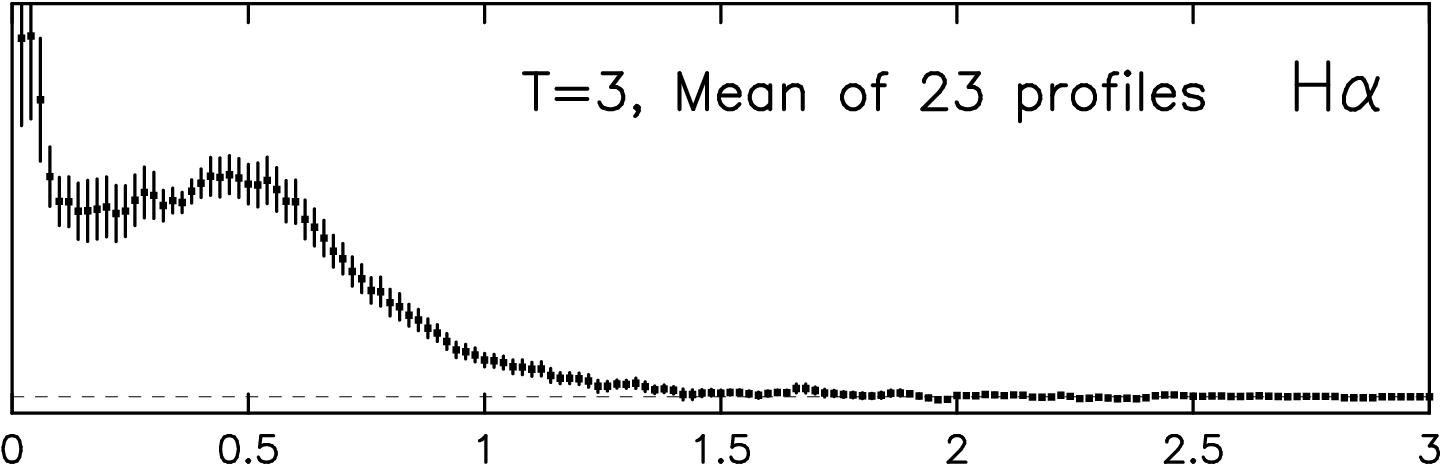}
}
\rotatebox{-90}{
\includegraphics[height=6.1cm]{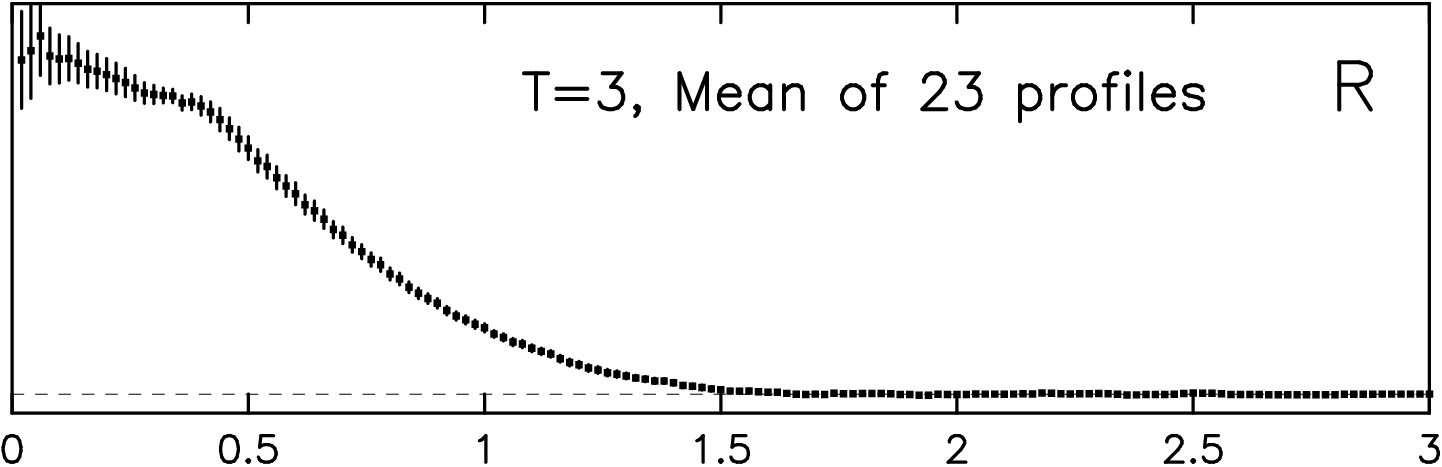}
}
\rotatebox{-90}{
\includegraphics[height=6.1cm]{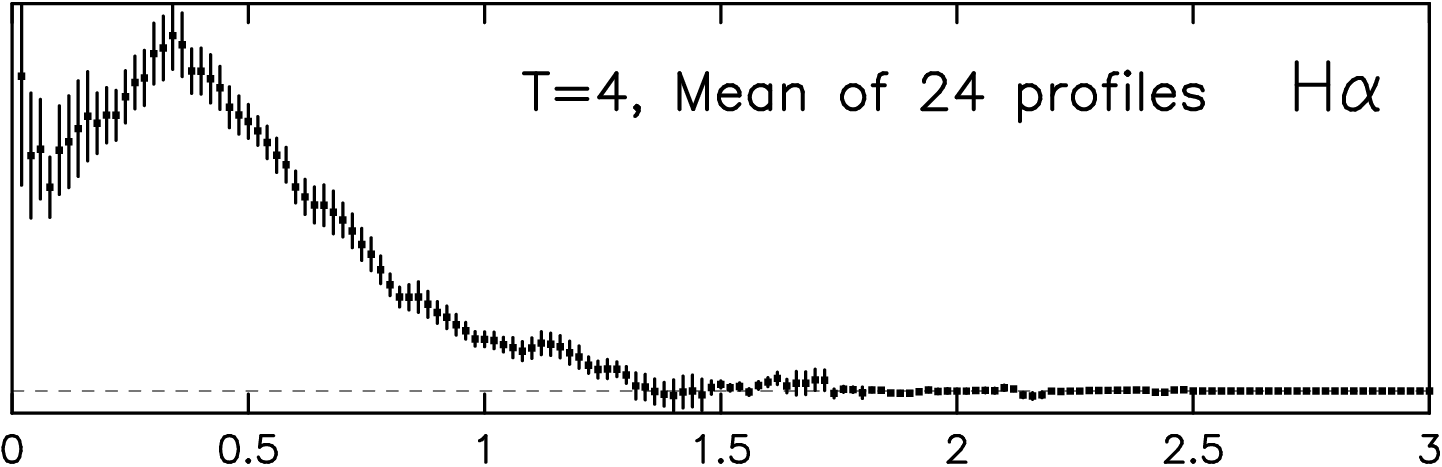}
}
\rotatebox{-90}{
\includegraphics[height=6.1cm]{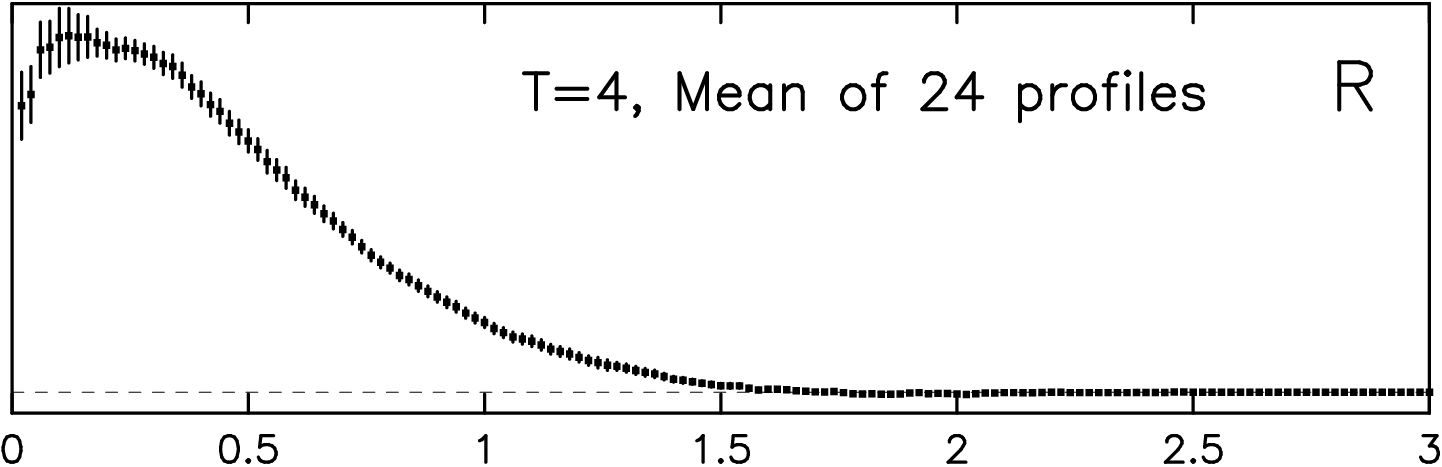}
}
\rotatebox{-90}{
\includegraphics[height=6.1cm]{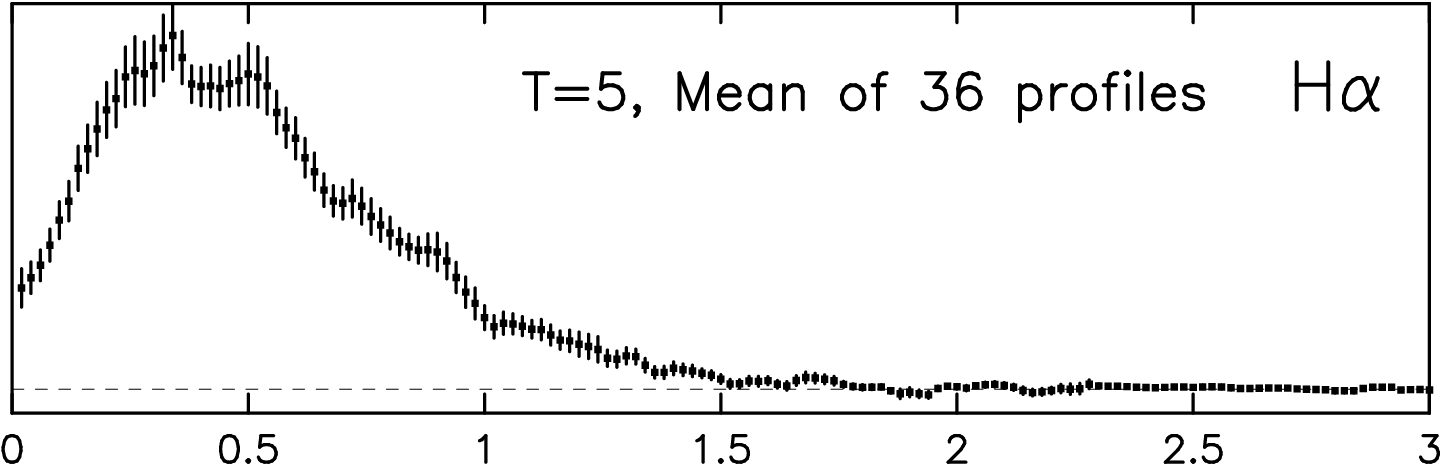}
}
\rotatebox{-90}{
\includegraphics[height=6.1cm]{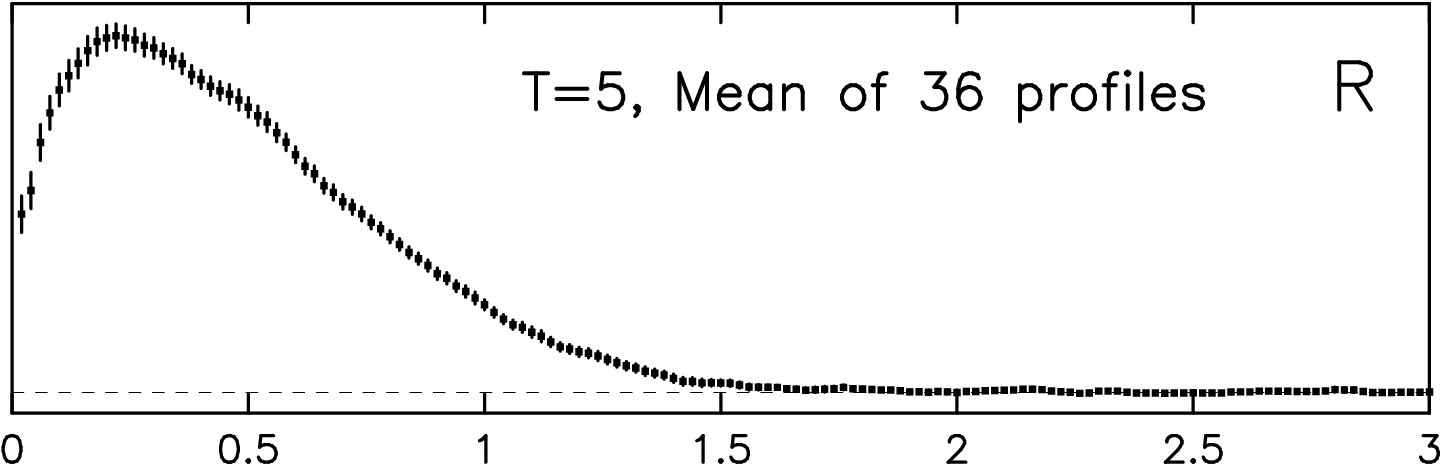}
}
\rotatebox{-90}{
\includegraphics[height=6.1cm]{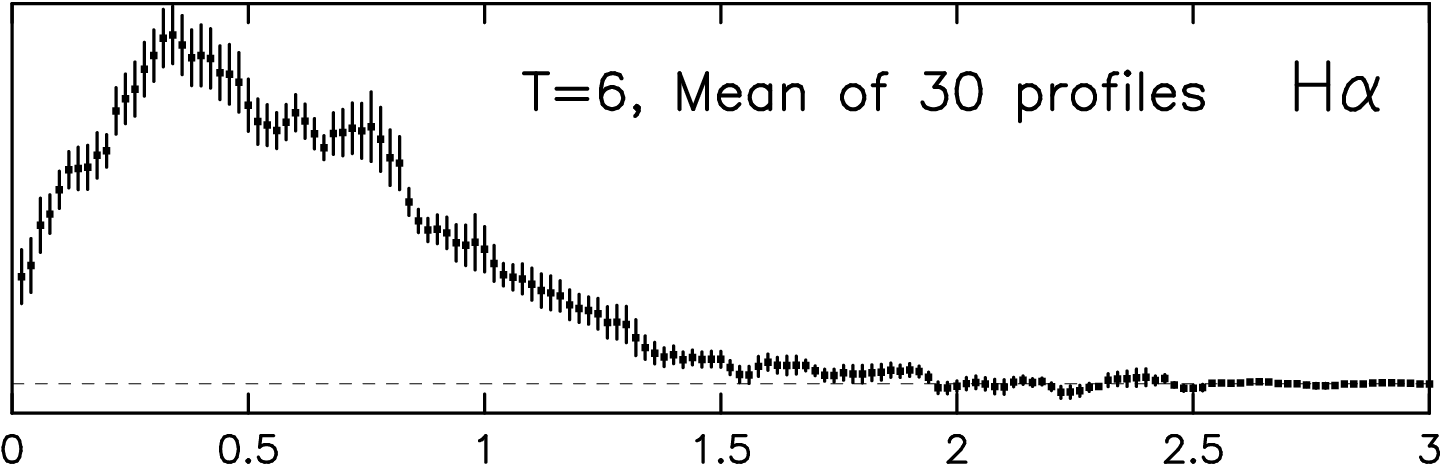}
}
\rotatebox{-90}{
\includegraphics[height=6.1cm]{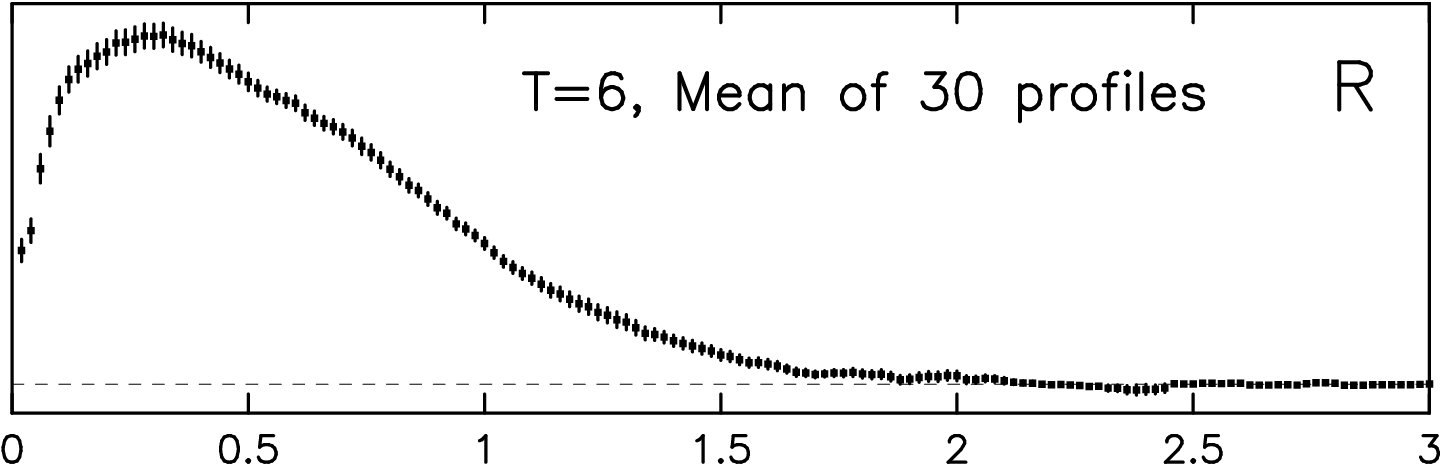}
}
\rotatebox{-90}{
\includegraphics[height=6.1cm]{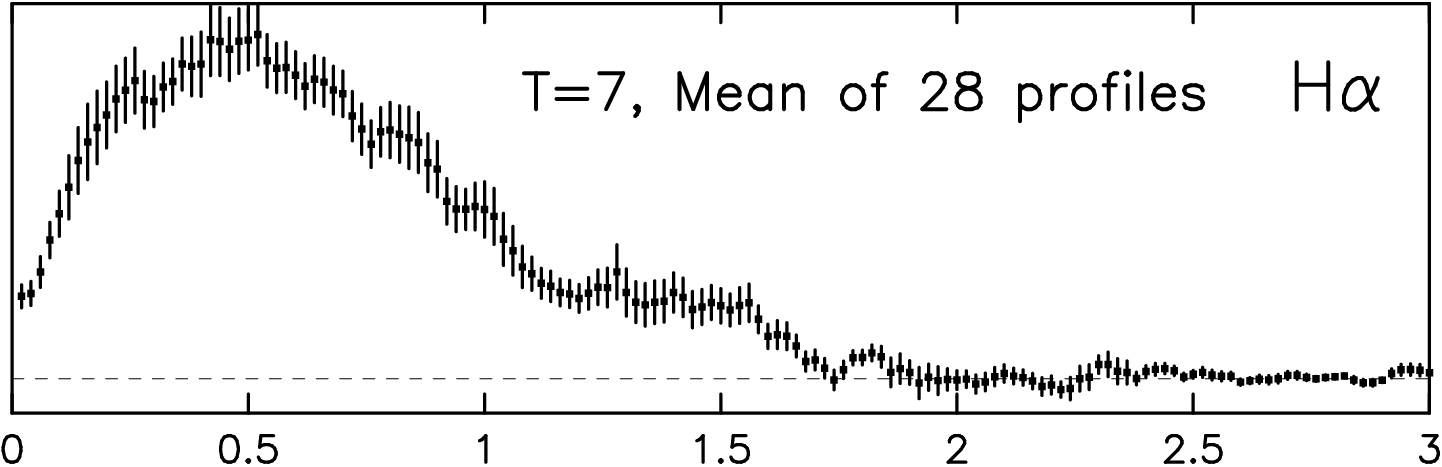}
}
\rotatebox{-90}{
\includegraphics[height=6.1cm]{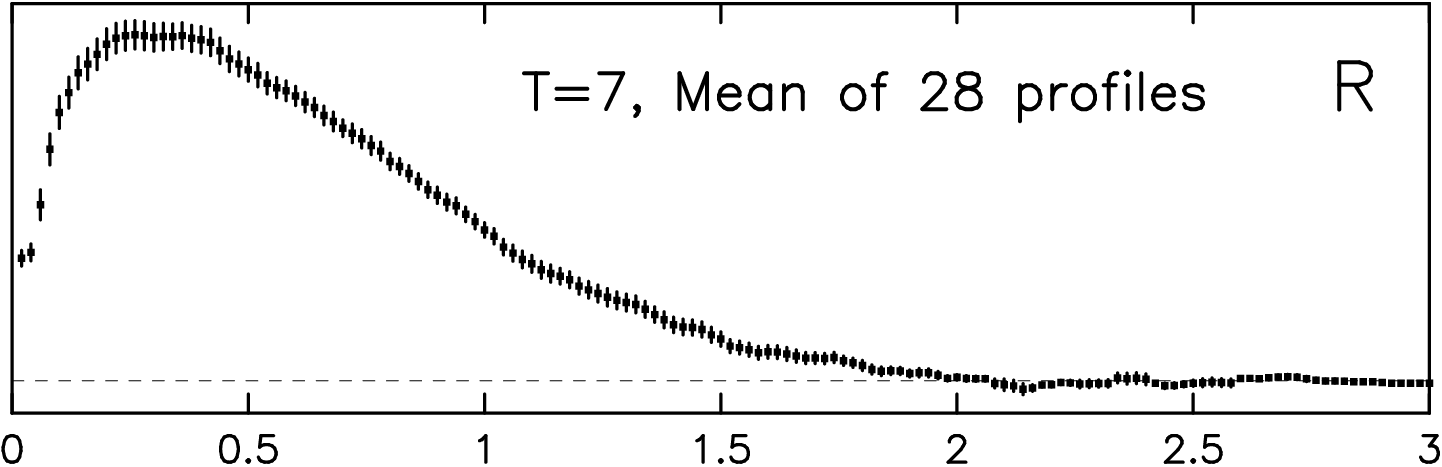}
}
\rotatebox{-90}{
\includegraphics[height=6.1cm]{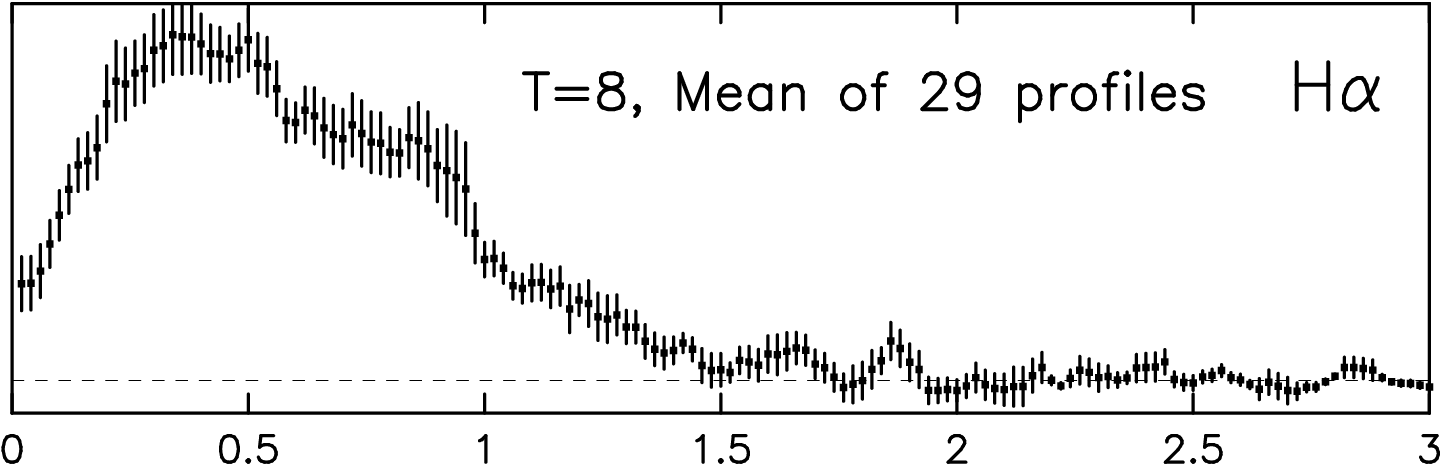}
}
\rotatebox{-90}{
\includegraphics[height=6.1cm]{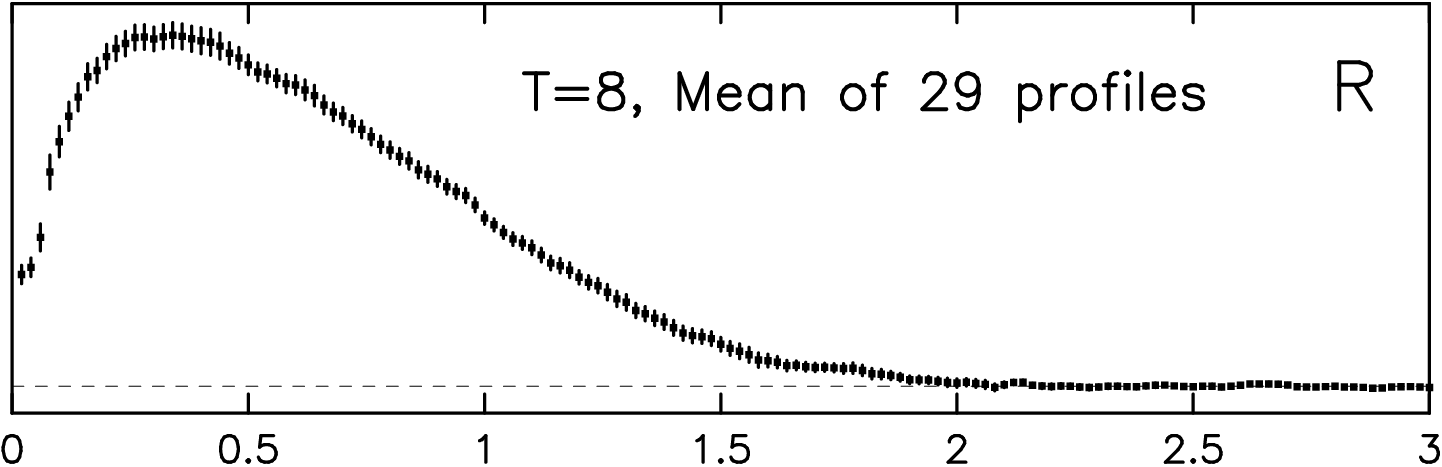}
}
\rotatebox{-90}{
\includegraphics[height=6.1cm]{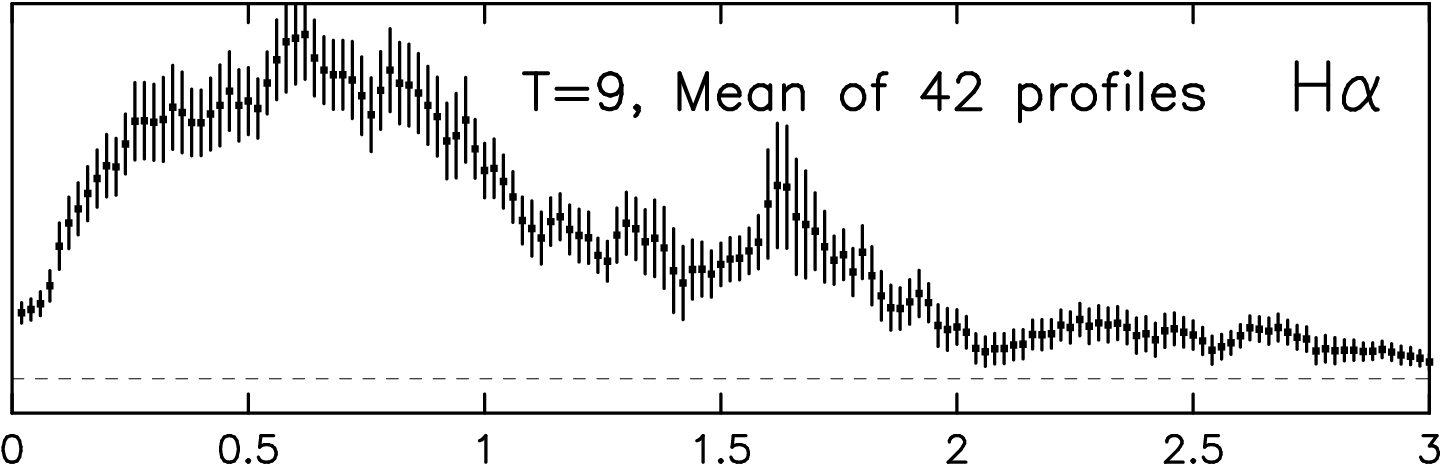}
}
\rotatebox{-90}{
\includegraphics[height=6.1cm]{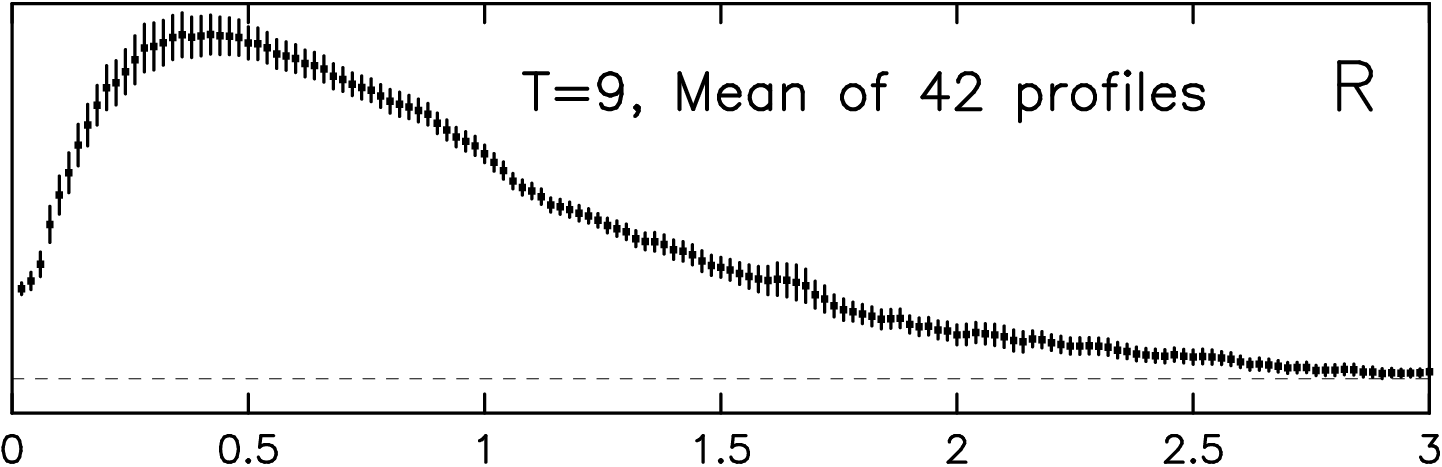}
}
\rotatebox{-90}{
\includegraphics[height=6.1cm]{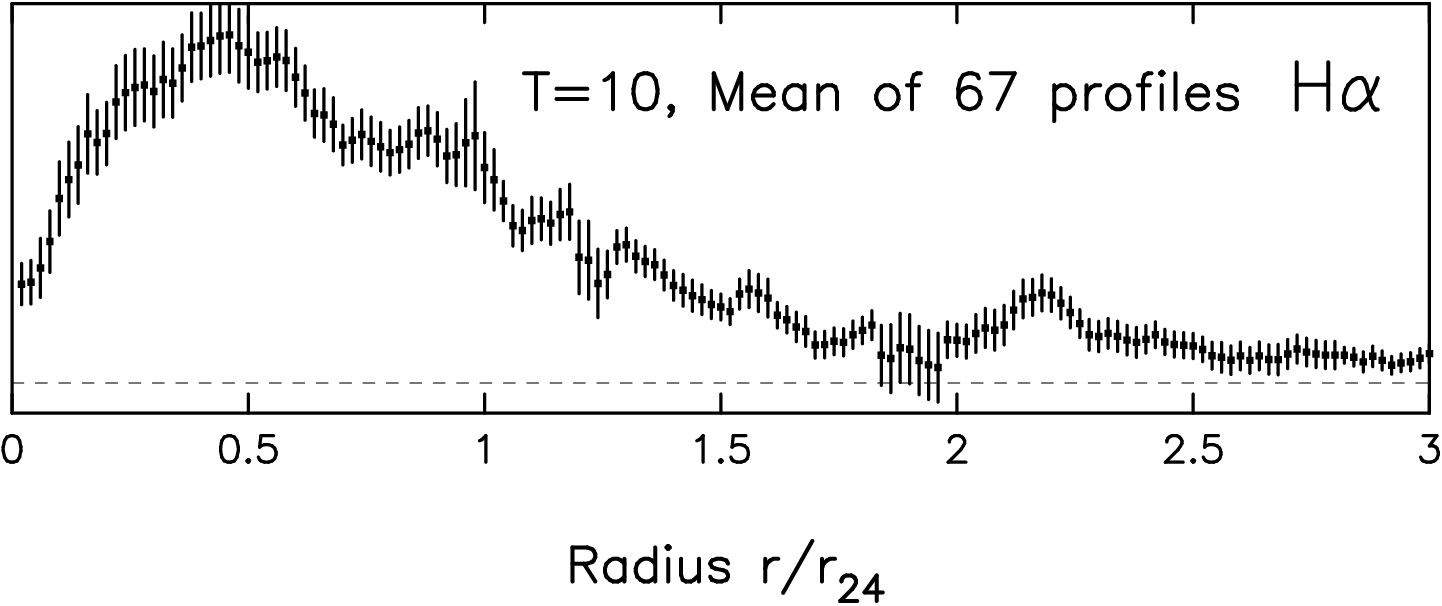}
}
\rotatebox{-90}{
\includegraphics[height=6.1cm]{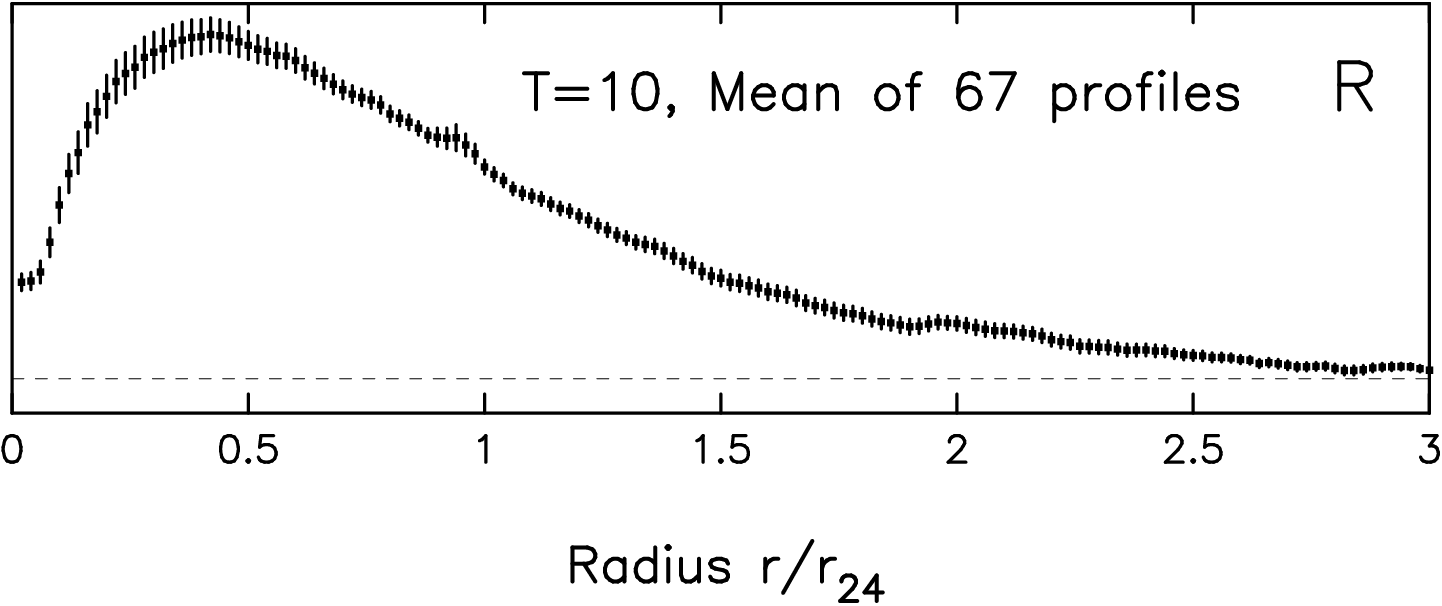}
}

\caption{Normalised mean profiles in \Ha\ (left) and $R$-band light 
(right), with types from T$=$ 0 to 10, top to bottom.}

\label{fig:meanprof}
\end{figure*}

\subsection{Calculation of mean normalised light profiles}

The concentration indices considered in the previous
section convey only a small fraction of the information on the spatial
distribution of luminosity that is contained in our images.  In this
section, we will present an analysis that attempts to extract more of
that information, whilst still enabling meaningful averages to be
taken, thus minimising the wide variations in properties that plague
studies of small numbers of galaxies.  This analysis makes use of normalised
light profiles, which we calculate for each galaxy based on the
elliptical aperture photometry that was used in the previous section to derive
concentration indices.  However, we choose not to present these in the
form of surface brightness profiles, in units of magnitudes per square
arcsec, but rather use the fluxes in the elliptical annuli,
uncorrected for the increasing area of the annulus as the apertures
grow.  Thus the area under such a profile within a given range in
radius is directly proportional to the amount of light contributed to
the total luminosity of the galaxy.  The
shape of the resulting profile is much more intuitively related to
concentration indices than is a surface brightness profile, because of
this link between flux and area under the profile.  For example, the
effective radius can be simply estimated from such a plot as the
radius which evenly divides the area under the profile.

It is necessary to normalise such profiles before combining them into,
for example, a mean profile for all galaxies of a given type.  A
simple average without normalisation will be dominated by the
brightest galaxies; and if a radial scale in kpc is adopted, then all
galaxies will contribute to the centre of the mean profile, but only
the largest galaxies to the outer regions, giving a distorted result
that is hard to interpret.  Thus we normalise the individual profiles
obtained from each galaxy in two ways.  Firstly, the radial scale is
expressed in units of the $r_{24}$ isophotal
semi-major axis of the galaxy.  Secondly, the area under each profile
is normalised to unity, giving each galaxy, bright or faint, equal
weighting in the mean profile.  This process was applied first to the
$R$-band image of each galaxy, and then to the \Ha\ image using
exactly the same positions, shapes and size of apertures, and scaling
again by the $R$-band isophotal size.  As a result of the normalisation,
the profiles have no calibration in flux or magnitude units on the
vertical axis; the profiles are simple shape functions, indicating
the fraction of light residing in a given radial range.

In the remainder of this section and in the next, we look at the mean 
profiles produced by this process, and examine the dependence of profile
shape on galaxy type, and on the presence or absence of bars.  The
mean profiles for the individual $T$-types contain between 8 and 67
profiles, as listed in Table \ref{tbl:ngal}, which also subdivides the 
numbers for each type according to bar classification. Some galaxies from 
the \Ha GS sample were omitted from this analysis due to contamination
by foreground stars or problems with varying background levels across
images.  Hence the total number of galaxies listed in Table \ref{tbl:ngal}
is 313, out of the full \Ha GS sample of 327.

\subsection{Mean profiles as a function of Hubble type}

The mean profiles for all galaxies as a function of Hubble $T$-type
are shown in Fig. \ref{fig:meanprof}, with the \Ha\ profiles on the
left and the $R$-band profiles on the right.  The number of galaxies
contributing to each profile is indicated; it is important to note
that all 313 profiles of sufficient cosmetic quality have
been included, so there has been no exclusion of outliers with unusual
profile shapes. Error bars indicate the standard error of the
individual scaled profiles about the mean at each radial point.  The
errors on the \Ha\ profiles are significantly larger than those on the
$R$-band profiles, originating from the clumpier nature of the
\Ha\ emission.

\subsubsection{$R$-band mean profiles}

\begin{table}
\begin{center}
\begin{tabular}{cccccc}
\hline \hline $T$-type & \Ha\ $r_{peak}$  & $R~r_{peak}$ & 
\Ha\ $r_{eff}$ & $R~r_{eff}$ & $<r_{24}>$\cr
\hline 
0  & 0.04 & 0.08 & 0.35 & 0.33 &  7.0 \cr 
1  & 0.12 & 0.08 & 0.26 & 0.31 & 10.5 \cr 
2  & 0.10 & 0.15 & 0.26 & 0.33 &  9.1 \cr 
3  & 0.04 & 0.06 & 0.41 & 0.37 &  9.0 \cr 
4  & 0.34 & 0.12 & 0.39 & 0.37 & 10.6 \cr 
5  & 0.34 & 0.22 & 0.47 & 0.41 &  8.2 \cr 
6  & 0.34 & 0.32 & 0.53 & 0.50 &  7.2 \cr 
7  & 0.52 & 0.26 & 0.59 & 0.53 &  5.8 \cr 
8  & 0.34 & 0.34 & 0.54 & 0.55 &  4.9 \cr 
9  & 0.62 & 0.42 & 0.86 & 0.73 &  3.0 \cr 
10 & 0.46 & 0.42 & 0.75 & 0.73 &  1.8 \cr 
\hline 
\end{tabular}
\caption[]{Peak and effective radii of the \Ha\ and $R$-band mean profiles,
in units of the $R_{24}$ isophotal radius; and the mean value of this
radius, in kpc, all listed as a function of $T$-type.}
\label{tbl:reff}
\end{center}
\end{table}

\begin{figure}
\centering
\rotatebox{-90}{
\includegraphics[height=8.2cm]{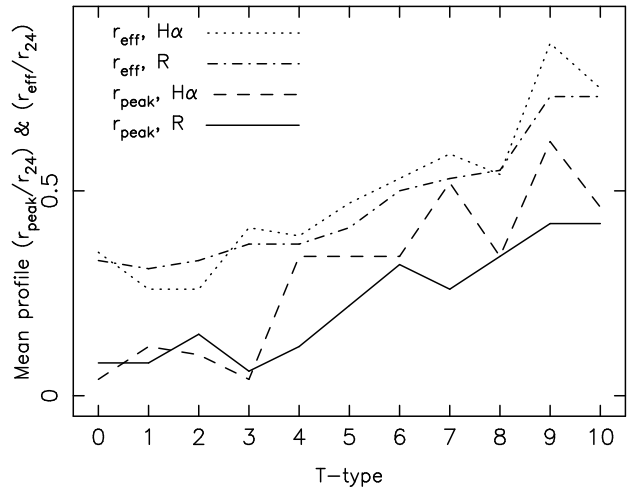}
}

\caption{Peak and effective radii of the \Ha\ and $R$-band mean
profiles, in units of the $R_{24}$ isophotal radius.}

\label{fig:reff}
\end{figure}

The main point to note from the $R$-band profiles in
Fig. \ref{fig:meanprof} is the degree of correlation between mean
profile shape and Hubble type, which is quite gratifying given the
subjective nature of galaxy classification.  The mean profiles of the
earliest types ($T$-types 0 and 1) are very centrally concentrated,
with the peak of the profiles, and hence the largest contribution to
the total galaxy luminosity, coming from $\sim$0.1 $r_{24}$.  The one
anomaly in an otherwise smooth sequence of $R$-band profiles is for
the Sb galaxies ($T=$ 3); the additional light in this mean profile
close to the centre and at 0.4 $r_{24}$ will be discussed in
Sect. \ref{sec:bars}.  For later types, the emission systematically
moves outwards and broadens, with the peak sitting at $\sim$0.4
$r_{24}$ for the Sm and Im galaxies ($T$-types 9 and 10).  This smooth
progression in profile properties is shown in Table \ref{tbl:reff},
which lists the peak and effective radii for the $R$-band and
\Ha\ mean profiles for each $T$-type, in units of $r_{24}$ (the mean
value of $r_{24}$ for each type is also given in kpc in the final
column, to enable approximate conversion of these values to physical
units, although it should be noted that there is a large range of
galaxy sizes present at each type). The same data are plotted in
Fig. \ref{fig:reff}.  Sizes in kpc are calculated using an
  effective Hubble constant of 75 km~s$^{-1}$~Mpc$^{-1}$, corrected
  for Virgo infall as explained in Paper I.

All four parameters plotted in Fig. \ref{fig:reff} are strongly
correlated with $T$-type, with Spearman $\rho$ values between 0.83 and
0.95.
Deviations from best-fit linear models have significances
of 64 - 94\%.

\subsubsection{\Ha\ and (\Ha\ -- $R$) difference profiles}

\begin{figure*}
\centering
\rotatebox{-90}{
\includegraphics[height=6.2cm]{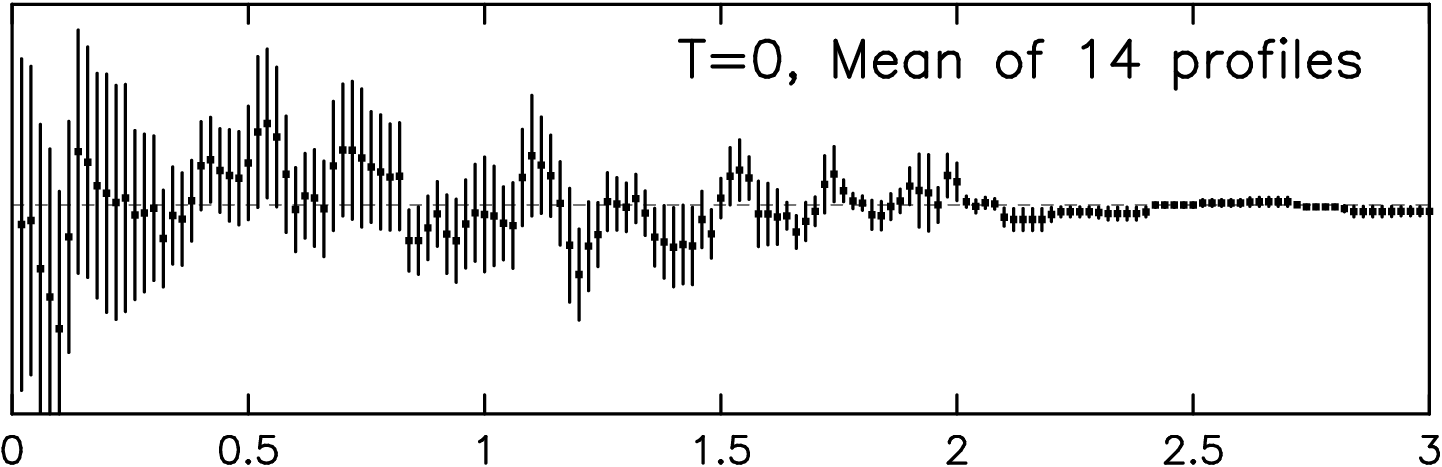}
}
\rotatebox{-90}{
\includegraphics[height=6.2cm]{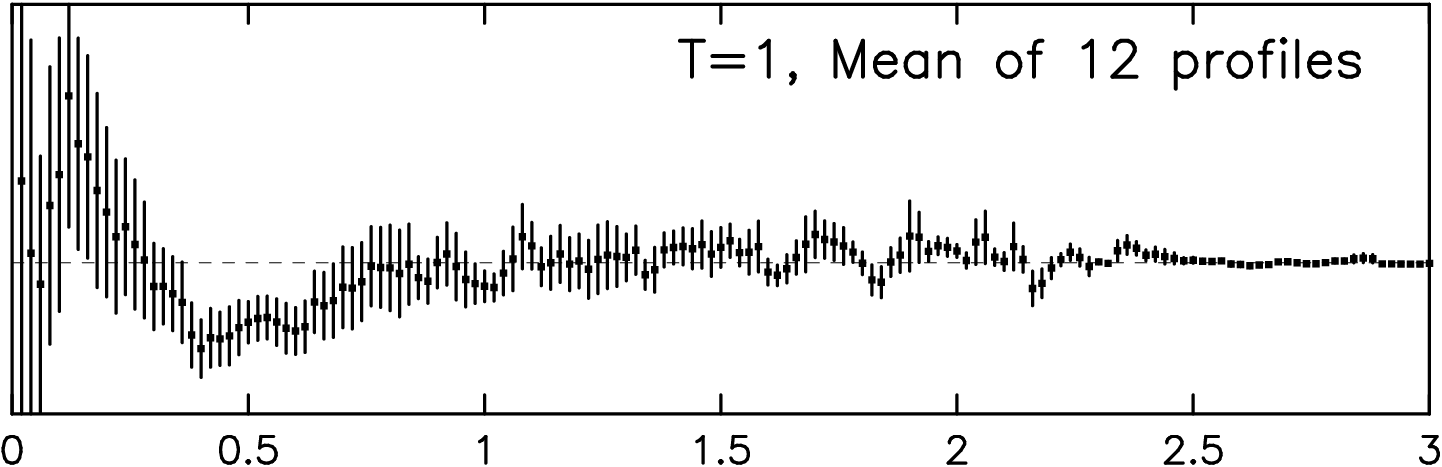}
}
\rotatebox{-90}{
\includegraphics[height=6.2cm]{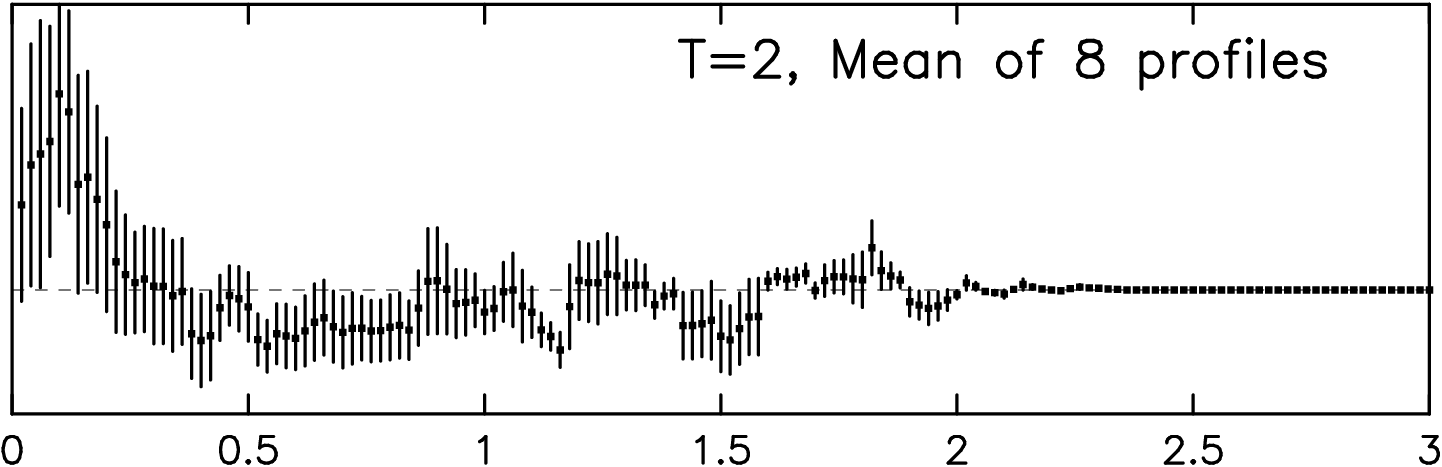}
}
\rotatebox{-90}{
\includegraphics[height=6.2cm]{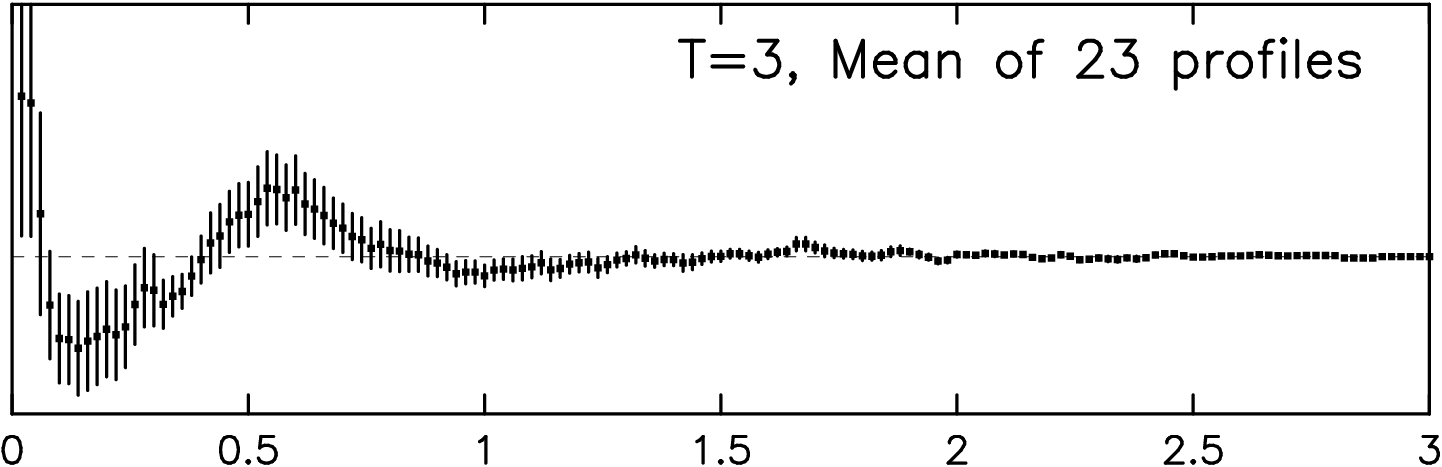}
}
\rotatebox{-90}{
\includegraphics[height=6.2cm]{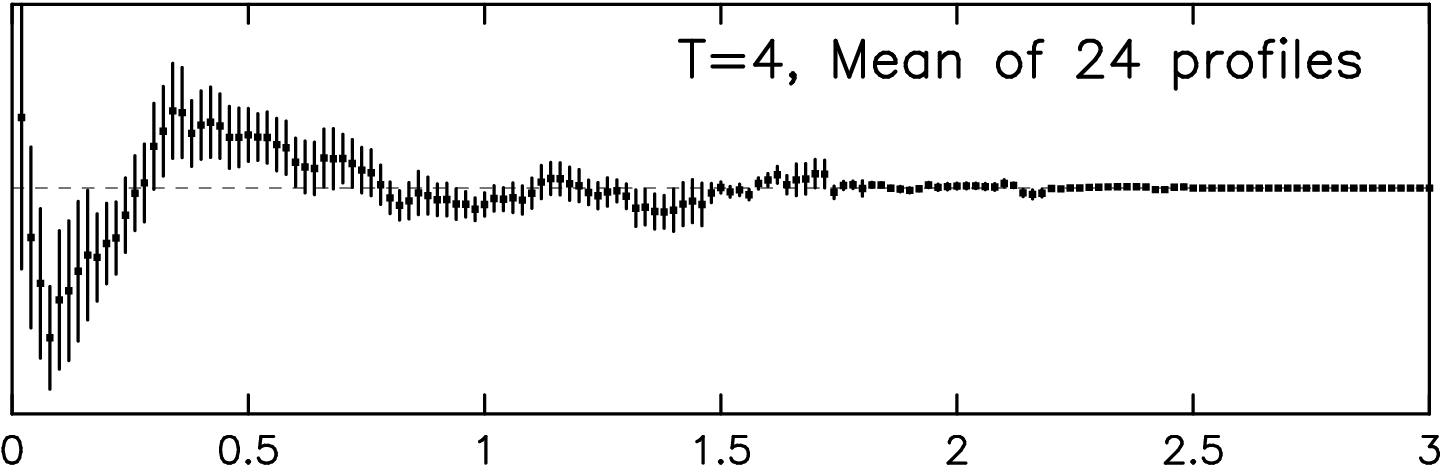}
}
\rotatebox{-90}{
\includegraphics[height=6.2cm]{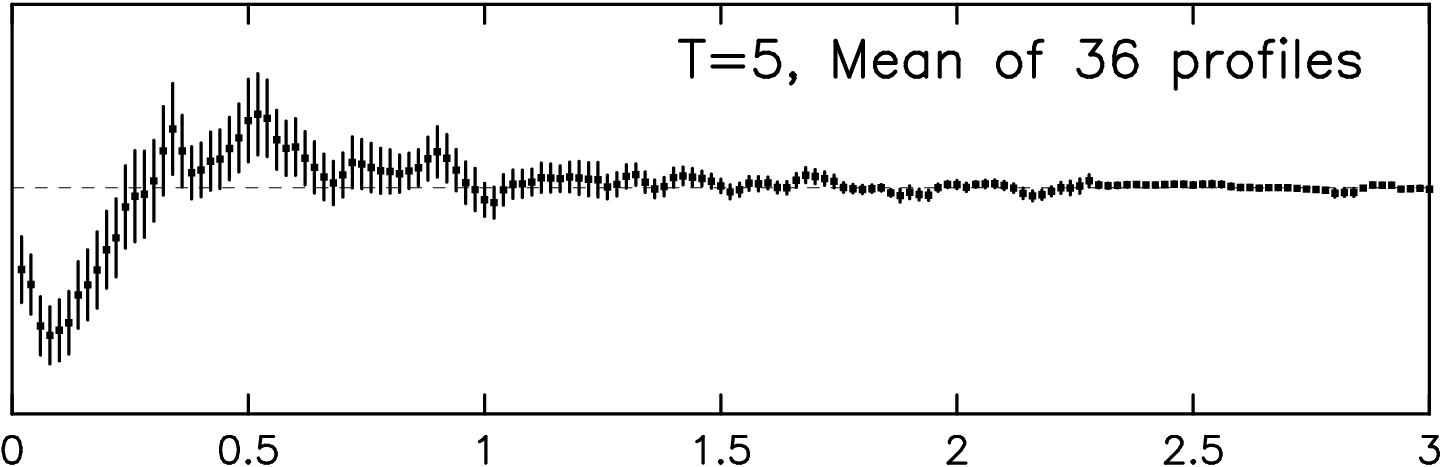}
}
\rotatebox{-90}{
\includegraphics[height=6.2cm]{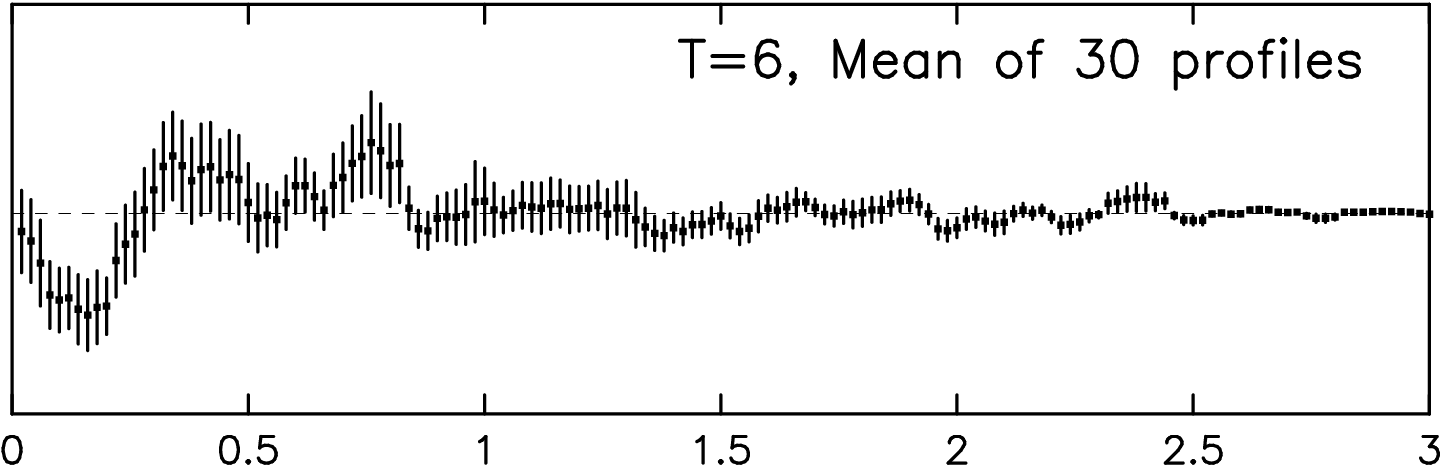}
}
\rotatebox{-90}{
\includegraphics[height=6.2cm]{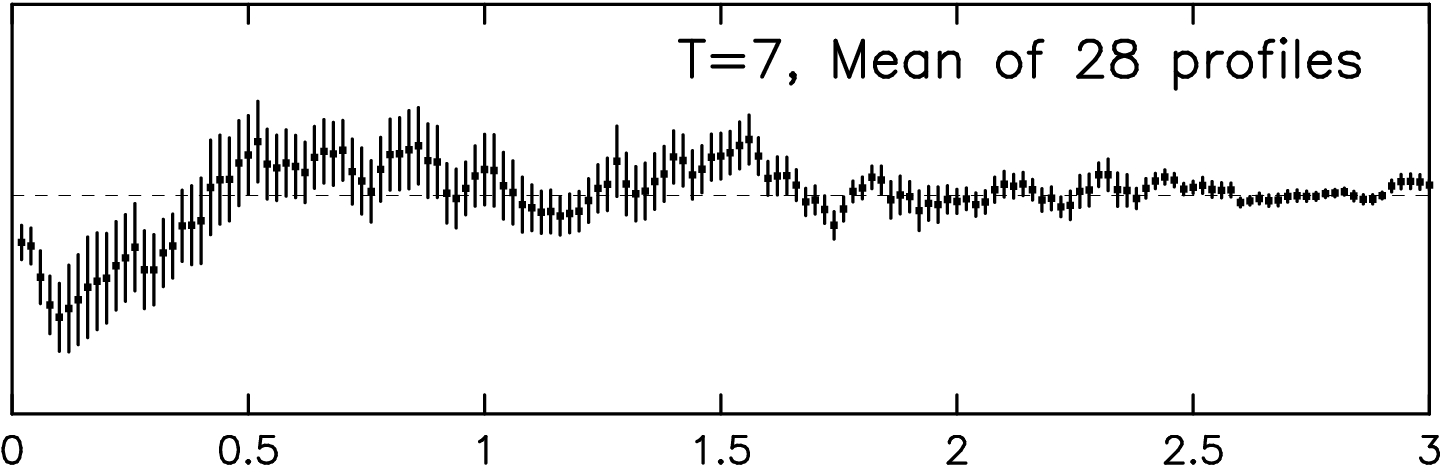}
}
\rotatebox{-90}{
\includegraphics[height=6.2cm]{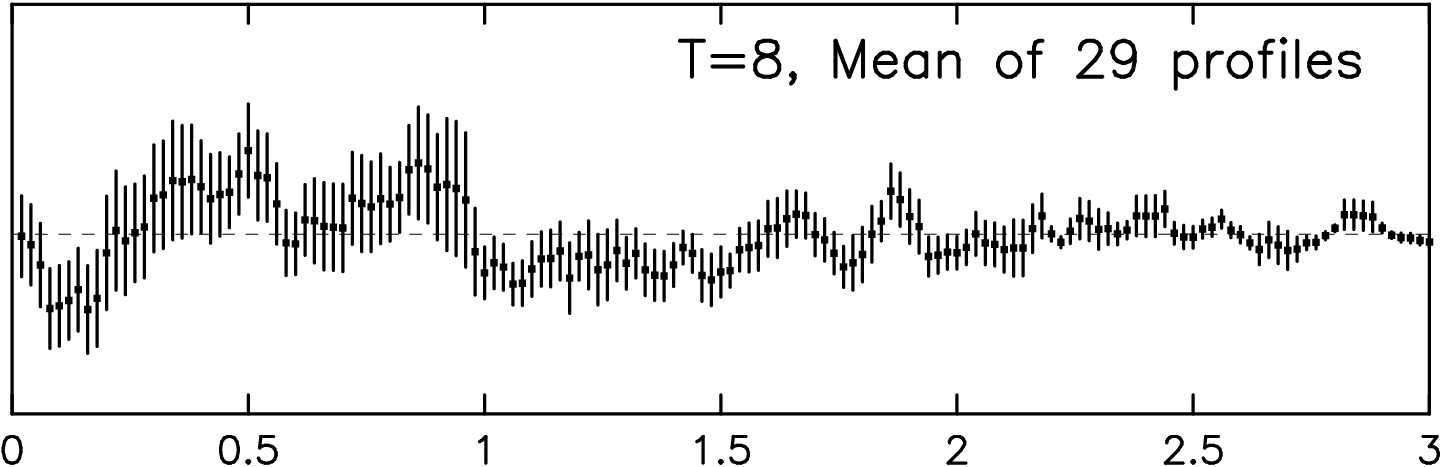}
}
\rotatebox{-90}{
\includegraphics[height=6.2cm]{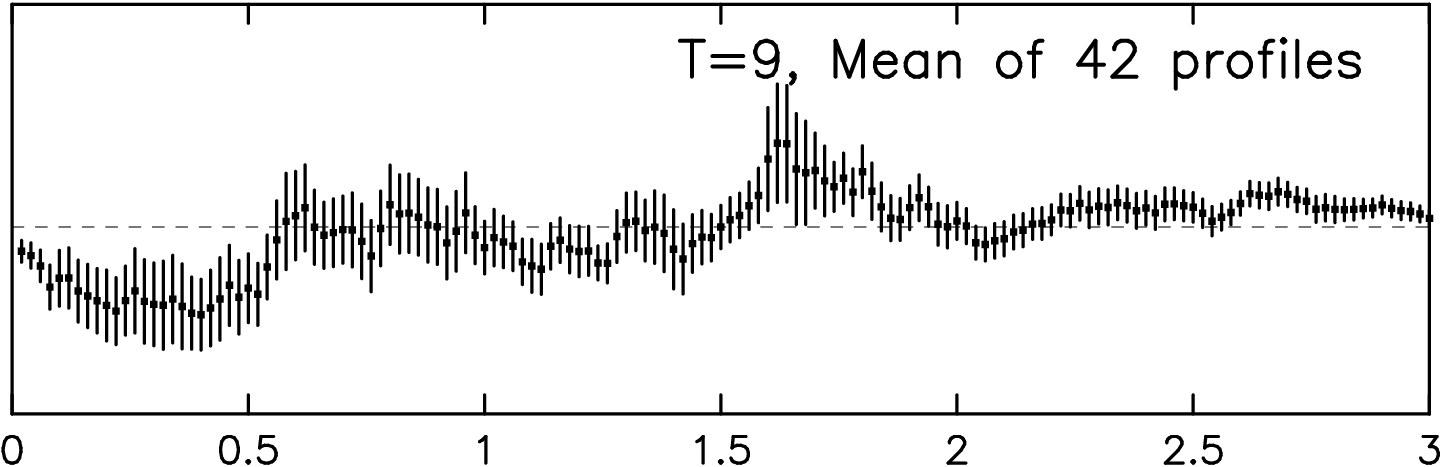}
}
\rotatebox{-90}{
\includegraphics[height=6.2cm]{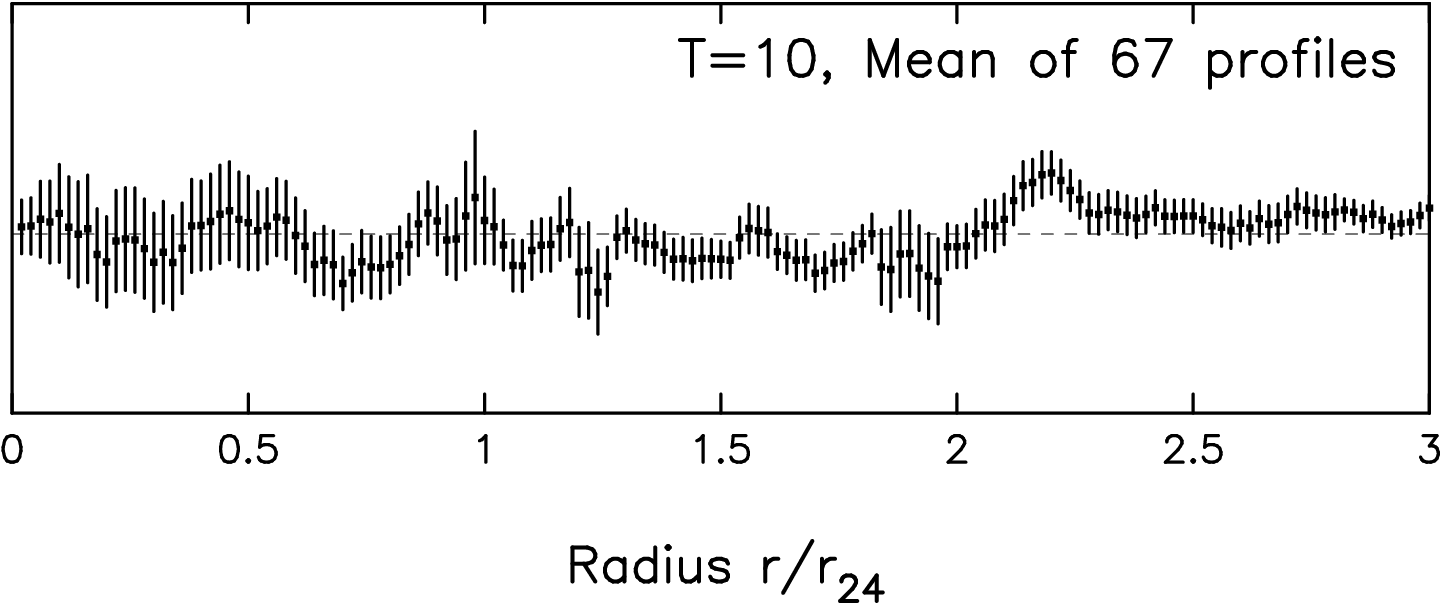}
}

\caption{Difference profiles, \Ha\ $-R$, for all T-types}
\label{fig:diffprof}
\end{figure*}

The mean \Ha\ profiles are predictably less smooth than those for the
$R$-band emission, but overall show a very similar trend of decreasing
central concentration from early to late types.  In order to emphasise
any differences between the \Ha\ and $R$-band mean profiles, we
constructed difference profiles, in the sense \Ha\ -- $R$, from those
shown in Fig. \ref{fig:meanprof}; the results are shown in
Fig. \ref{fig:diffprof}.  For the late-type spirals and Magellanic
irregulars, these plots have the expected form and are simple to
interpret: late-type spirals ($T$ between 4 and 9 inclusive) show
statistically similar distributions of $R$-band and \Ha\ emission
outside $\sim$0.5~$r_{24}$, but within this radius the relative
strength \Ha\ emission dips, presumably due to the influence of
predominantly old stellar populations associated with
central bulge or bar components.  This `bulge dip' in the
\Ha\ emission appears to be the explanation for the lower
\Ha\ concentration indices found for $T$-types 3 - 9 in the right hand
plot of Fig. \ref{fig:c30_v_t}.

For the Magellanic irregulars ($T=$ 10), the star formation traces the
$R$-band light at all radii, and there is no trace of a central older
population (previously described in Paper V); this is clear evidence
that these are bulge-free galaxies.  It is important to note here the
need to average over many tens of galaxies for this agreement to
emerge. Individual \Ha\ profiles of individual Im galaxies are very
broken and spikey, since there are typically only a few \HII\ regions
per galaxy, so a statistical approach is needed to reach this
conclusion.  It is somewhat surprising that the central suppression of
SF is clearly seen in the latest spiral types, even the $T=9$ Sm
galaxies; these might also have been thought to be `bulge-free' prior
to this analysis.

\begin{figure*}
\centering
\rotatebox{-90}{
\includegraphics[height=6.2cm]{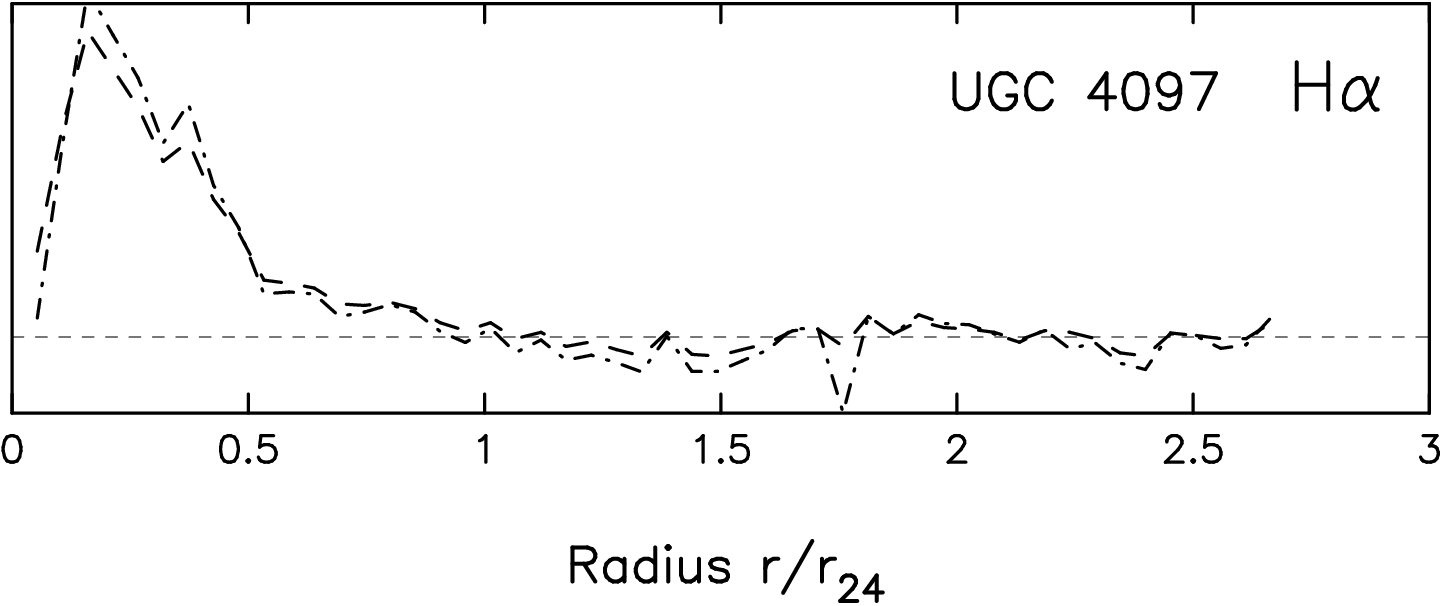}
}
\rotatebox{-90}{
\includegraphics[height=6.2cm]{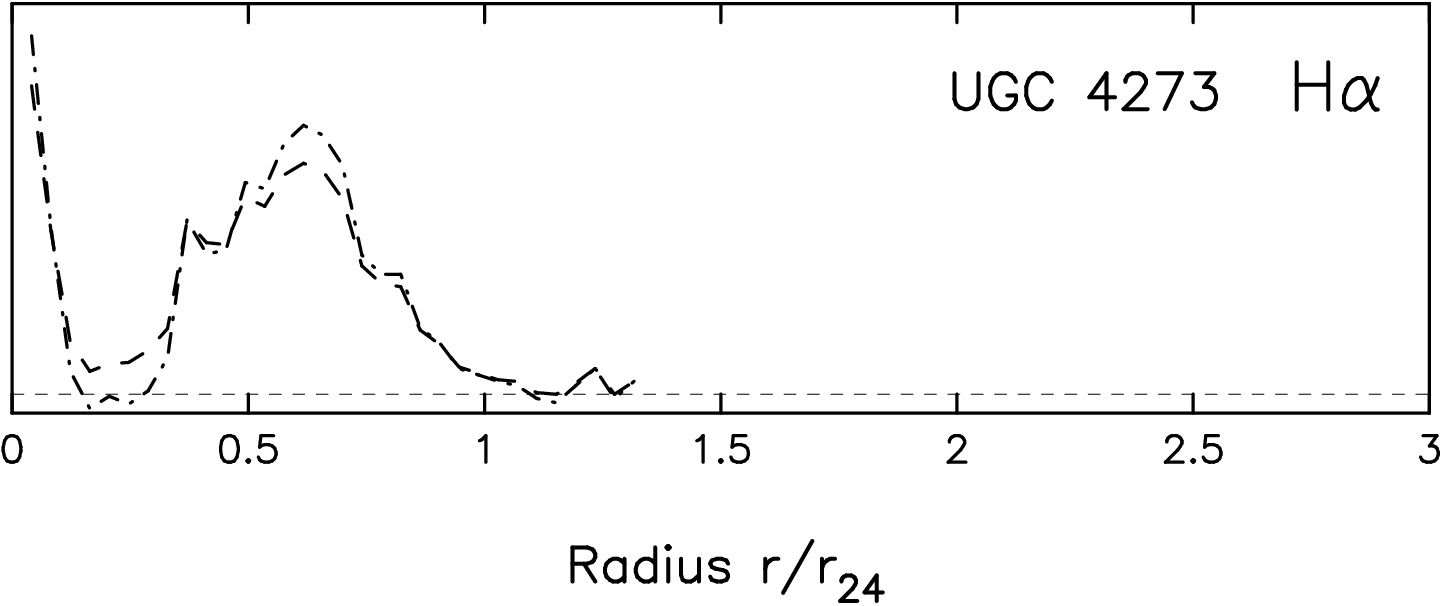}
}
\rotatebox{-90}{
\includegraphics[height=6.2cm]{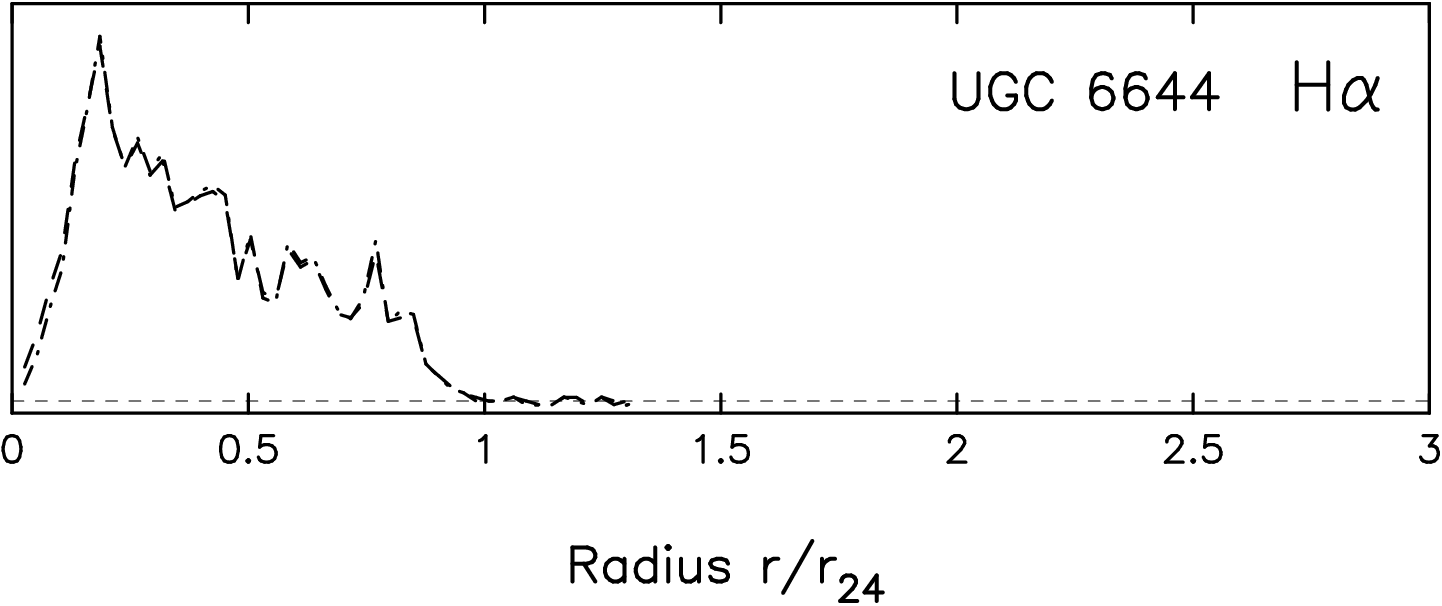}
}
\rotatebox{-90}{
\includegraphics[height=6.2cm]{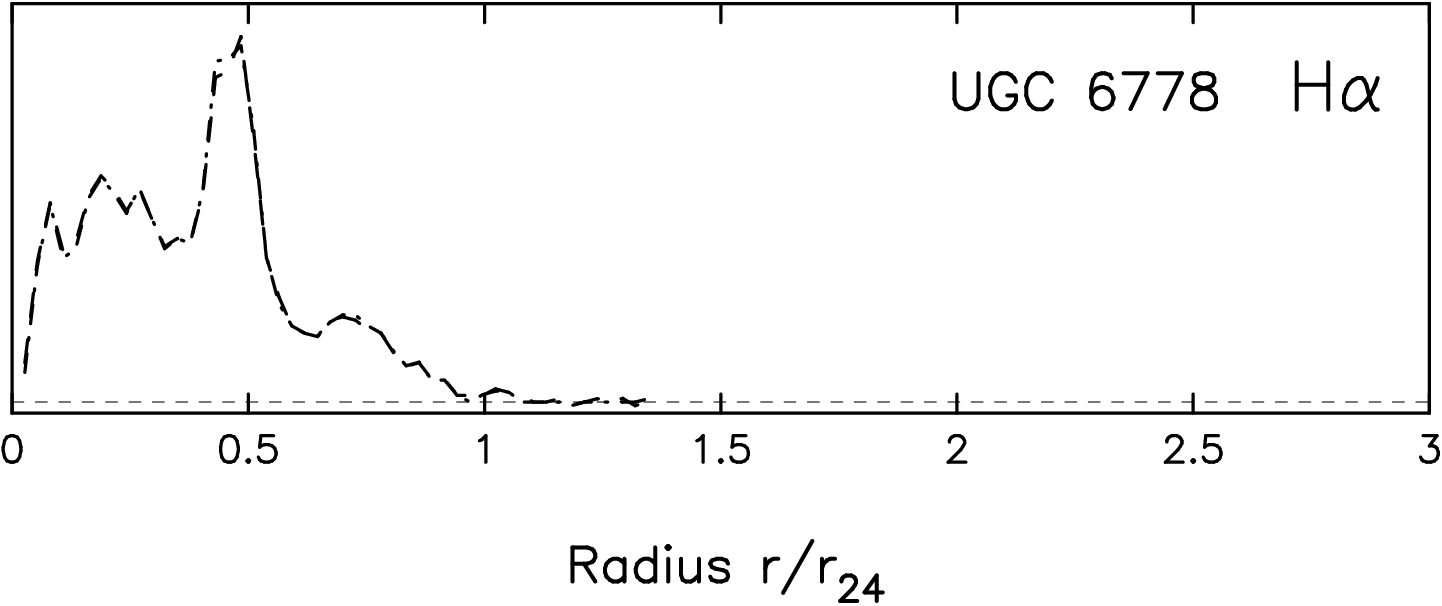}
}
\rotatebox{-90}{
\includegraphics[height=6.2cm]{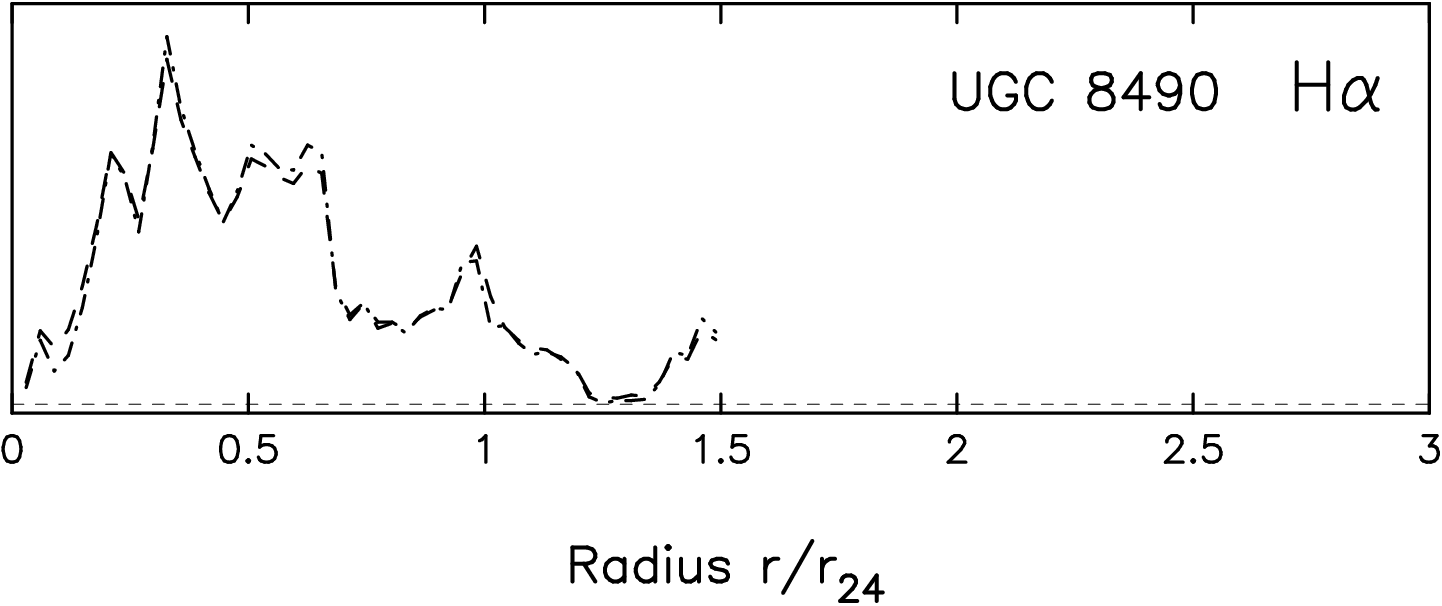}
}

\caption{Profiles made with the continuum light deliberately over- and under-
  subtracted (dot-dashed and dashed curves respectively), for
  five representative sample galaxies: UGC~4097 (Sa), UGC~4273 (SBb),
  UGC~6644 (Sc), UGC~6778 (SABc) and UGC~8490 (Sm).}
\label{fig:contsub}
\end{figure*}

The most surprising finding from Fig. \ref{fig:diffprof} is that the
earliest types, $T=$ 0, 1 and 2, which should contain the most
dominant bulges, do not show the expected central depression in their
\Ha\ -- $R$ profiles.  For the $T=$ 0 (S0a) galaxies the difference
profile is consistent with being flat, within the rather large error
bars, whereas the Sa and Sab profiles ($T=$ 1 and 2) show emission
line excesses, relative to the $R$-band light, out to radii of
$\sim$0.3 $r_{24}$, which corresponds to $\sim$3~kpc for the average
size of galaxy contributing to these profiles.  The nature of such
extended emission is not clear.  Active Galactic Nuclei (AGN) and
Low-Ionization Nuclear Emission-line Regions (LINERS) are common in
galaxies of these types \citep{ho97a}, but the line emission from such
nuclei should be unresolved in our profiles.  \citet{hame99} report
the finding of what they termed Extended Nuclear Emission-line Regions
(ENERs) in 7 out of 27 Sa and Sab galaxies for which they had
narrow-band \Ha\ $+$ \NII\ imaging.  They speculated that this
emission arises from gas excited by post-Asymptotic Giant Branch
stars, which would lead to the expectation that such emission should
be distributed like the bulge stellar luminosity.  \citet[][Paper
  II]{paper2} confirmed that central emission-line components with the
same smooth morphology reported by \citet{hame99} are present in \Ha
GS galaxies, and demonstrated that in at least one case the emission
is dominated by the \NII\ line and not \Ha .  Such emission is
qualitatively very different in appearance from the \Ha\ emission from
SF regions; the former has a smooth, diffuse and centrally symmetric
structure, where as SF regions are characteristically very clumpy and
irregular. In the present sample, extended central emission regions
which may be ENERs are seen in 9 out of 62 galaxies with $T=$ 0 to 3,
and are very rarely detected in later-type galaxies, possibly due to
the dominance of the emission by that from SF.  Circumnuclear rings of
SF also predominantly occur in galaxies of types 3--4. They occur in
up to 20\% of local spiral galaxies \citep{knap05}, and at higher
frequencies in types 3--4.  With radii varying from a few hundred
parcsec to some 2~kpc, they may also make a substantial contribution
to a central peak in H$\alpha$ emission in individual cases.

Outside the bulge regions, the mean \Ha\ and $R$-band profiles of each
type are essentially identical in their overall morphologies.  Thus
the disk SF appears to have a very similar radial distribution to the
older stellar population traced by the $R$-band light.  Minor
differences are apparent: the \Ha\ profiles for $T>$ 3 peak at slightly larger
radii than the $R$-band profiles, and the effective radii of the \Ha\ profiles
also tend to be larger (see Table \ref{tbl:reff}).  Both can
be accounted for by the central suppression of \Ha\ emission in bulge
regions.  Outside these regions, in the disk-dominated parts of the
profiles, the \Ha\ and $R$-band light distributions are identical
within the errors.  Thus there is no evidence of, for example, the
radial truncation of \Ha\ emission found by \citet{koop04} for a
sample of 55 Virgo cluster spirals.  This is consistent with their
interpretation that the truncation is a consequence of the cluster
environment (they found no truncation for a comparison sample of 29
isolated spiral galaxies), since our sample is predominantly composed
of field galaxies.  There is also no evidence from these profiles to
support inside-out theories of disk formation; the old stellar
population could have been produced by historical SF distributed like
that occurring at the present epoch.  The possibility of large-scale
radial migration of disk material driven by, for example, bar torques
obviously complicates this conclusion.

\subsubsection{The effect of continuum subtraction errors on \Ha\ profile shape}

One of the most important, and problematic, stages in the reduction of
narrow-band imaging is the removal of the continuum light which passes
through the narrow-band filter in addition to the desired line
emission.  For \Ha GS this was done using additional imaging through
either a broad $R$ or an intermediate-width continuum filter. The
procedures and associated (significant) errors on derived total fluxes
are explained in Paper I.  The possibility of errors leading to
under- or over-subtraction of the continuum light is particularly
important for the present paper as, for example, systematic
under-subtraction could easily lead to a spurious apparent agreement
between the shapes of $R$-band and ``\Ha'' profiles due to possible
red-light contamination in the latter.  A test was
performed to determine the impact of continuum subtraction errors on
normalised \Ha\ profile shapes, for 10 galaxies from \Ha GS covering the 
full range of types.  For each one, the scaling factor applied to the 
continuum image was varied by $\pm 3$ times the statistical error on this
parameter, such that the residuals of
unsaturated stellar images on the subtracted frame would be all negative or
all positive in the two resulting frames.  
Thus we took conservative cases of clearly over- and under-subtracted 
images for this analysis.
\Ha\ profiles were then produced
for each of these pairs of frames, using the methods described above. 

The results of this analysis are quite reassuring.  Even though
continuum subtraction errors can alter total \Ha\ fluxes by
$\sim$30\%, the effects on the shapes of profiles, after renormalising
the profile area to unity, is much smaller.  Figure \ref{fig:contsub}
shows, for a selection of the 10 galaxies studied, that the strongest
effects are unsurprisingly found in the central regions of early-type galaxies,
since these are dominated by high surface brightness bulges.  However, even
in these cases the overall shapes of the derived \Ha\ profiles are not
greatly affected.  The \Ha\ effective radii for UGC~4097 and UGC~4273,
which show the greatest variation in profile shape in
Fig. \ref{fig:contsub}, both change by $\pm2.5$\% as a result of this
degree of continuum over- and under-subtraction.  For galaxies of type
Sc or later, the effects of continuum subtraction errors on profile shape are
negligible. 

\subsubsection{Dependence of profiles on galaxy distance, size and luminosity}

\begin{figure*}
\centering
\rotatebox{-90}{
\includegraphics[height=6.4cm]{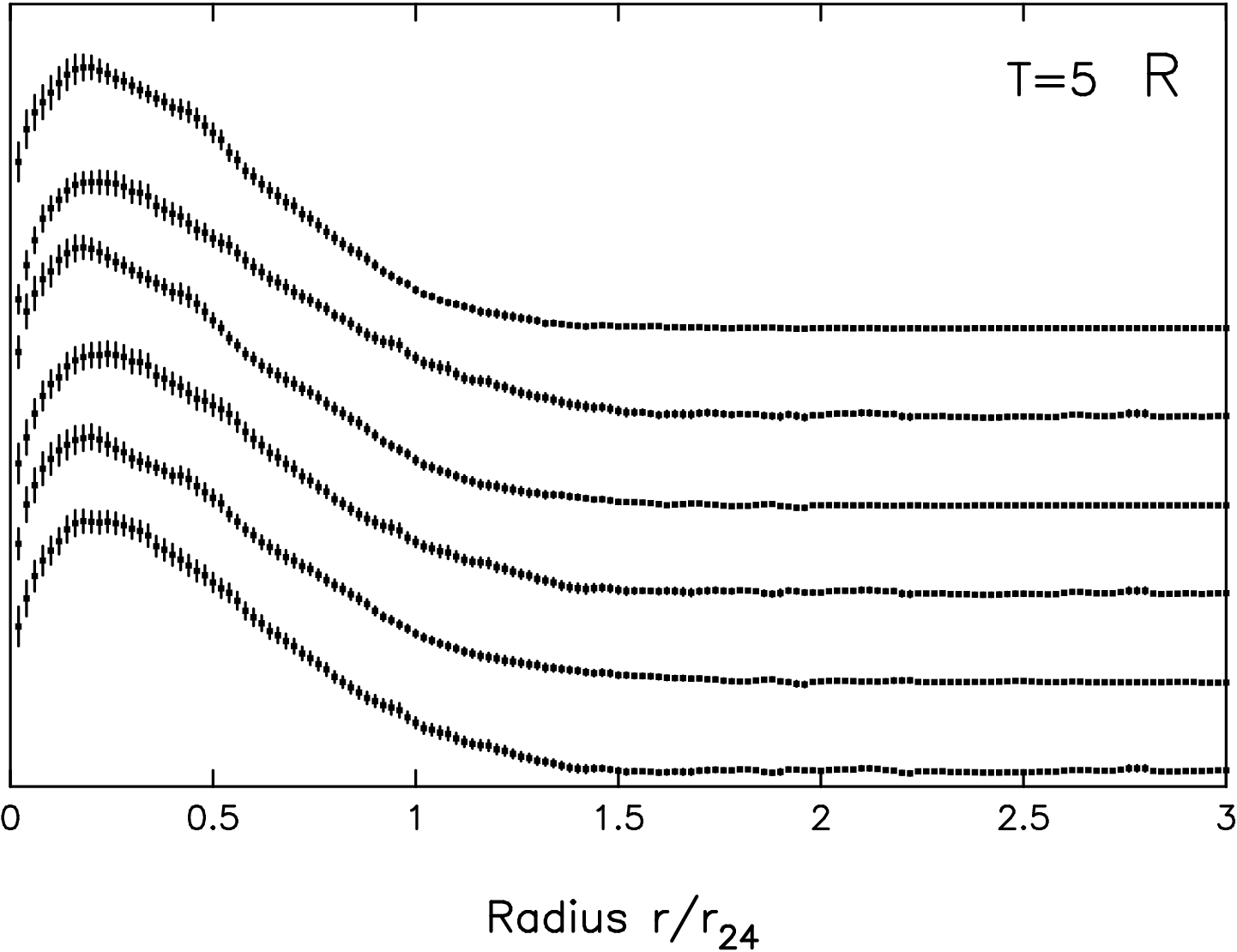}
}
\rotatebox{-90}{
\includegraphics[height=6.4cm]{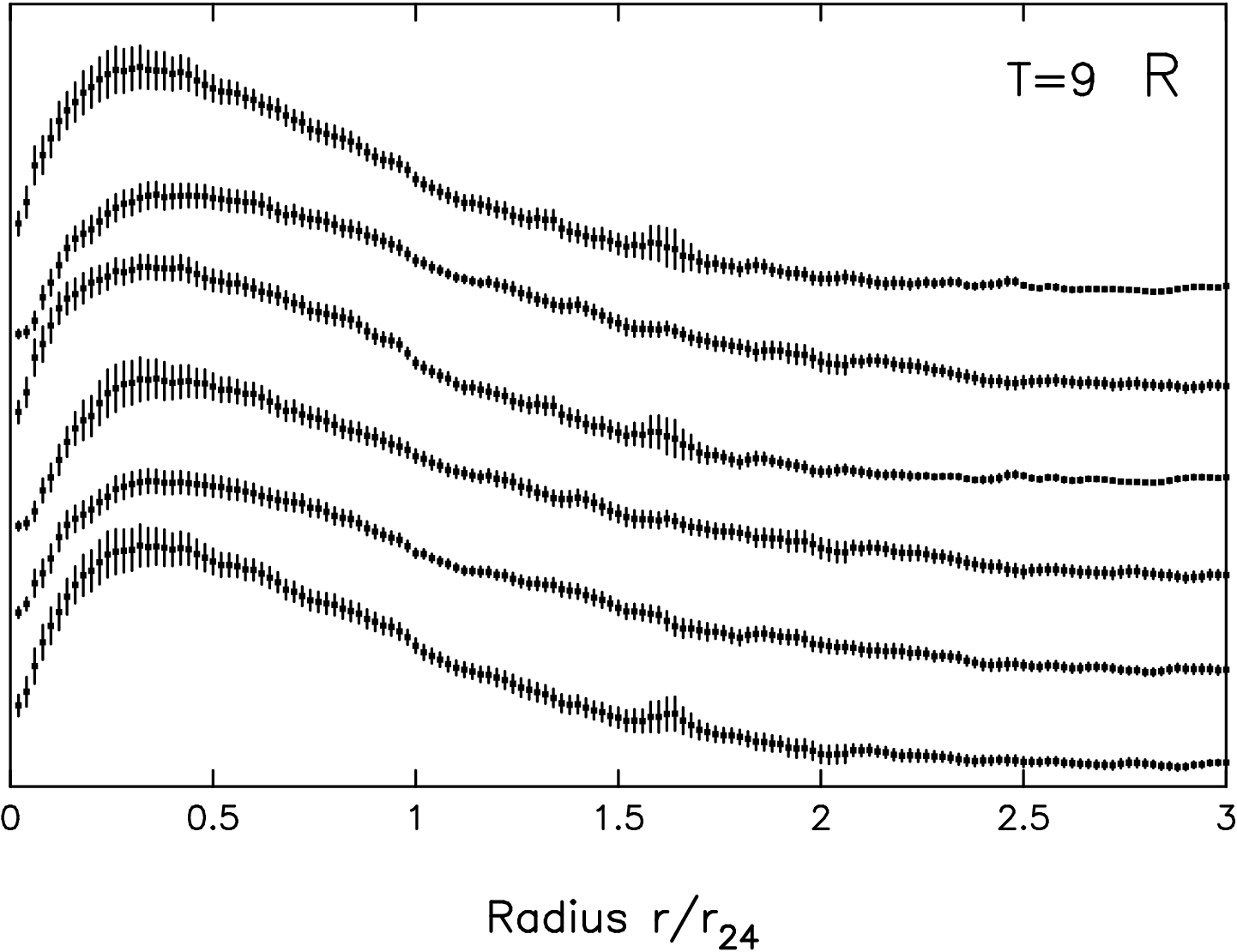}
}

\caption{
Mean $R$-band light profiles for $T=$ 5 galaxies (left)
and $T=$ 9 galaxies (right), with the 6 profiles in each frame
showing the bright, faint, large, small, far and near halves of the 
total sample from top to bottom.}

\label{fig:t5split}
\end{figure*}

The strong dependence of $R$-band profile shape on $T$ type, and the
generally small error bars on the $R$-band profile points in
Fig. \ref{fig:meanprof}, indicate a high degree of uniformity of light
profiles which is somewhat surprising given the range of properties of
the galaxies contributing to each profile.  The 36 galaxies
contributing to the $T=$ 5 profiles lie at distances ranging from 2.1
to 34 Mpc, and have $R$-band absolute mags from --16.1 to --21.6,
corresponding to a factor of 160 in luminosity; for the 42 $T=$ 9
galaxies, the distance range is similar, and the luminosity range
still larger at a factor of 360 ($M_R$ from --12.9 to --19.3).  We
have thus checked for any dependence of the mean profile shape on
distance, galaxy radius in kpc, and luminosity, using these two types
as test cases due to the large numbers of profiles available and the
large range of properties included in each type.  The results are
shown in Fig. \ref{fig:t5split}.  Each panel shows the mean $R$-band
profiles for one half of the galaxies of the appropriate type, with
the split being done by distance, galaxy radius and $R$-band
luminosity.  $T=$ 5 galaxy profiles are on the left, those for $T=$ 9
on the right.  For $T=$ 5, the only effects seen are the appearance of
a subtle `shoulder' at 0.5~$r_{24}$ in the bright, large and distant
halves of the sample, but the overall shapes of the profiles are all
very similar.  For $T=$ 9, the bright and large halves of the sample
have mean profiles that are somewhat more centrally concentrated than
the faint half.  The bright half matches the overall $T=$ 8 profile in
peak position (0.34~$r_{24}$), and has an effective radius
(0.64~$r_{24}$) between those for the $T=$ 8 and $T=$ 9 profiles.
Thus the luminosity dependence appears slight, and overall we conclude
that profile shape is largely determined by galaxy classification.


\section{The effects of bars on distributions of star formation and 
continuum light}
\label{sec:bars}

\subsection{Mean light profiles for barred and unbarred galaxies}

\begin{figure*}
\centering
\rotatebox{-90}{
\includegraphics[height=5.7cm]{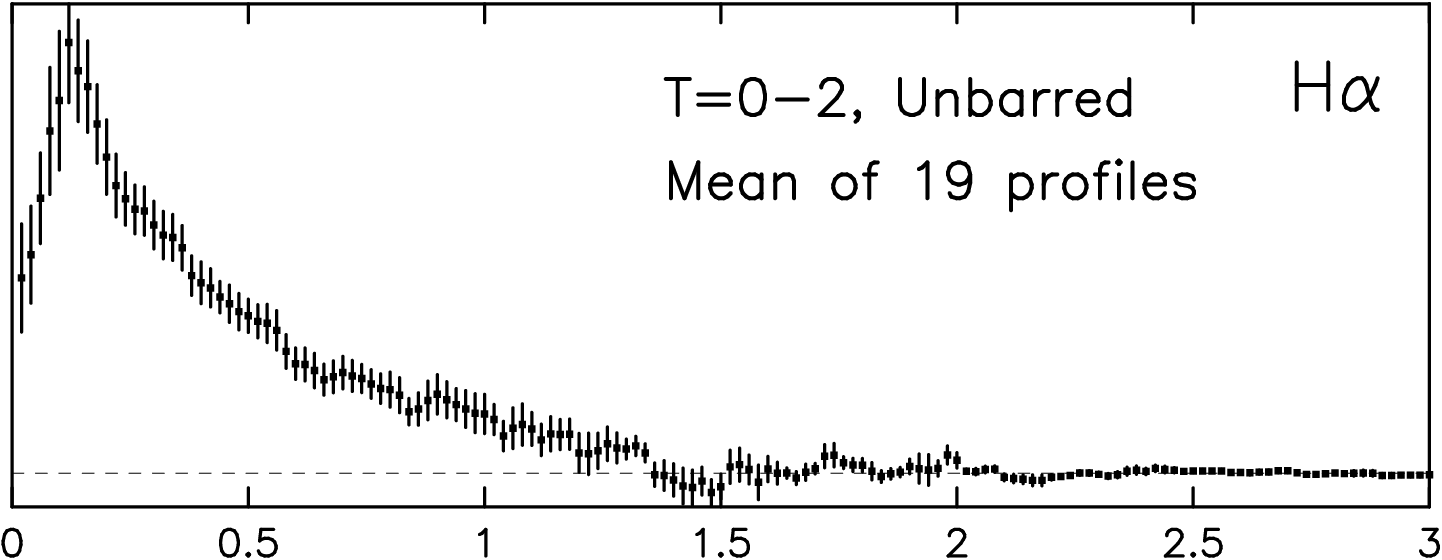}
}
\rotatebox{-90}{
\includegraphics[height=5.7cm]{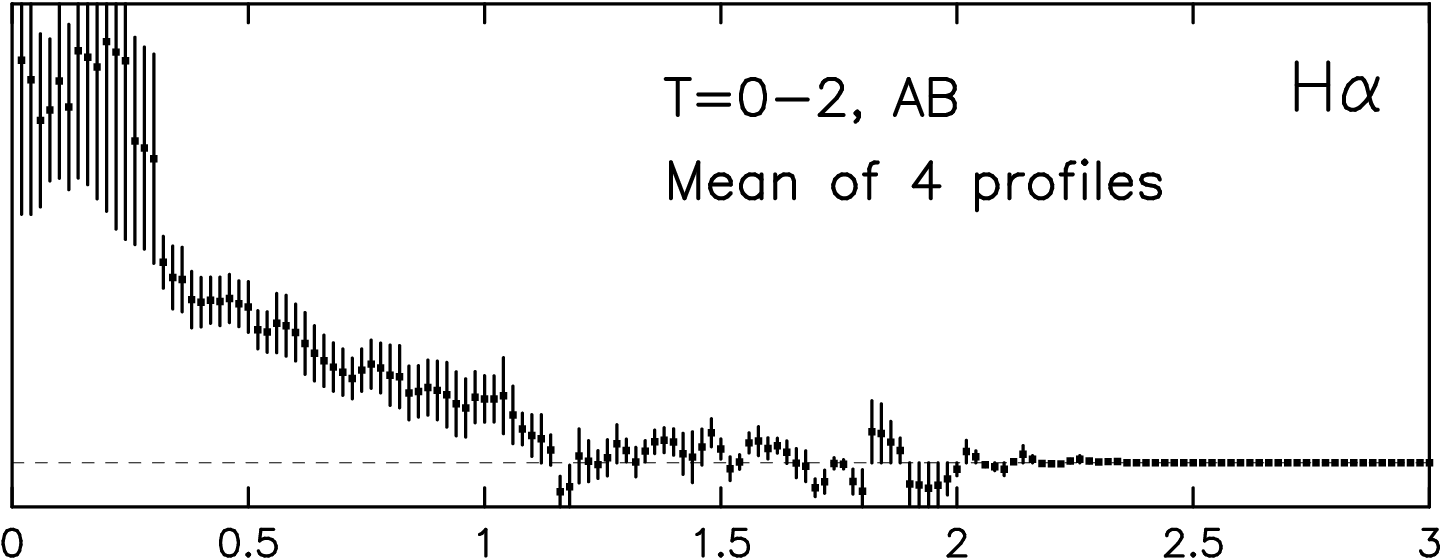}
}
\rotatebox{-90}{
\includegraphics[height=5.7cm]{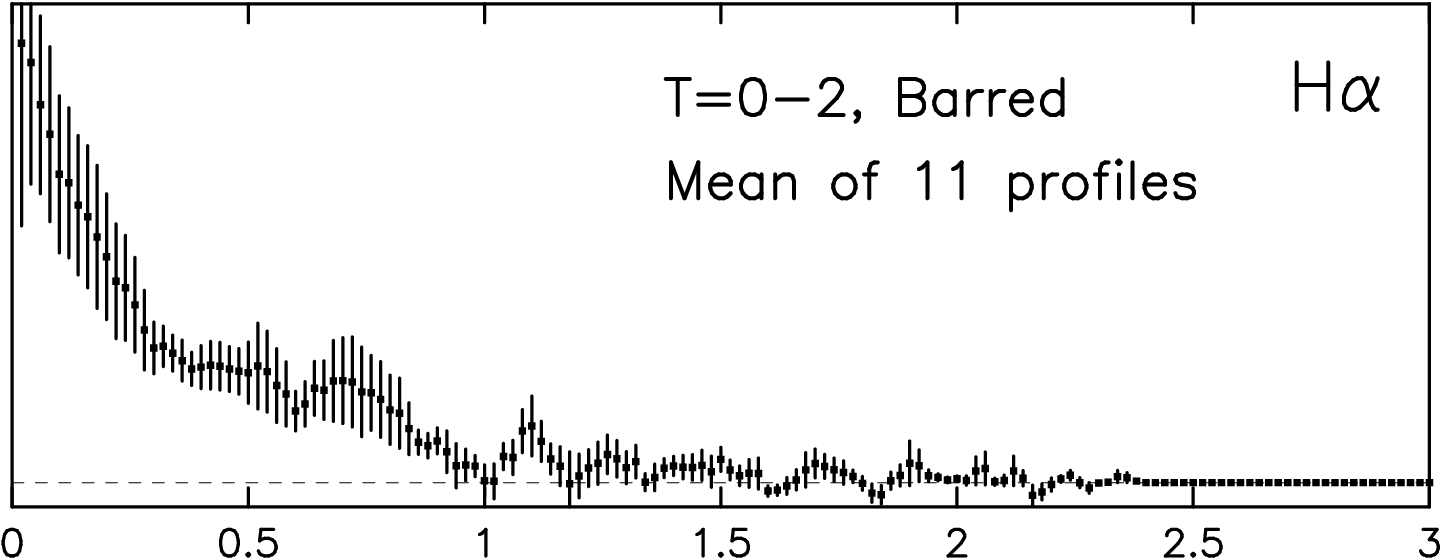}
}
\rotatebox{-90}{
\includegraphics[height=5.7cm]{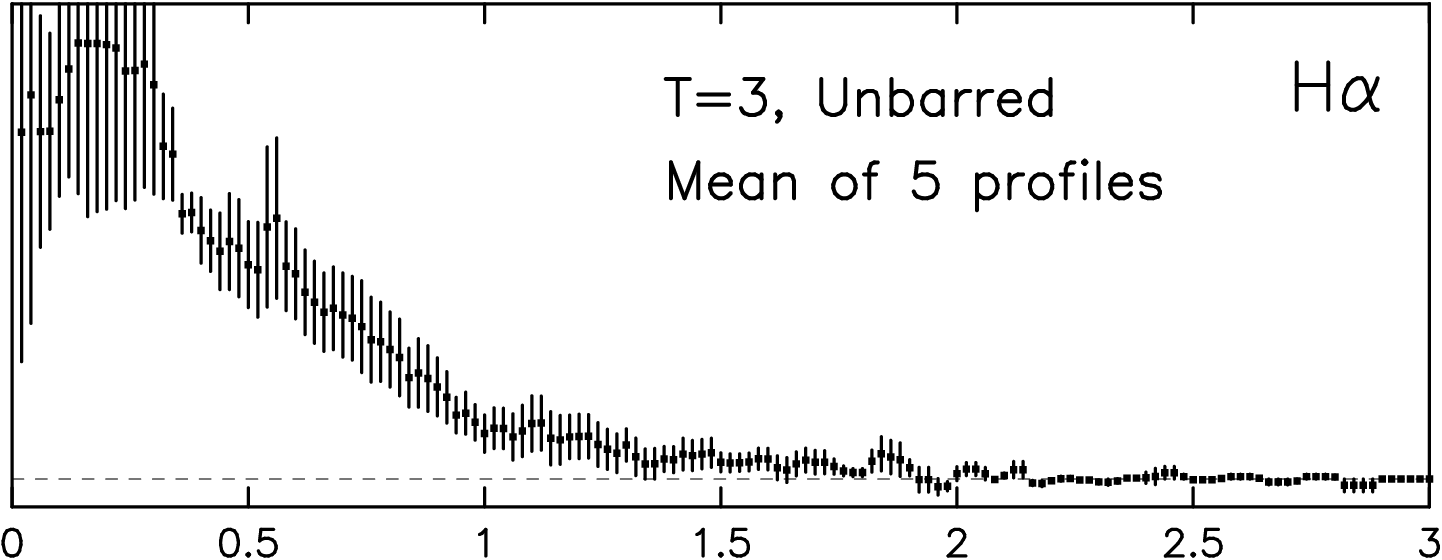}
}
\rotatebox{-90}{
\includegraphics[height=5.7cm]{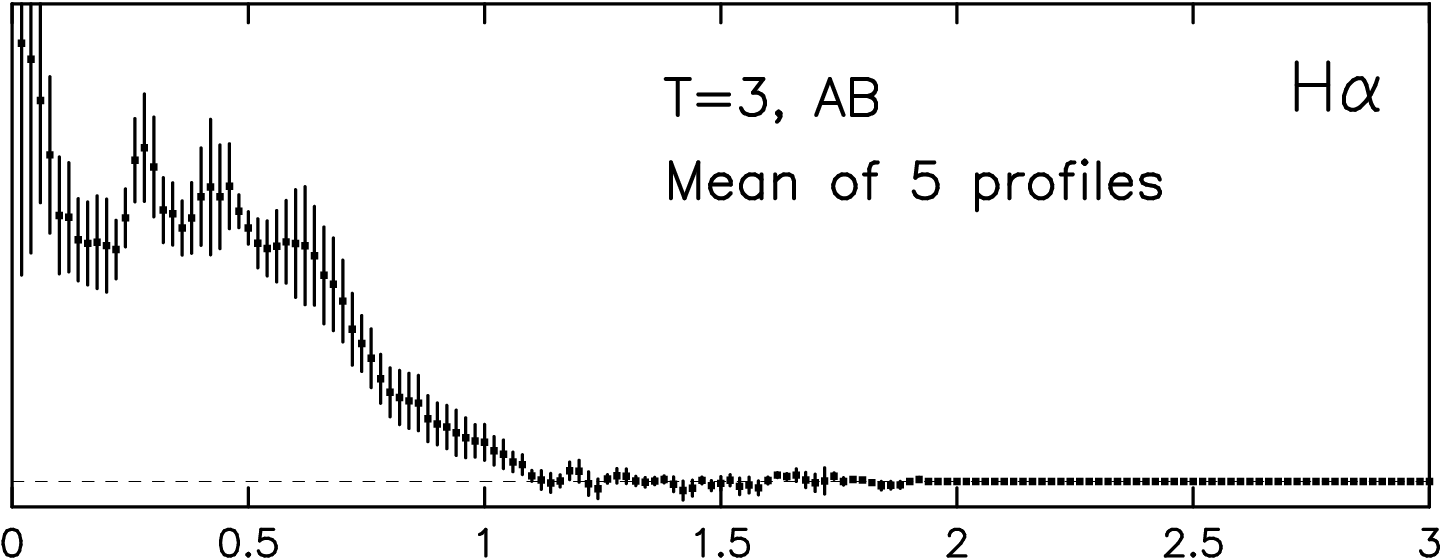}
}
\rotatebox{-90}{
\includegraphics[height=5.7cm]{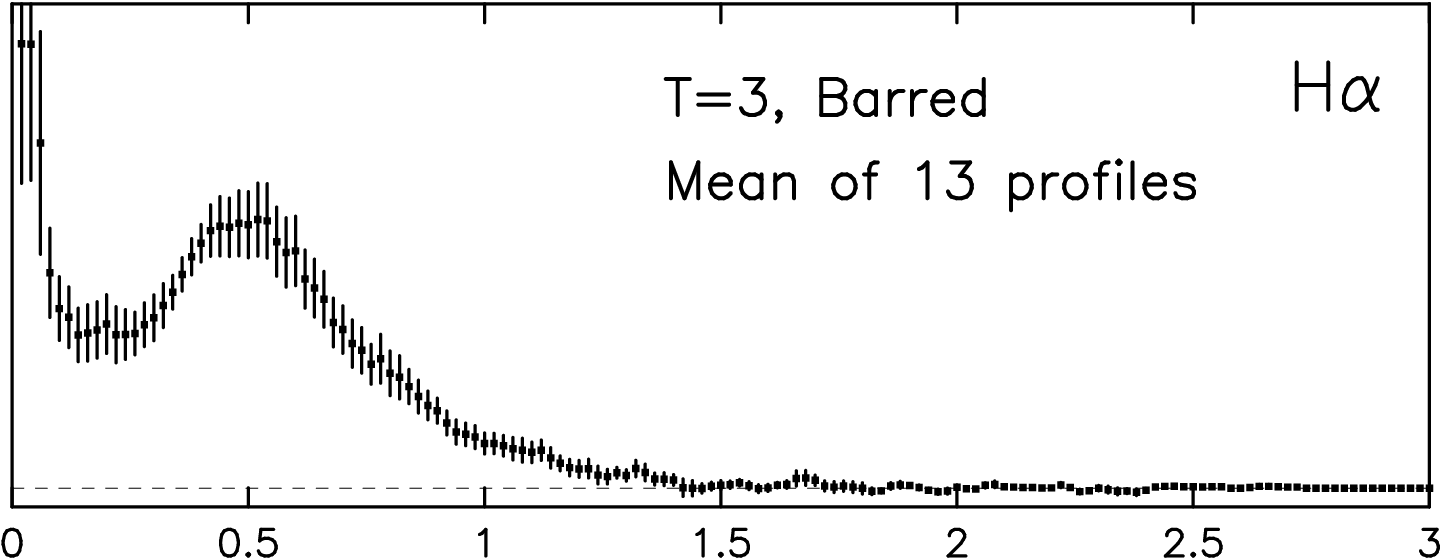}
}
\rotatebox{-90}{
\includegraphics[height=5.7cm]{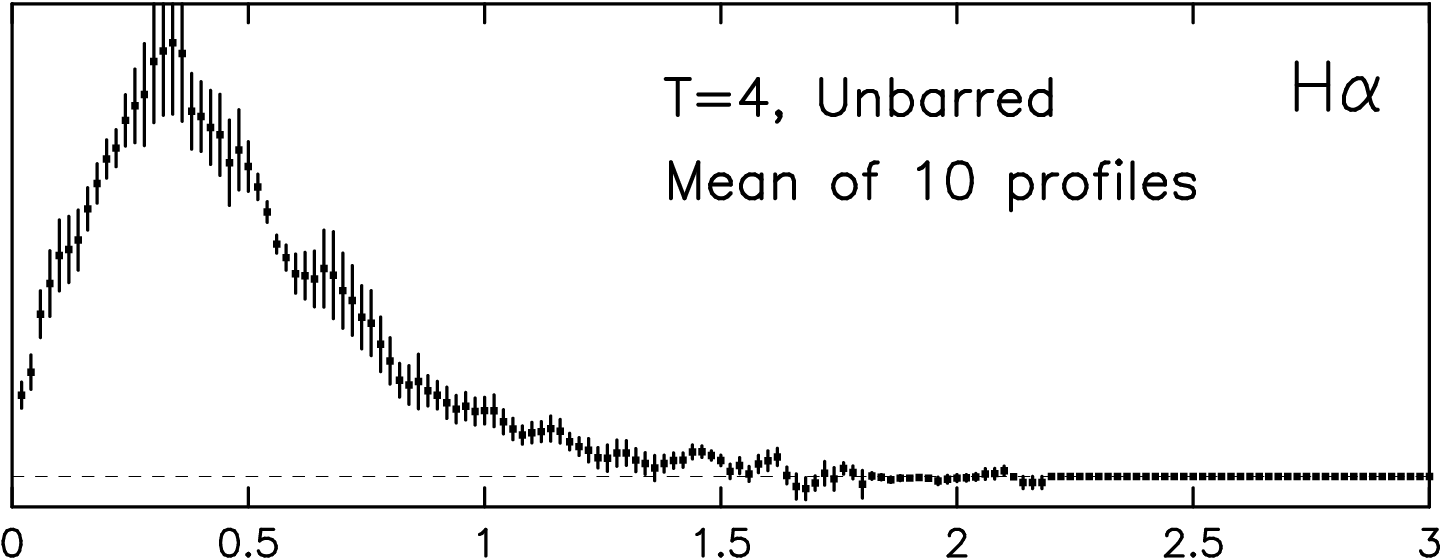}
}
\rotatebox{-90}{
\includegraphics[height=5.7cm]{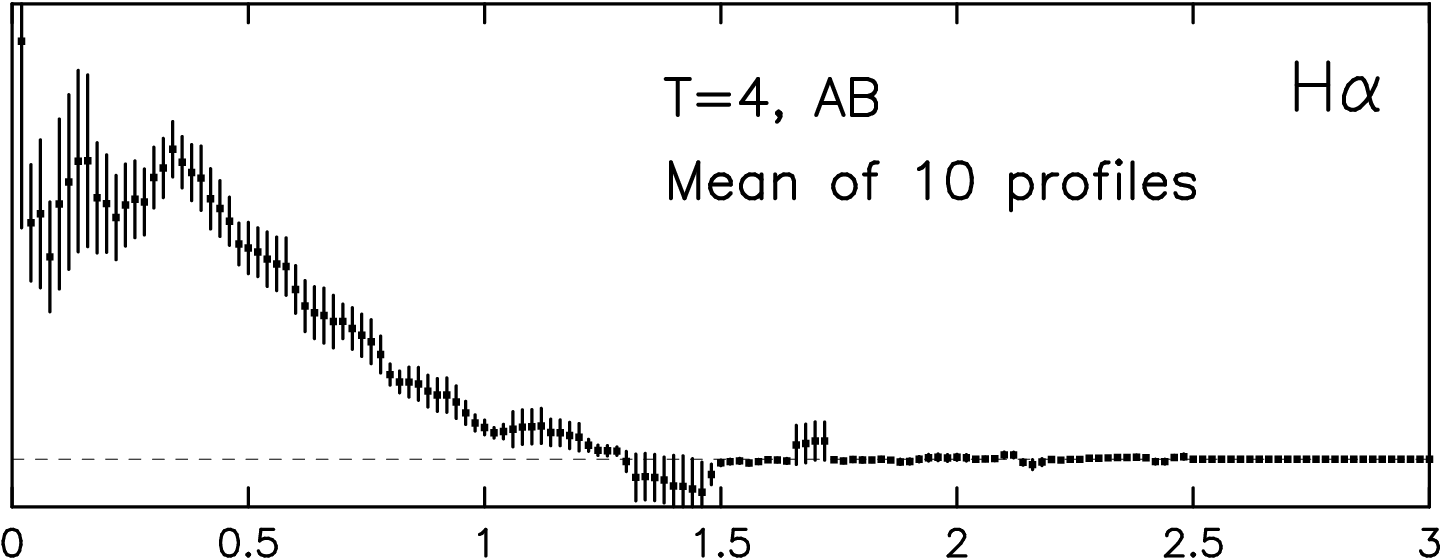}
}
\rotatebox{-90}{
\includegraphics[height=5.7cm]{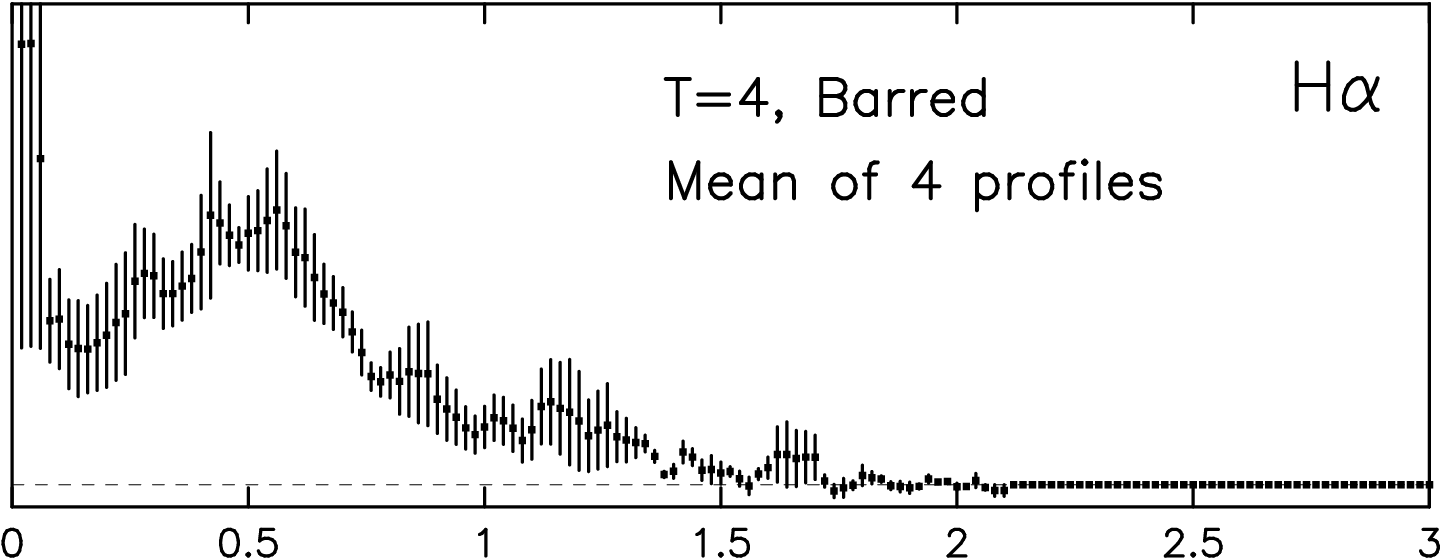}
}
\rotatebox{-90}{
\includegraphics[height=5.7cm]{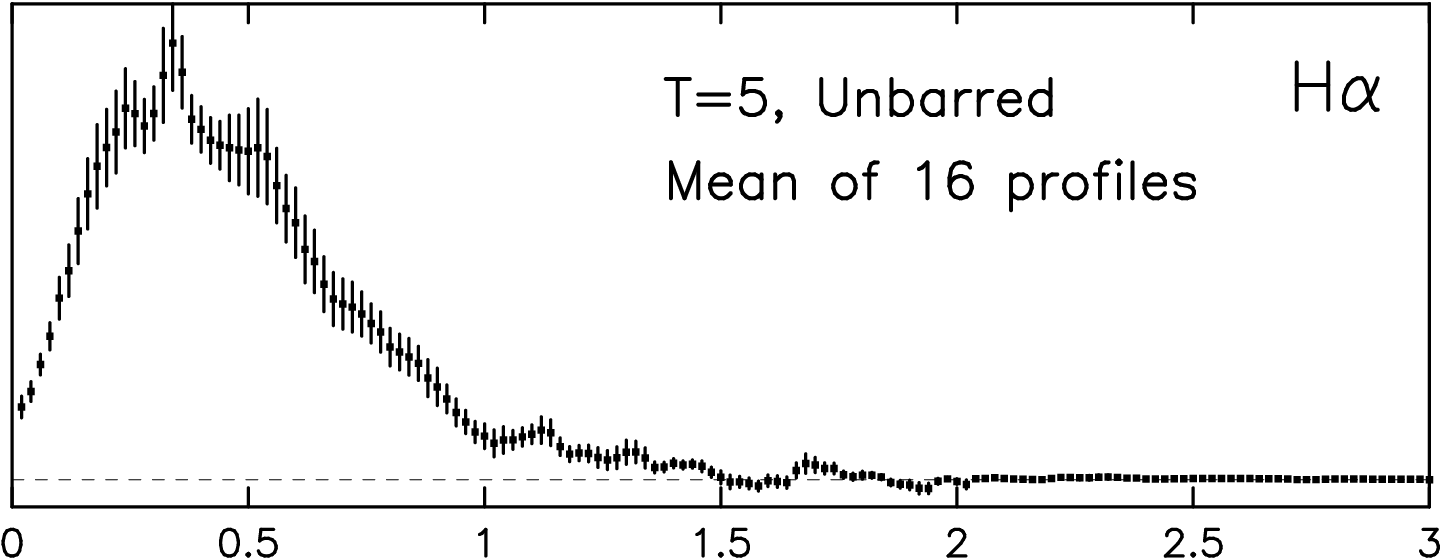}
}
\rotatebox{-90}{
\includegraphics[height=5.7cm]{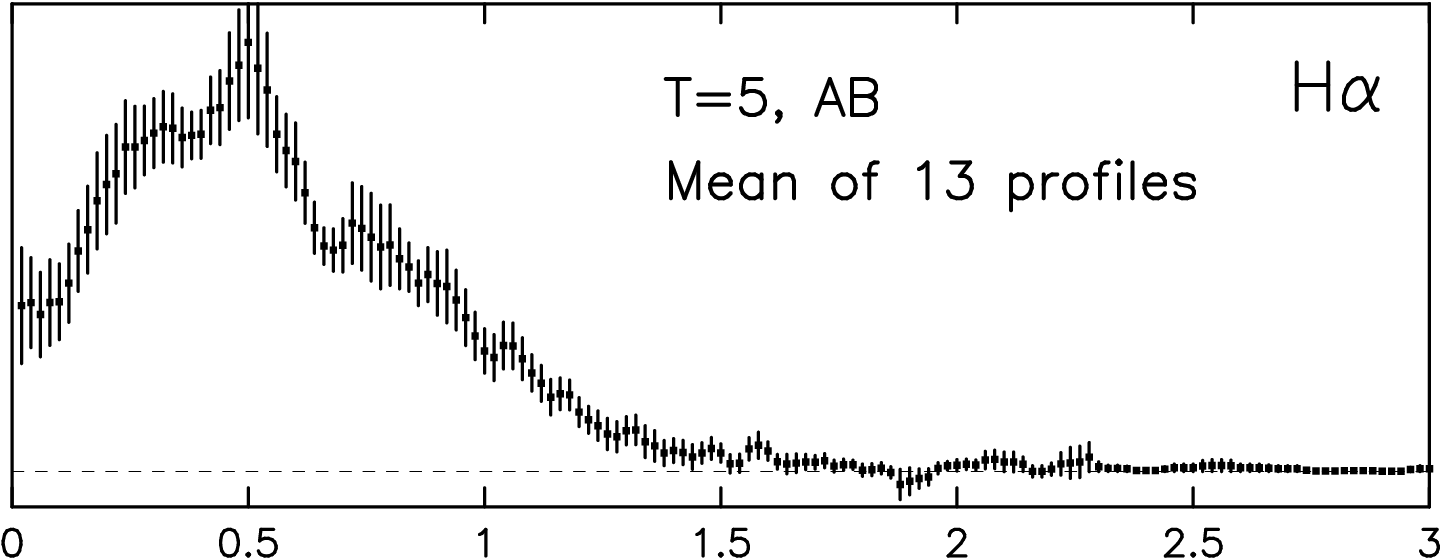}
}
\rotatebox{-90}{
\includegraphics[height=5.7cm]{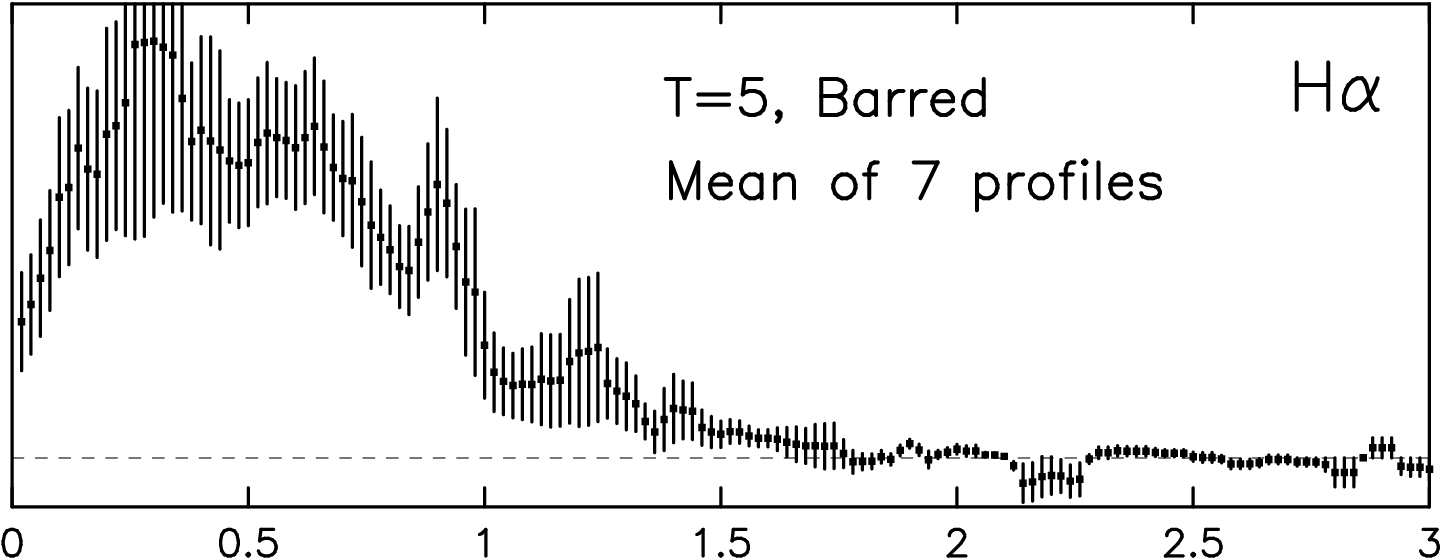}
}
\rotatebox{-90}{
\includegraphics[height=5.7cm]{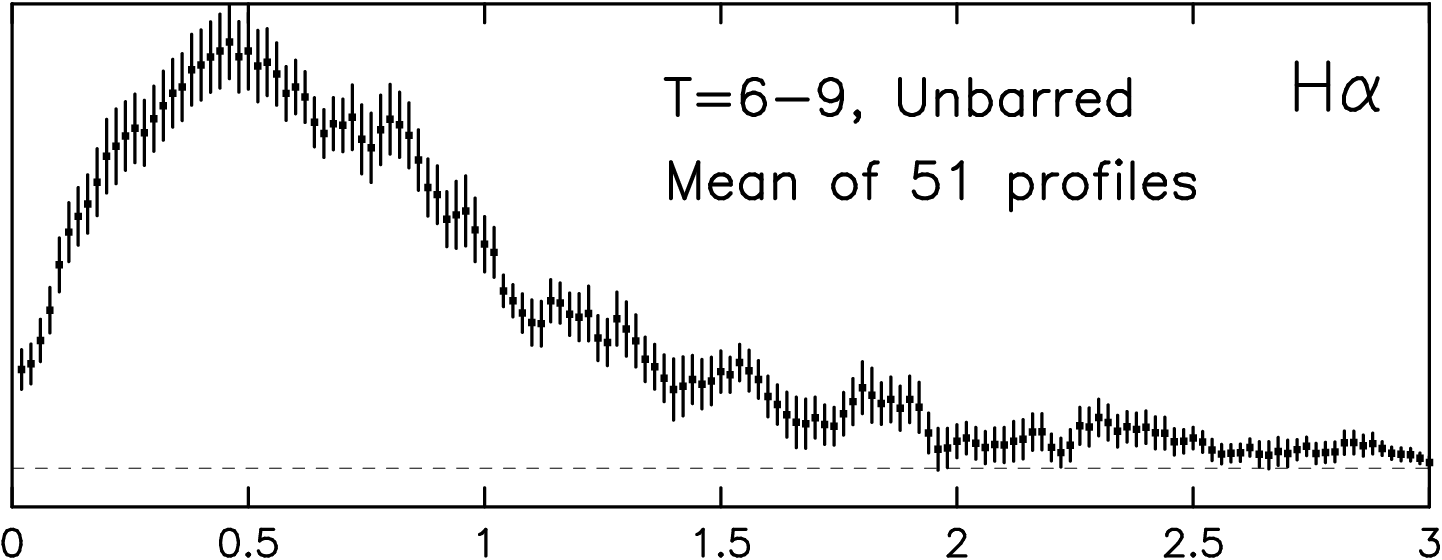}
}
\rotatebox{-90}{
\includegraphics[height=5.7cm]{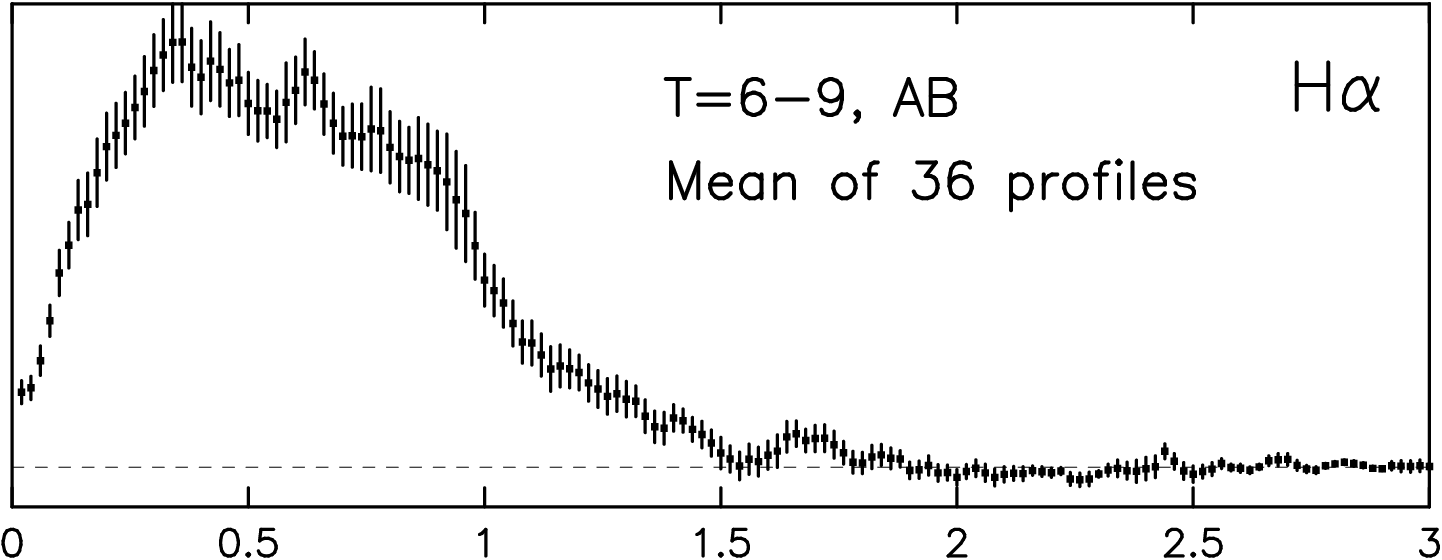}
}
\rotatebox{-90}{
\includegraphics[height=5.7cm]{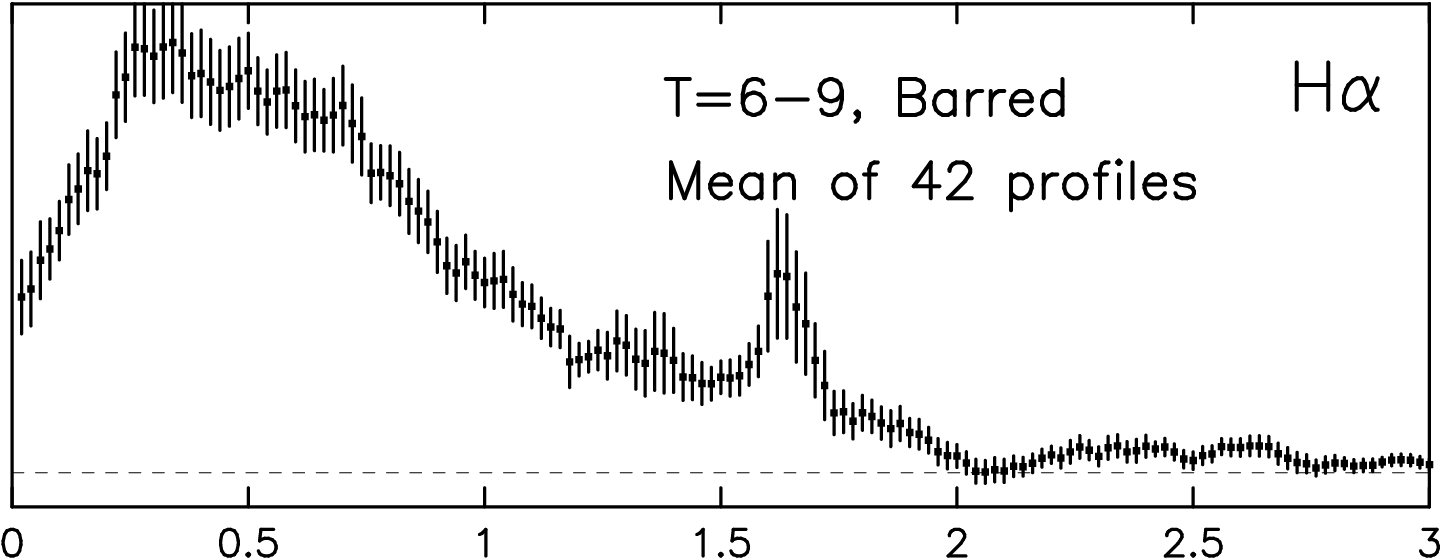}
}
\rotatebox{-90}{
\includegraphics[height=5.7cm]{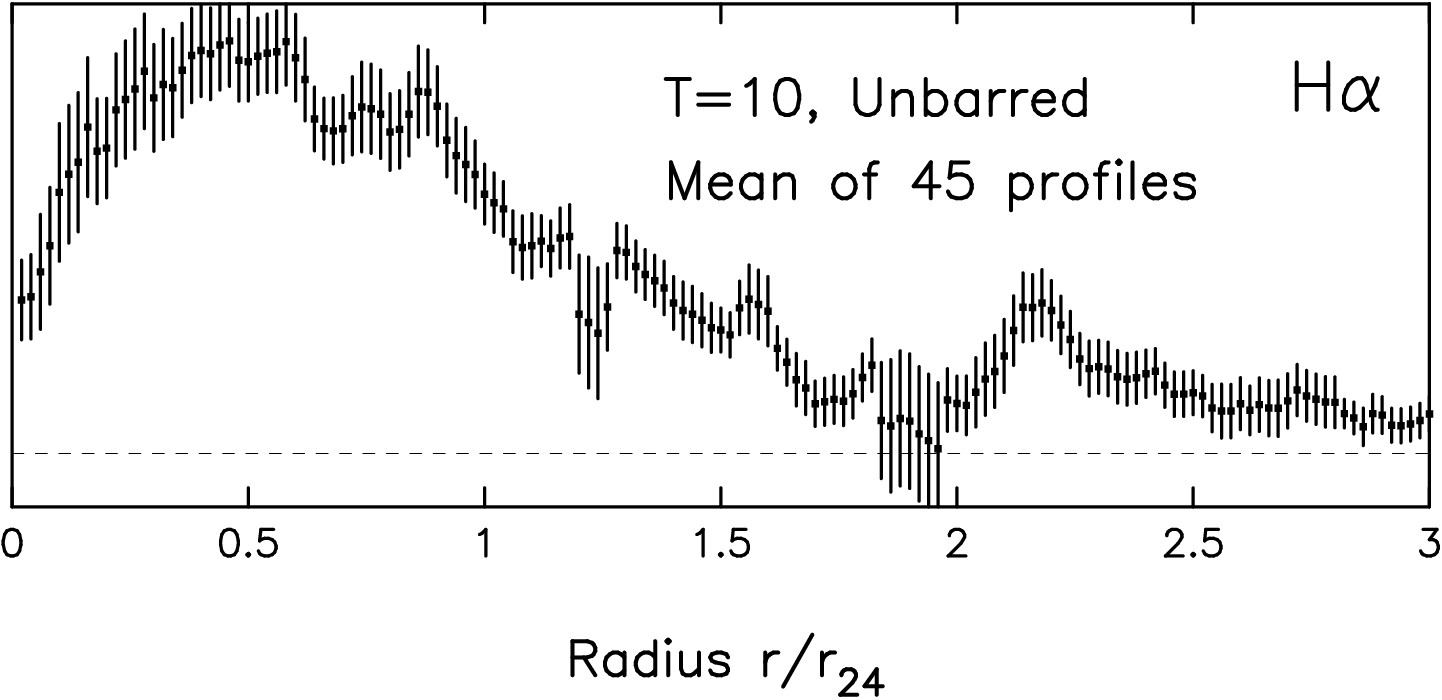}
}
\rotatebox{-90}{
\includegraphics[height=5.7cm]{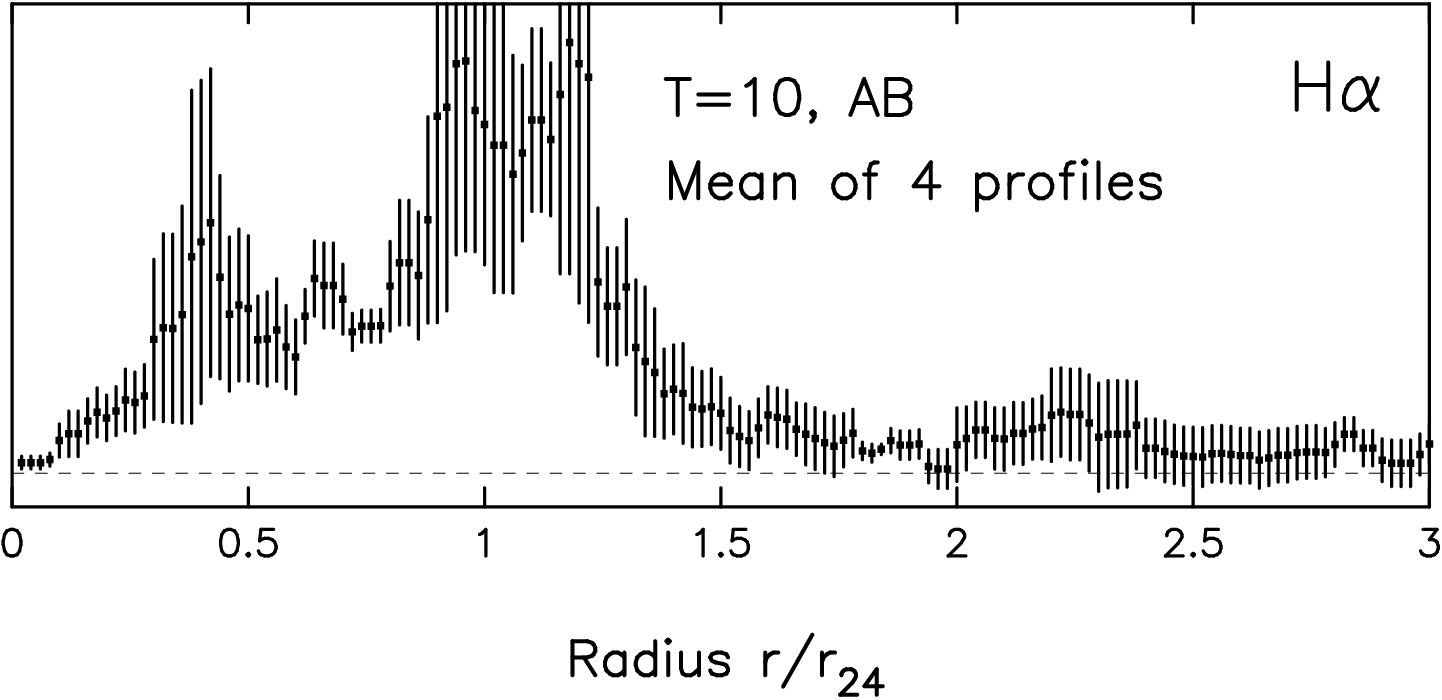}
}
\rotatebox{-90}{
\includegraphics[height=5.7cm]{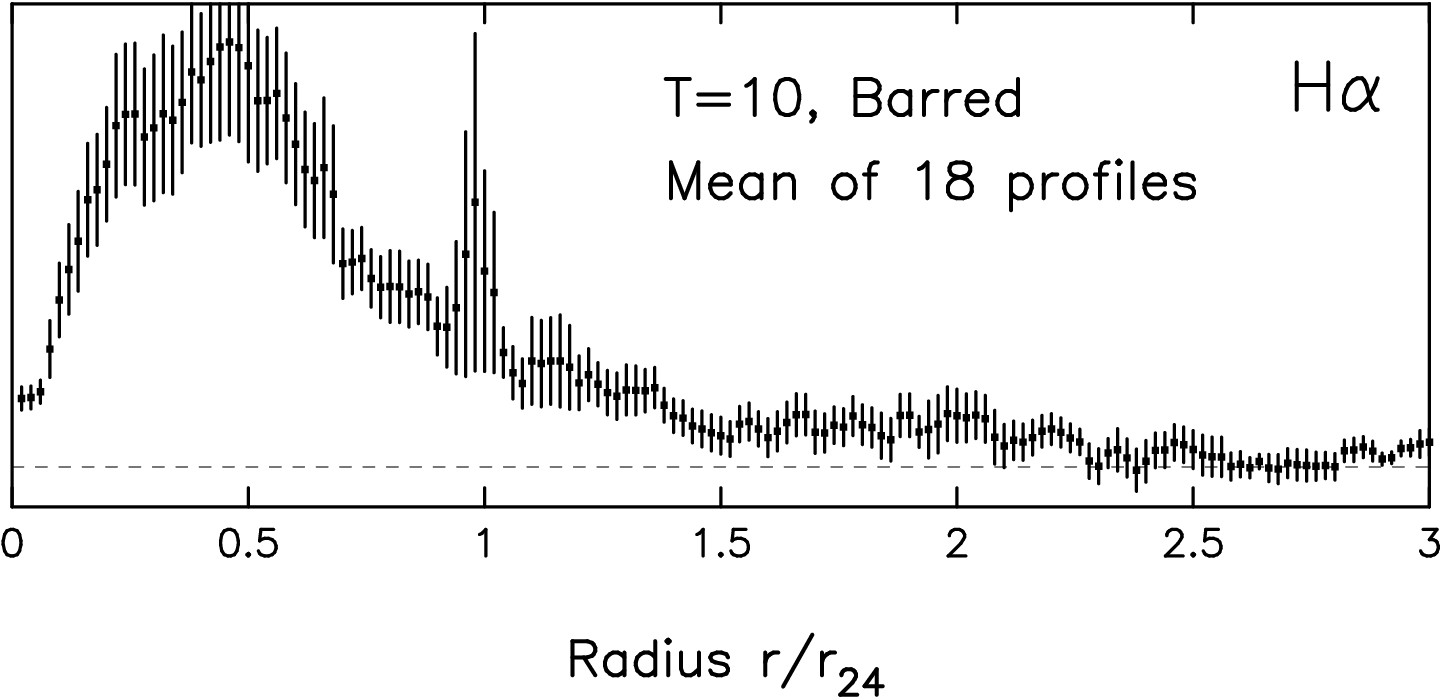}
}

\caption{Mean \Ha\ light profiles, as shown in Fig. \ref{fig:meanprof}, 
but subdivided
into unbarred, AB and barred galaxies, in left, middle and 
right frames respectively.}
\label{fig:habarprof}
\end{figure*}

\begin{figure*}
\centering
\rotatebox{-90}{
\includegraphics[height=5.7cm]{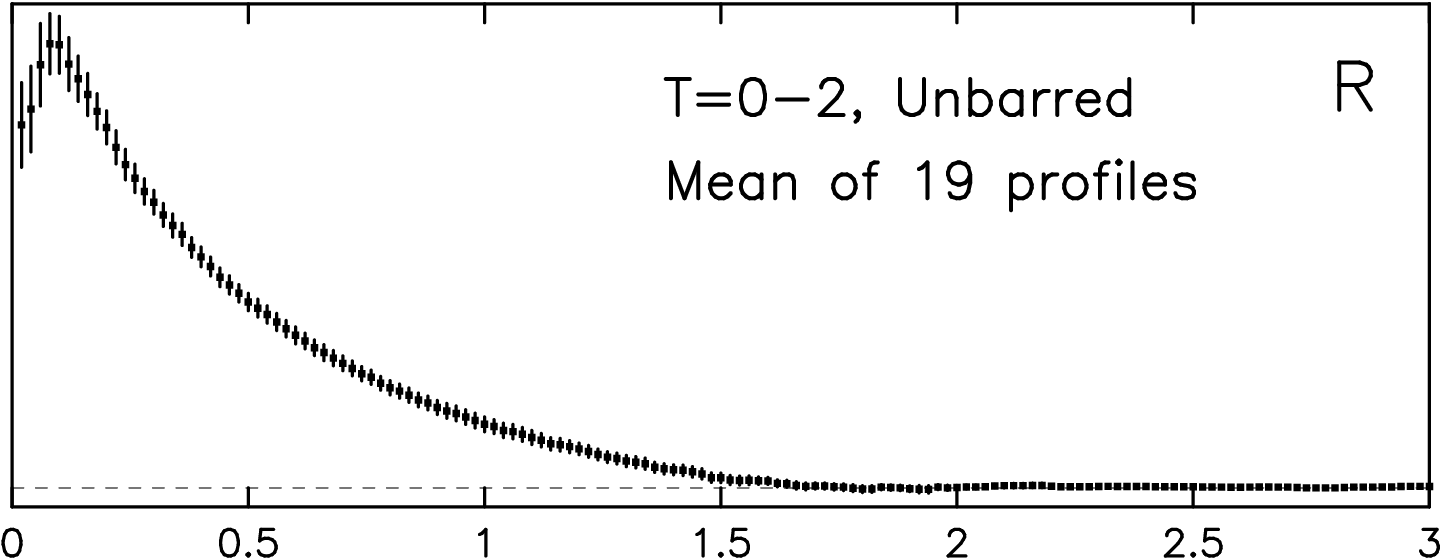}
}
\rotatebox{-90}{
\includegraphics[height=5.7cm]{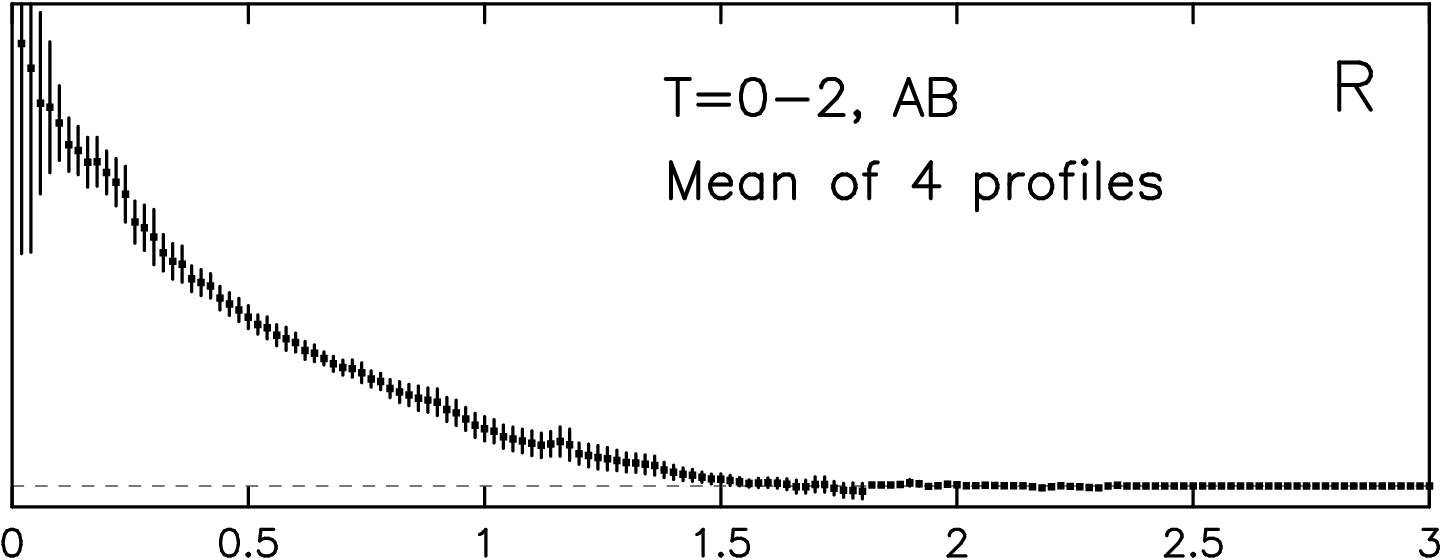}
}
\rotatebox{-90}{
\includegraphics[height=5.7cm]{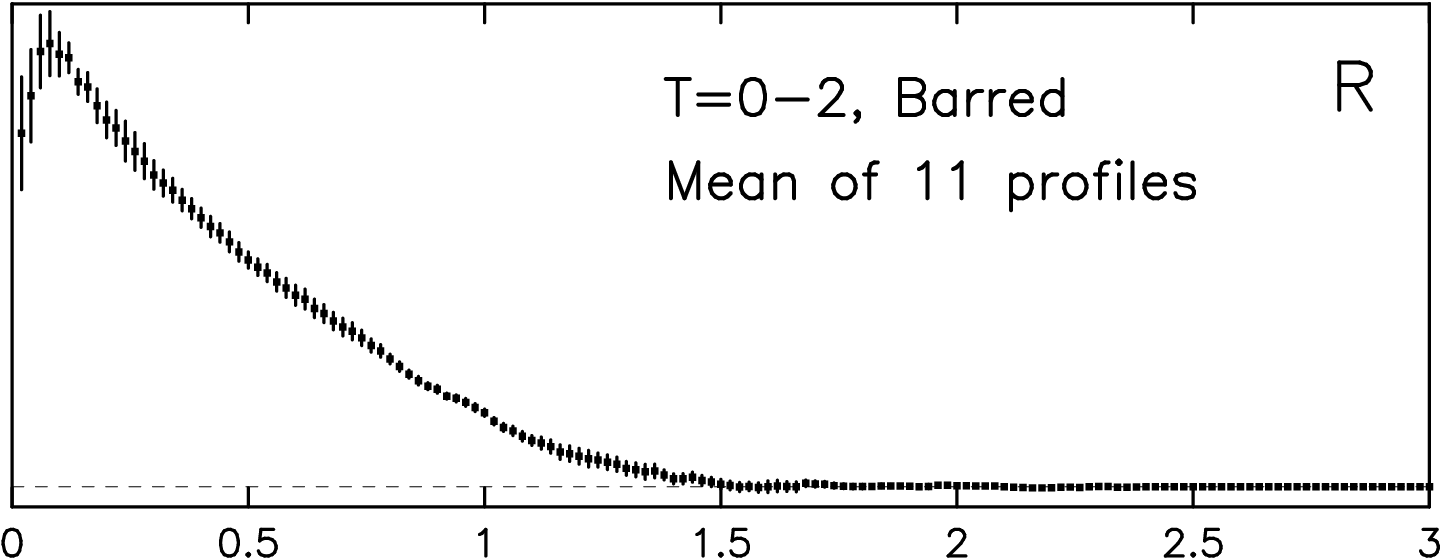}
}
\rotatebox{-90}{
\includegraphics[height=5.7cm]{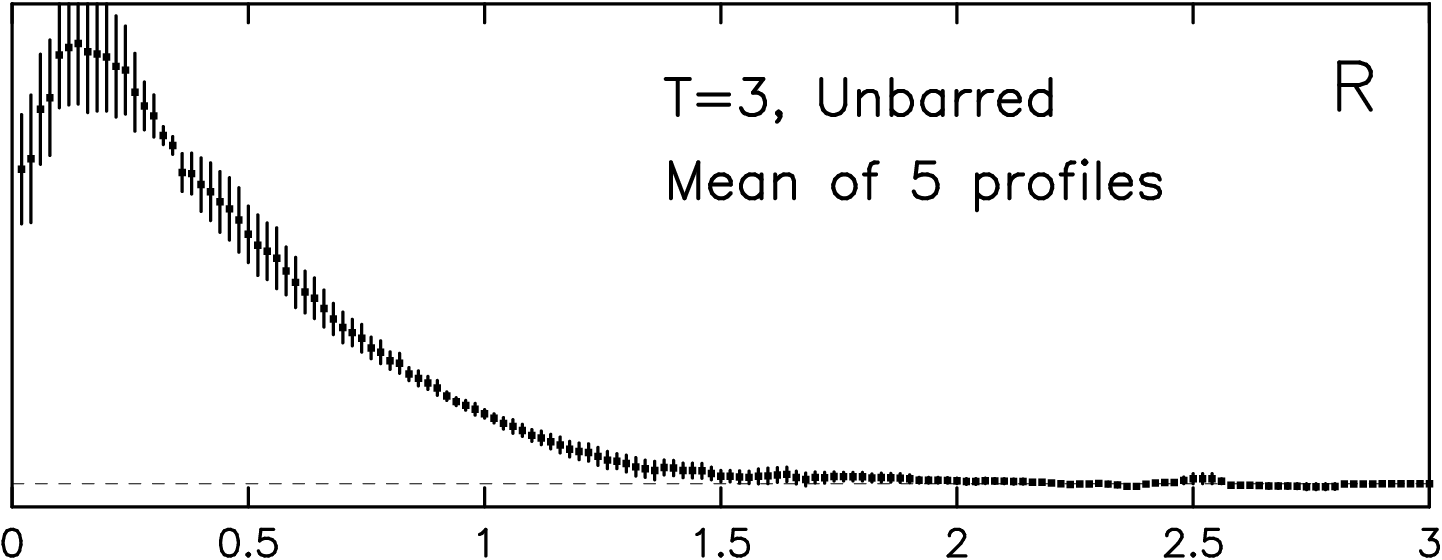}
}
\rotatebox{-90}{
\includegraphics[height=5.7cm]{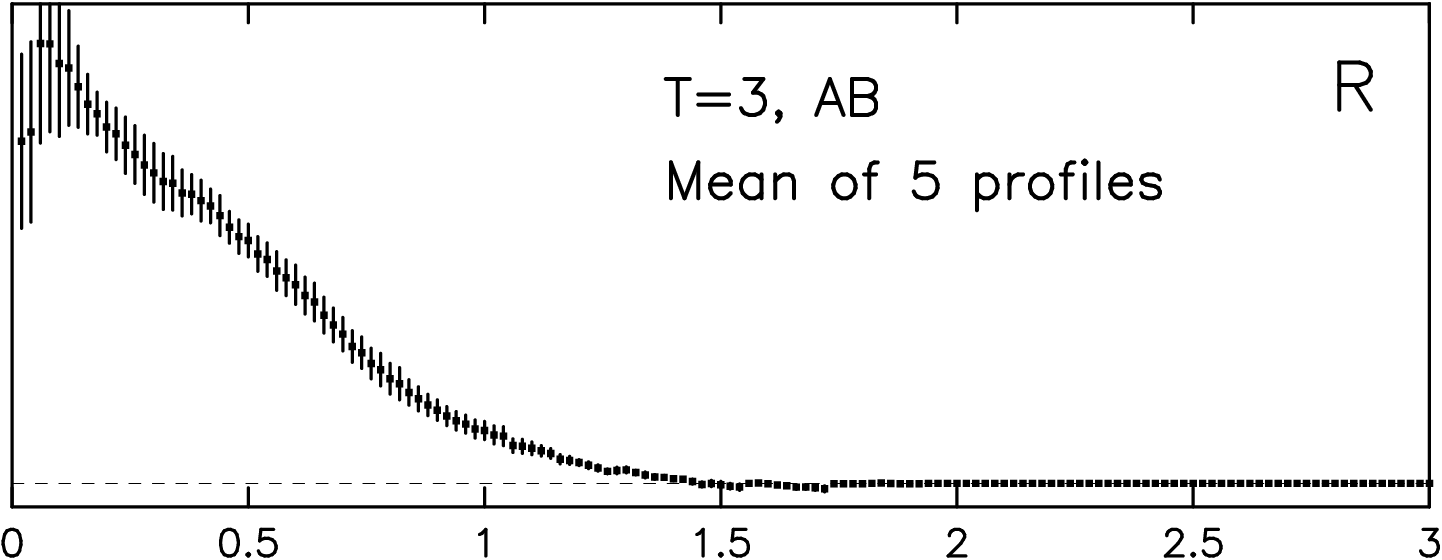}
}
\rotatebox{-90}{
\includegraphics[height=5.7cm]{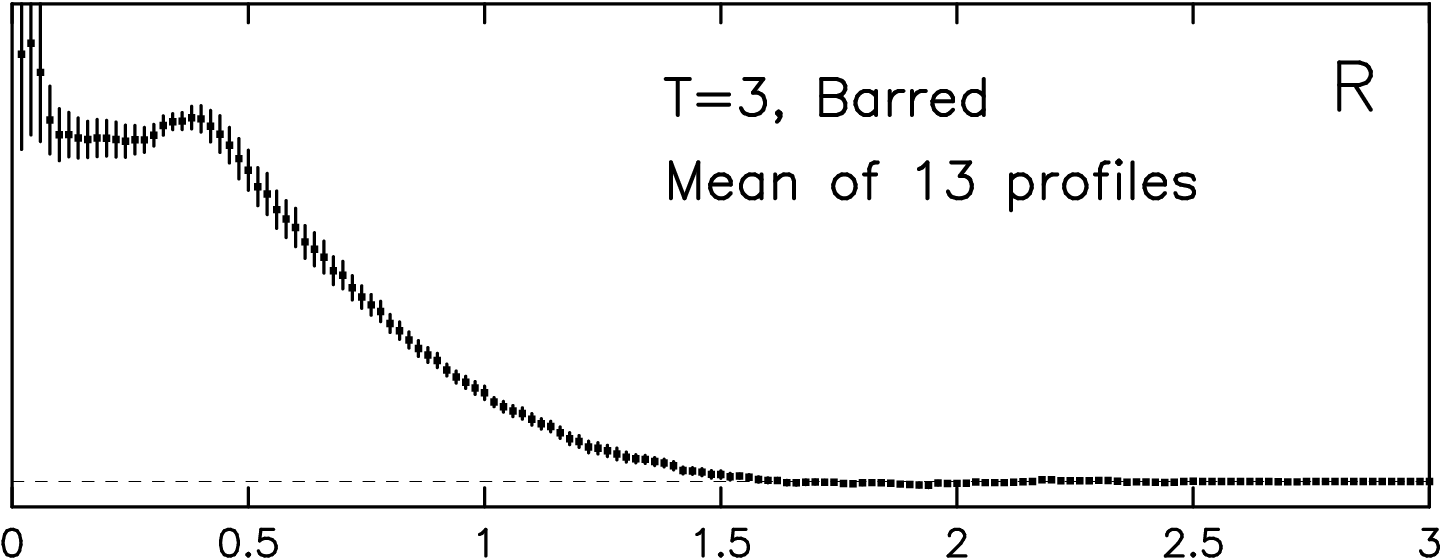}
}
\rotatebox{-90}{
\includegraphics[height=5.7cm]{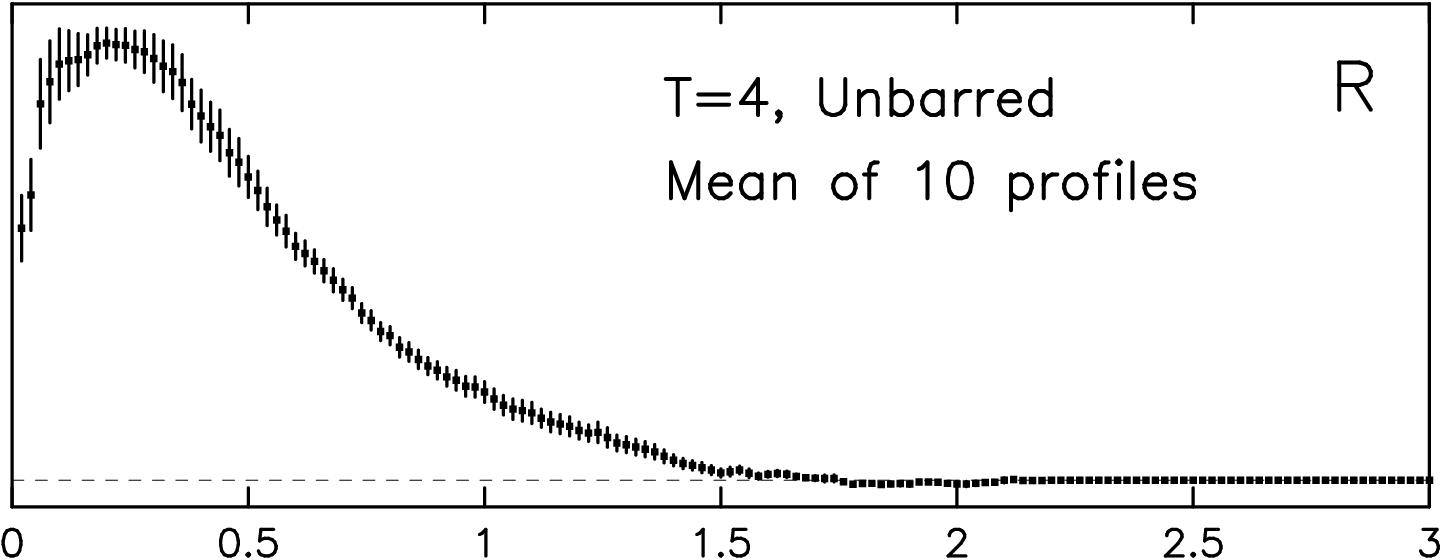}
}
\rotatebox{-90}{
\includegraphics[height=5.7cm]{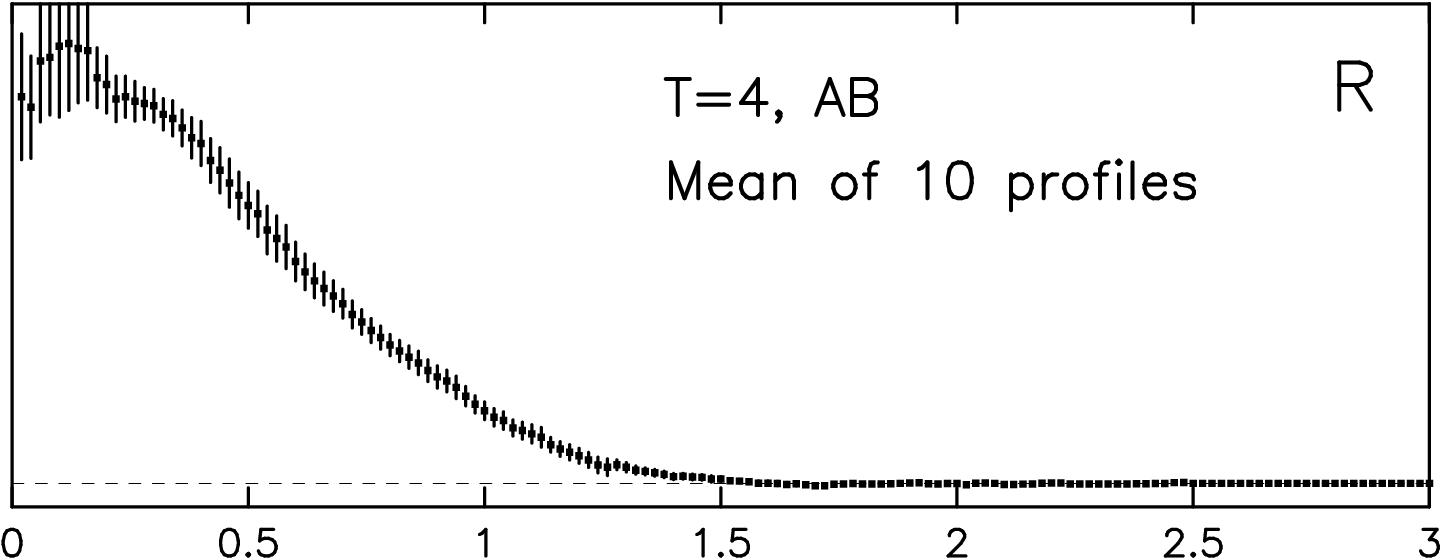}
}
\rotatebox{-90}{
\includegraphics[height=5.7cm]{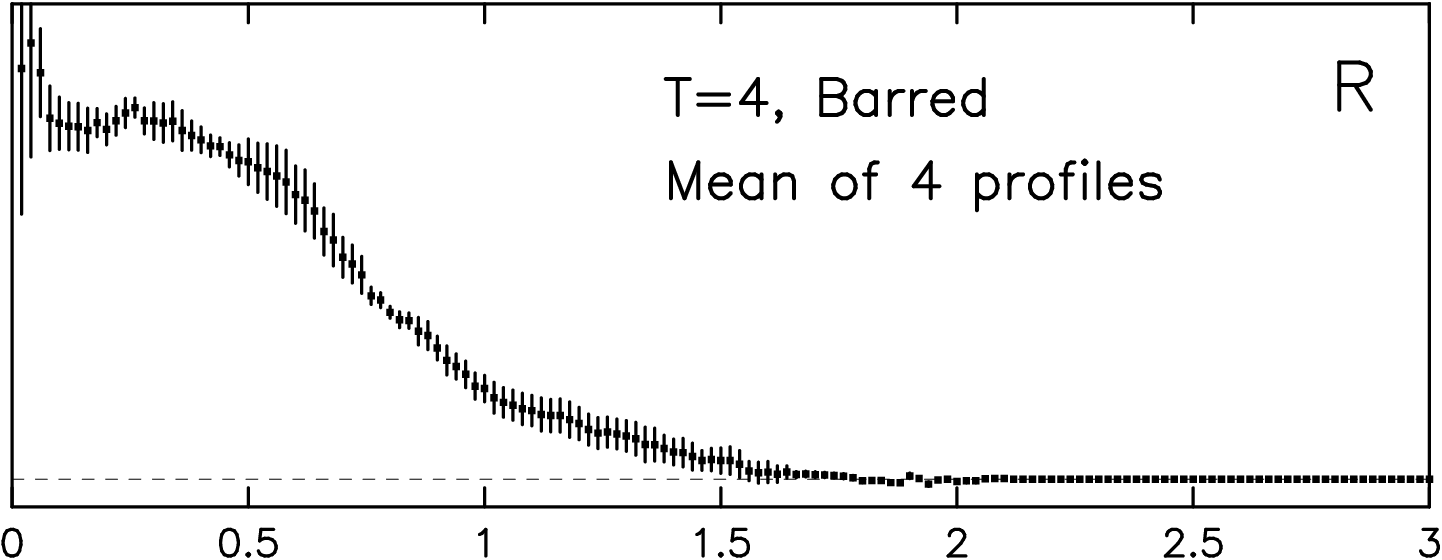}
}
\rotatebox{-90}{
\includegraphics[height=5.7cm]{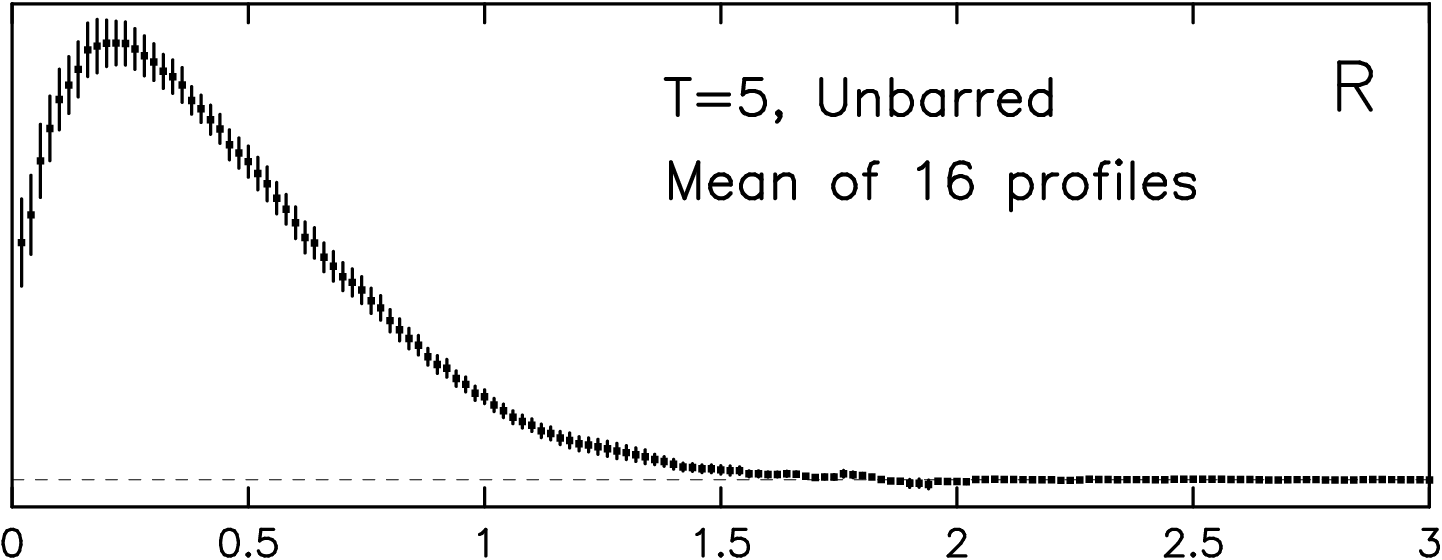}
}
\rotatebox{-90}{
\includegraphics[height=5.7cm]{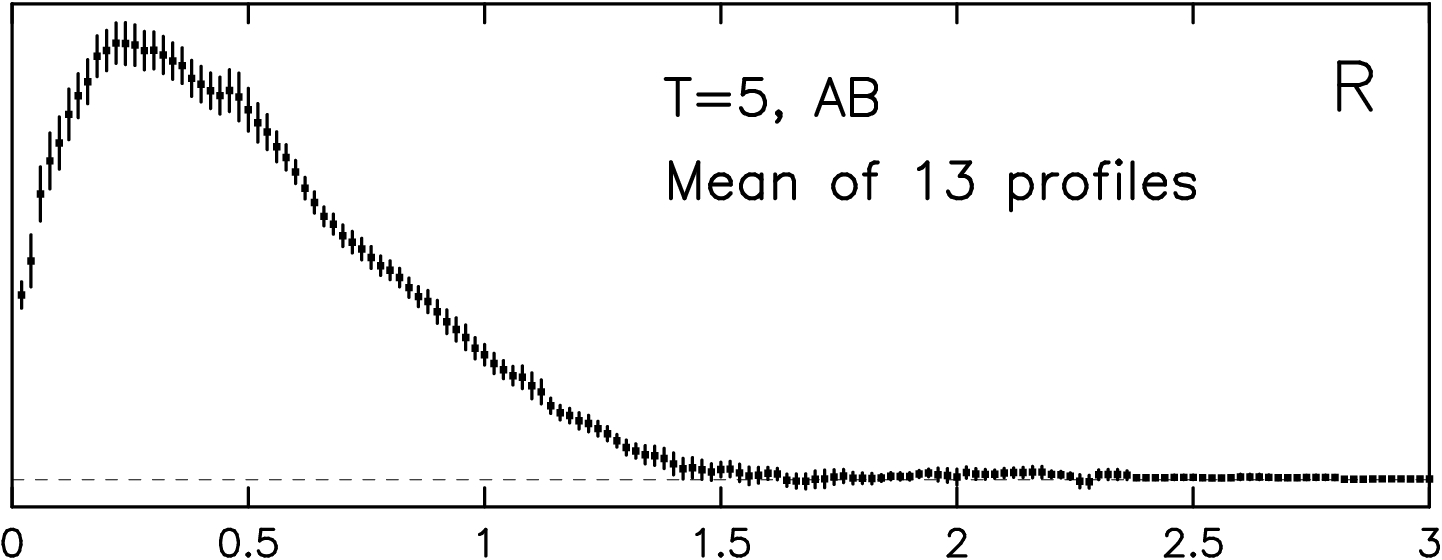}
}
\rotatebox{-90}{
\includegraphics[height=5.7cm]{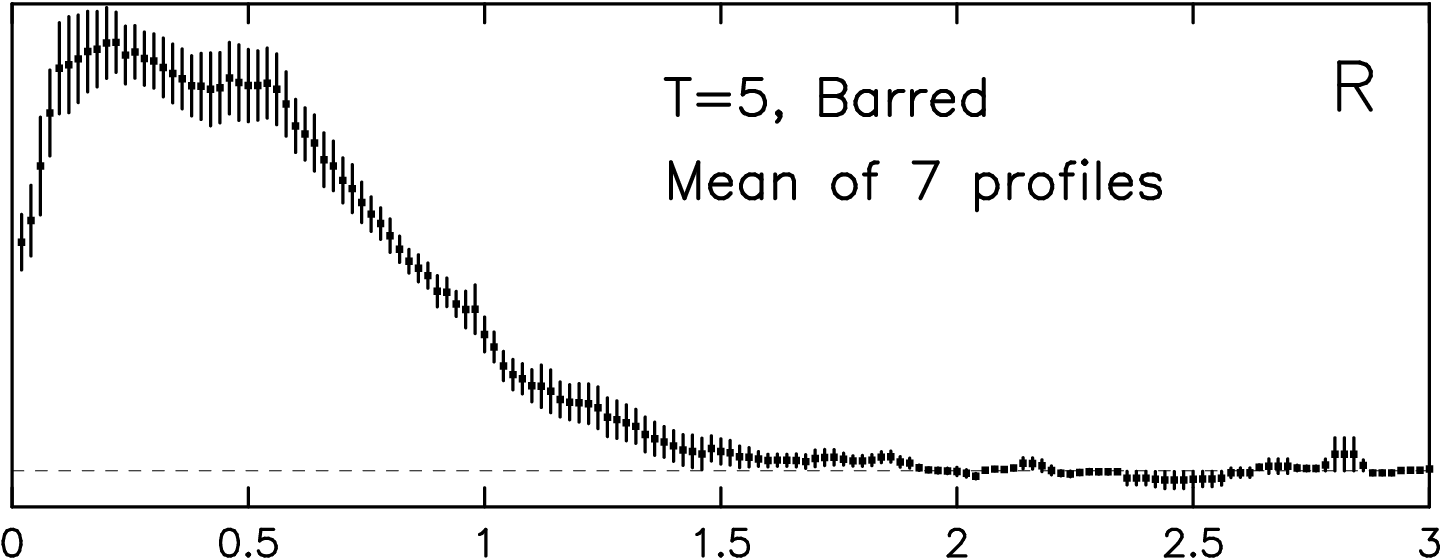}
}
\rotatebox{-90}{
\includegraphics[height=5.7cm]{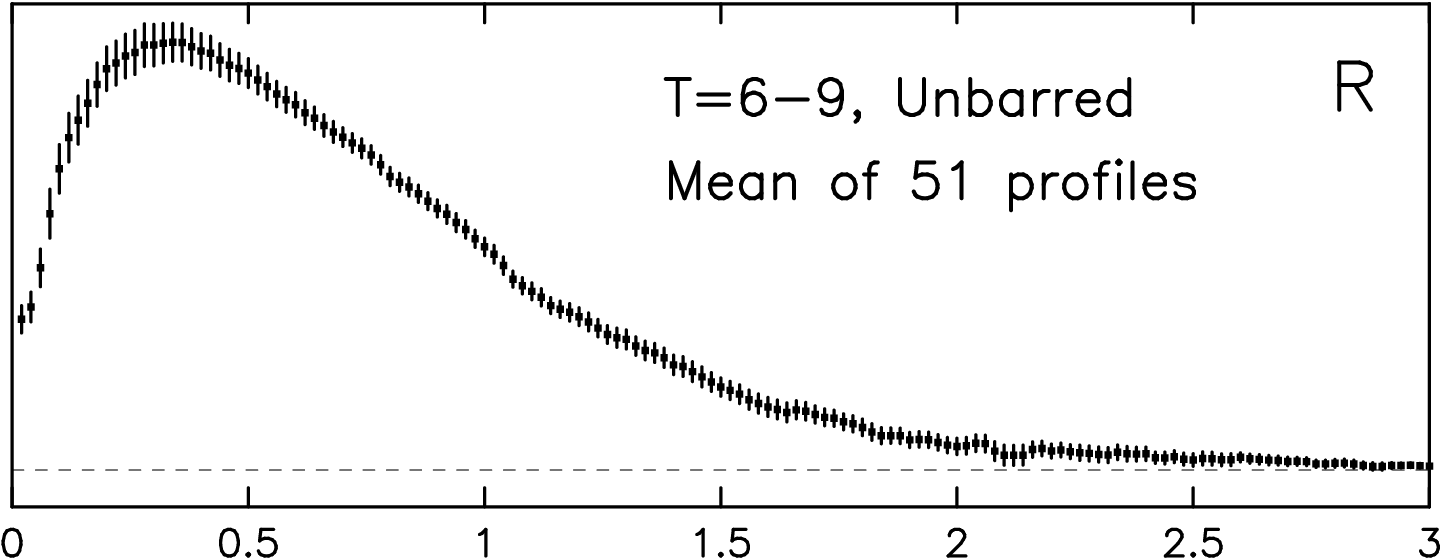}
}
\rotatebox{-90}{
\includegraphics[height=5.7cm]{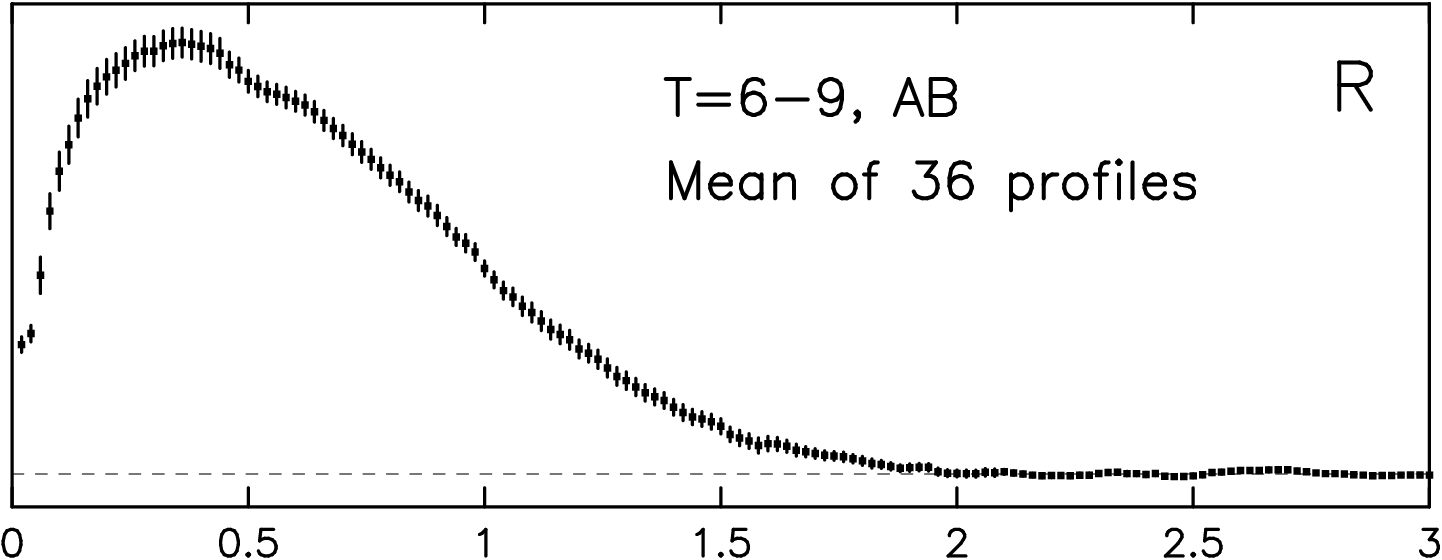}
}
\rotatebox{-90}{
\includegraphics[height=5.7cm]{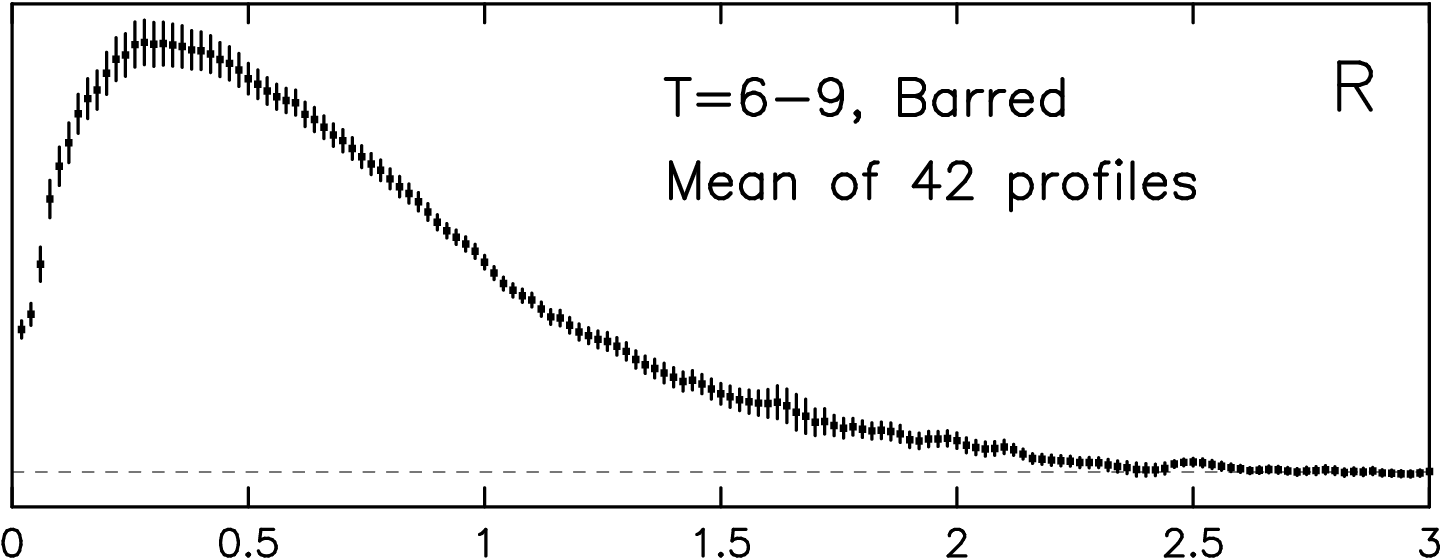}
}
\rotatebox{-90}{
\includegraphics[height=5.7cm]{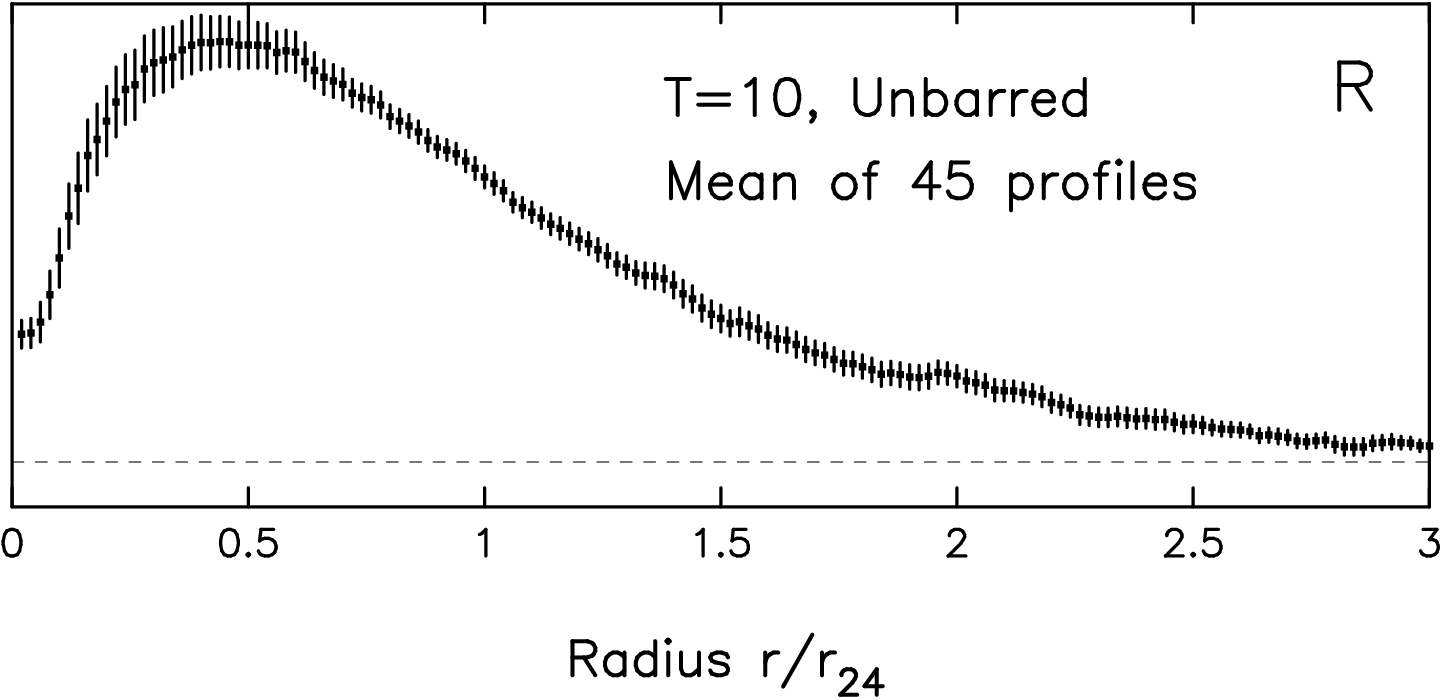}
}
\rotatebox{-90}{
\includegraphics[height=5.7cm]{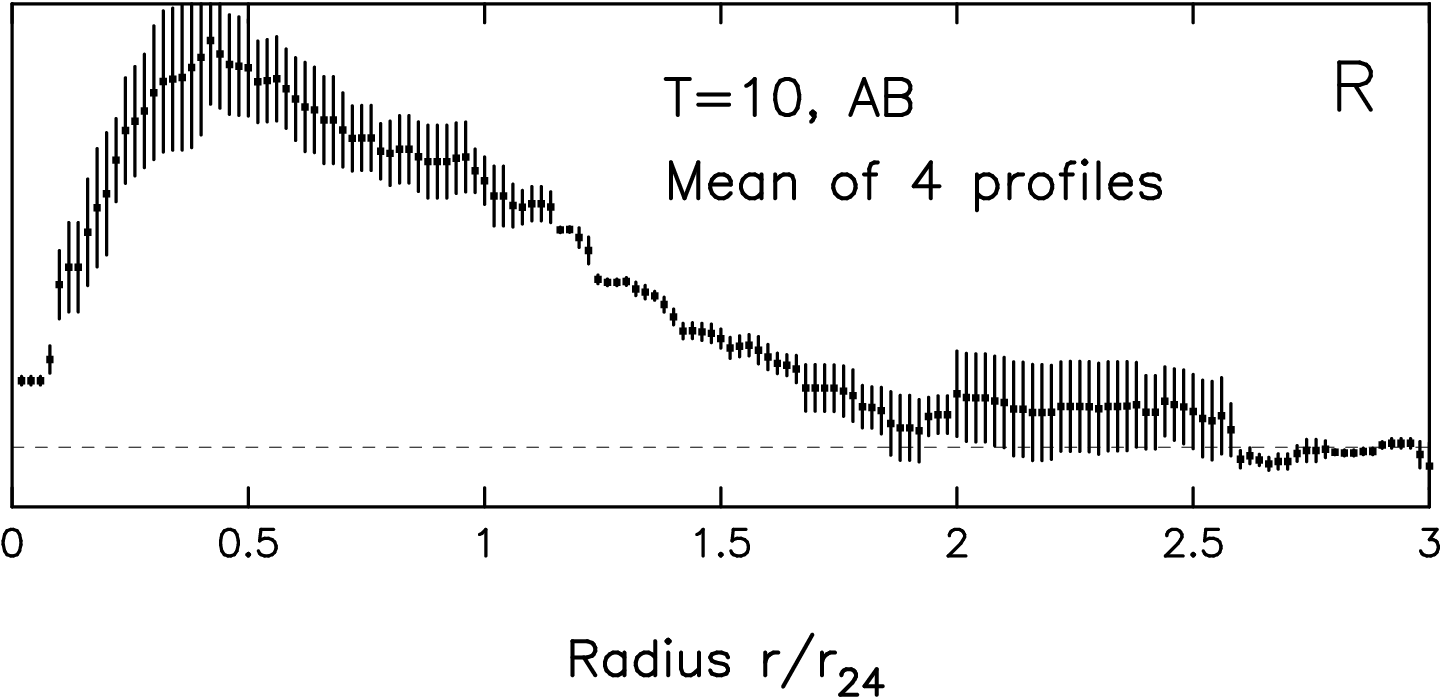}
}
\rotatebox{-90}{
\includegraphics[height=5.7cm]{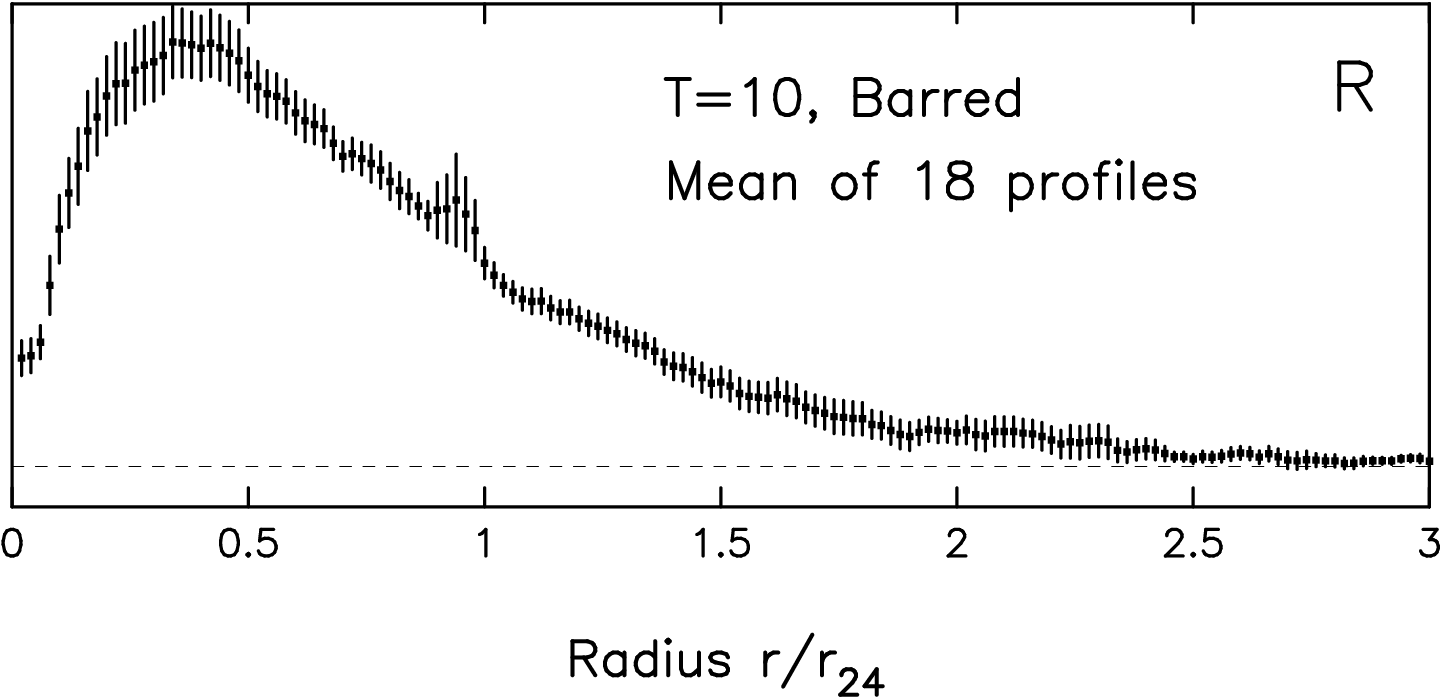}
}

\caption{As Fig. \ref{fig:habarprof} but showing the $R$-band profiles,
again subdivided by bar classification}
\label{fig:rbarprof}
\end{figure*}

The mean $R$-band profile for Sb galaxies ($T=$ 3) in Fig.
\ref{fig:meanprof} shows a distinctive `shoulder' on the central
emission peak, extending to a radius of $\sim$0.5~$r_{24}$.  The same
effect is seen, albeit less strongly, in the mean Sbc profile.  The
mean \Ha\ profiles of these two types are also distinctive, with a
narrow central peak, a {\it minimum} in the profile (not seen for any
other types), and an outer maximum which is at an anomalously large
radius given the relatively early types (see Table \ref{tbl:reff}).
This is particularly marked for the Sb mean profile.  The size
corresponding to the `shoulder' radius for the Sb profile is 4.8~kpc,
which immediately suggests that the excess light may be due to bars.
Indeed, 13 of the 23 galaxies contributing to this mean profile are
classified SBb.  To investigate this possibility, in
Figs. \ref{fig:habarprof} and \ref{fig:rbarprof} we have replotted the
\Ha\ and $R$-band mean light profiles of Fig. \ref{fig:meanprof}
subdivided according to bar type.  In each figure, the left-hand plots
show mean profiles for galaxies classified as showing no evidence for
bars in their optical morphologies (S$t$/SA$t$, or Im); the central
plots show those for intermediate or possible bar types (SAB$t$ or
IABm); and the right-hand plots the profiles for galaxies with clear 
optical bars (SB$t$ or IBm).  Bar classifications are taken from 
\citet{deva91} who used optical images; it should be noted that near-IR
imaging can reveal weak or small bars not seen in the optical. 
In order to maintain reasonable numbers in 
each mean profile, early ($T=$ 0 - 2) and late ($T=$ 6 - 9) spiral types 
have been combined.  Types 3 - 5 are plotted separately as these are the 
most important for the following discussion.

Figures \ref{fig:habarprof} and \ref{fig:rbarprof} confirm that the
distinctive profiles seen for the $T=$ 3 galaxies are indeed due to
the barred SBb galaxies.  The mean \Ha\ profile for the SBb galaxies
shows a very strong central peak of emission, and a clear outer peak
at 0.5~$r_{24}$.  Both are absent in the mean profile of the Sb
(unbarred) galaxies, and indeed from the mean profiles of the unbarred
galaxies of any $T$ type.  The profile for the $T=$ 3 SABb galaxies is
noisy, given that only 5 galaxies contribute, but appears intermediate
between those of the Sb and SBb types, with some evidence of a central
peak.  The $R$-band mean profiles for Sb, SABb and SBb types show the
same pattern, with the characteristic bar features being somewhat less
strong in the older stellar population.  The SBb $R$-band profile
still has a central peak, and an outer peak at 0.5~$r_{24}$, but both
represent a much smaller fraction of the overall flux than is the case
for the mean \Ha\ profile.  Bars thus have a substantial effect on the
$R$-band light distribution for $T=$ 3 galaxies, with the outer peak
pushing a large fraction of the flux further out in these galaxies, an
effect which explains the lower concentration indices for barred
galaxies illustrated in Fig. \ref{fig:dc30_v_t}.  Indeed, the effect
on the concentration indices would presumably be even more marked were
it not for the central peaks in the mean barred profiles, which must
act to increase the concentration indices.

These characteristic bar profiles occur for $T=$ 3 galaxies where they
are most apparent, but also for $T=$ 4 galaxies, and possibly for
those of $T=$ 5. (In comparing the $T=$ 3 and 4 barred mean profiles,
it should be noted that there are 13 SBb galaxies contributing to the
mean profiles against only 4 of type SBbc, so inevitably the mean
profile for the latter type is more noisy.)  The mean \Ha\ profile
for the SBbc ($T=$ 4) galaxies is clearly different from that for the
SBc ($T=$ 5) galaxies, with the former again having a central peak and
an outer peak at $\sim$0.5$r_{24}$, like those in the SBb mean
profile.  The SBbc $R$-band profile shows only a weak central peak,
and the outer profile shows a `shoulder' of flux pushed to larger
radius, but no outer peak as such.  For $T=$ 5, the only effect is the
broadening of the mean profiles, with flux again pushed to larger
radii.  The \Ha\ profile is noisy at all radii, but the $R$-band
profile shows evidence for this additional flux causing a `shoulder'
in the profile.  No central or outer peaks are seen in the $T=$ 5 mean
profiles, or indeed in any of the profiles for types later than this.

Thus we have identified a clear effect of bars on the pattern of SF as
a function of radius within galaxy disks, which results in strongly
enhanced \Ha\ emission, and moderately enhanced $R$-band emission, in
both the central regions and at 0.5 $r_{24}$.  This effect seems to
apply most strongly to those galaxies classified as SBb or SBbc ($T=$
3, 4), where the overall distributions of SF are profoundly different
from their unbarred counterparts.  It should be recalled that the mean
profiles we show here are constructed such that areas under the
profiles are directly proportional to fractions of total flux.  Thus
inspection of the SBb mean \Ha\ profile shows that a significant
fraction of the total \Ha\ $+$ \NII\ emission from these galaxies is
associated with the outer peak feature.

The greater strength of the bar-induced features in the \Ha\ mean
profiles than in the $R$-band profiles is important, since this
confirms that the bars are inducing SF that would not otherwise be
happening, and are not merely redistributing pre-existing stellar
populations.  If the latter were the case, the amplitude of the effect
would be the same in $R$-band light and \Ha\ (see \citet{seig02} for a
similar argument applied to the triggering of SF by spiral arms).

Having identified this characteristic pattern of SF in SBb galaxies,
we next investigate the nature of both nuclear and outer
emission peaks, by looking in detail at the continuum-subtracted \Ha\
images for all galaxies of this type in the \Ha GS sample.

\subsection{Individual SBb galaxies: nuclear emission peaks}
\label{sec:cen_peak}

\begin{table}
\begin{center}
\begin{threeparttable}
\begin{tabular}{lcccclc}
\hline 
\hline 
UGC            & NGC & Classn. & Dist. & Peak & Res & FWHM \cr
               &     &         & Mpc &       &          & kpc\cr
\hline 
3685           &  --  &    SB(rs)b    & 26 & Y & U    & $<$0.17 \cr 
4273           & 2543 &    SB(s)b     & 35 & Y & R    &    0.65 \cr 
4484           & 2608 &    SB(s)b     & 32 & Y & R    &    0.34 \cr 
4574           & 2633 &    SB(s)b     & 31 & Y & R    &    0.62 \cr 
4705\footnote  & 2710 &    SB(rs)b    & 36 & Y & R$+$ &    1.10 \cr
4708           & 2712 &    SB(r)b:    & 29 & Y & R$+$ &    1.20 \cr 
6077           & 3485 &    SB(r)b:    & 27 & Y & R    &    0.29 \cr 
6123\footnote  & 3507 &    SB(s)b     & 21 & Y & R?   &     --  \cr 
6595\footnote  & 3769 &    SB(r)b:    & 12 & Y & R$+$ &    0.32 \cr 
7002           & 4037 &    SB(rs)b:   & 18 & Y & R?   &    0.24 \cr 
7523\footnote  & 4394 & (R)SB(r)b:    & 18 & Y & R$+$ &    0.46 \cr 
7753\footnote  & 4548 &    SB(rs)b    & 18 & Y & R$+$ &    0.45 \cr 
9649\footnote  & 5832 &    SB(rs)b?   &  8 & N & --   &     --  \cr 
12699\footnote & 7714 &    SB(s)b:p   & 31 & Y & R?   &    0.32 \cr 
\hline 
\end{tabular}
\begin{tablenotes}
\item[1] Emission elongated along bar axis
\item[2] LINER; foreground star contamination, so not in mean profiles
\item[3] Interacting with NGC~3769A, type Sm
\item[4] LINER; Smooth extended emission, possible ENER
\item[5] M91; LINER/Seyfert; possible ENER
\item[6] Classified as SBc in UGC
\item[7] \HII/LINER; tidally distorted
\end{tablenotes}
\caption[]{Properties of the central emission peaks.  For each galaxy, we
list the UGC and NGC catalogue numbers, the full galaxy classifications from 
\citet{deva91}, the distance in Mpc from Paper~I, whether or not the 
galaxy exhibits a central peak in \Ha\ emission, the resolution of the 
peak as explained
in Sect. \ref{sec:cen_peak}, and the FWHM of the emission, in kpc.
All LINER/Seyfert classifications in the footnotes are taken from
the NASA/IPAC Extragalactic Database (NED).}
\label{tbl:cenpeak}
\end{threeparttable}
\end{center}
\end{table}

To study in more detail the origin of the central peak in \Ha\ in SBb
($T=$ 3) galaxies and, in the next subsection, that of the peak near
0.5 $r_{24}$, we consider here the properties of the 14 individual 
galaxies that make up this subsample.
The properties of the central emission-line peaks seen in these 14 
galaxies are listed in Table \ref{tbl:cenpeak}
(UGC~6123, which was omitted from the mean profiles because of a
bright superposed star, is reinstated in this table; note that this galaxy
does show both nuclear and bar-end emission).  The major
conclusion to draw from Table \ref{tbl:cenpeak} is the near-ubiquity
of nuclear peaks, which are found in 13 of the 14 SBb galaxies (denoted
by `Y' in Col. 5). 

Table \ref{tbl:cenpeak} also contains information on the sizes
of the central peaks.
That for UGC~3685 is unusual in that it is unresolved in our
\Ha\ image (`U' in the Col. 6), i.e. the FWHM of the peak is
consistent with that of stellar images in the same image before
continuum subtraction.  This favours an AGN interpretation for this
emission, as a typical SF complex would be resolved at the distance of
this galaxy. All the other central profiles are marginally resolved (`R?') or
resolved (`R'; `R$+$' implies the source is several times
larger than the seeing disk).  The approximate diameter of the
emission line region is given in Col. 7.  For the resolved regions
this is typically a few hundred pc, which would imply the presence of
SF complexes or nuclear rings of SF.  However, this does not exclude
the possibility of AGN emission in many or all of these galaxies, and
indeed four of those with extended emission are known to have
LINER-type nuclei.  In two of these (UGC~7523 $=$ NGC~4394 and
UGC~7753 $=$ M~91) the extended emission has the smooth morphology
characteristic of ENERs.

Two of the galaxies in Table \ref{tbl:cenpeak} will be discussed in more
detail in Paper VIII; these are the luminous interacting galaxy
NGC~7714, and NGC~3769.

\subsection{Individual SBb galaxies: outer emission peaks}
\label{sec:out_peak}

\begin{table}
\begin{center}
\begin{threeparttable}
\begin{tabular}{rcccccl}
\hline 
\hline 
~UGC & NGC & Classn. & ~Peak~ & ~$r$~     & $r$  & Morph \cr
     &     &         &        & ~$r_{24}$~ & kpc  &       \cr           
\hline 
3685  &  --  & SB(rs)b    & Y & 0.58 & 4.5 & RB\footnote[1]   \cr 
4273  & 2543 & SB(s)b     & Y & 0.60 & 8.3 & BE2              \cr 
4484  & 2608 & SB(s)b     & Y & 0.56 & 5.8 & BE2              \cr 
4574  & 2633 & SB(s)b     & Y & 0.40 & 5.2 & BE2              \cr 
4705  & 2710 & SB(rs)b    & Y & 0.46 & 4.8 & RB\footnote[1]   \cr
4708  & 2712 & SB(r)b:    & Y & 0.36 & 4.4 & BE1              \cr 
6077  & 3485 & SB(r)b:    & Y & 0.54 & 4.0 & RB\footnote[1]   \cr 
6123  & 3507 & SB(s)b     & Y &  --  &  -- & BE1\footnote[2] \cr 
6595  & 3769 & SB(r)b:    & N &  --  &  -- & --               \cr 
7002  & 4037 & SB(rs)b:   & Y & 0.54 & 3.1 & BE1              \cr 
7523  & 4394 & (R)SB(r)b: & Y & 0.44 & 3.6 & R\footnote[1]    \cr 
7753  & 4548 & SB(rs)b    & Y & 0.48 & 6.4 & BE2              \cr 
9649  & 5832 & SB(rs)b?   & Y & 0.54 & 1.7 & BE2              \cr 
12699 & 7714 & SB(s)b:p   & N &  --  &  -- & --               \cr 
\hline 
\end{tabular}
\begin{tablenotes}
\item[1] Ring at same radius as bar ends
\item[2] SF at one bar end; other behind foreground star
\end{tablenotes}
\caption[]{Properties of the outer emission peaks.  Columns 1 to 3
are as in Table \ref{tbl:cenpeak}, and the remaining entries denote whether
the galaxy exhibits an outer peak in the \Ha\ profile, give the radius 
of the peak centre in units of the $r_{24}$ radius and in kpc, and finally
describe the morphology of the peak, as explained in 
Sect. \ref{sec:out_peak}.}
\label{tbl:outpeak}
\end{threeparttable}
\end{center}
\end{table}

The properties of the regions contributing to the line emission at a
radius of $\sim$0.5~$r_{24}$ are summarised in Table
\ref{tbl:outpeak}.  Here the second column indicates whether the
individual \Ha\ profile for the galaxy concerned exhibits such an
outer peak, Cols. 3 and 4 give the radius of the peak in units of the
$r_{24}$ radius and kpc respectively; and the final column contains
descriptors of the morphology of the emission-line regions leading to
this peak, obtained through inspection of the continuum-subtracted
\Ha\ images.  `R' denotes a ring morphology, `RB' a broken or partial
ring, `BE1' emission predominantly from one end of the bar, and `BE2'
emission from both bar ends.

In this case we find that 11 of the 13 galaxies contributing to the mean 
SBb profile have clear outer peaks in their mean light profiles.  The 
contaminated image of UGC~6123 shows emission from the one end of the 
bar that is clearly visible, but the foreground star makes it impossible to 
decide between `BE1' and `BE2' designations for this galaxy.

\section{Discussion}

\subsection{Concentration indices}

Three different concentration indices have been investigated in
Sect. \ref{sec:concind}.  All three methods show a strong
correlation with Hubble type, with the early-type, bulge-dominated
galaxies having concentration indices close to those expected for an
$r^{1/4}$ profile and the late types approximating to those of
exponential profiles.  Investigation of the relative concentrations
finds that most galaxies are more centrally concentrated in the
continuum light than in the \Ha. Strong bars tend to decrease the
central concentration of $R$-band light, and to a smaller extent the
\Ha\ light, over most disk types. For the latest type barred galaxies
(SBdm - IBm) the reverse trend is seen.

\subsection{Radial profiles}

From our radial flux profiles, we identify two features that are seen
in barred galaxies of morphological types $T=$ 3 and 4 but not in
un-barred galaxies of those types, which are a central peak of
emission and a second peak at $0.5~r_{24}$. As was was discussed in
Sect. \ref{sec:cen_peak}, the nuclear peak is most probably due to
nuclear emission not related to SF, ENERs, and/or circumnuclear
rings. The second peak occurs just outside the end of the bar, and may
be related to the presence of an inner (pseudo-ring) as known to exist
in many galaxies.

From the mean H$\alpha$ profile, we can estimate the overall
importance of the SF related to these features.  We find that
$\sim$50\% of the current massive SF activity in type $T=3$ and 4
galaxies occurs in the radial range of the outer peak.  The {\it
  excess} SF is somewhat harder to estimate, but flattening the `hump'
between 0.26 and 0.72 $r_{24}$ reduces the overall SFR by
$\sim$20\%. This implies that inner (pseudo)-rings or other features
at the radial position of the end of the bar contain a significant
fraction of the total massive SF activity in the galaxies under
consideration.  The equivalent number for the central peak is $\sim$10\%.

We thus confirm that barred galaxies of $T$-types 3 and 4 have strong
concentrations of SF both in their central regions (with the caveat
that some part of the emission we base this statement on may be due to
non-stellar processes) and in the regions at the ends of the bar.

\subsection{Inner rings}

The galaxy classifications in the Col. 3 of Table
\ref{tbl:outpeak} show that four have an `(r)' classification,
indicating an inner ring, and five `(rs)' indicating a pseudo-ring.
This frequency is similar to or slightly higher than the fraction
expected from the numbers of such rings found in $T=$ 3 galaxies by
\citet{buta96}.  The radial size of the outer peaks found in the
present study is similar to that of inner rings, but the match is not
perfect.  \citet{deva80} give a formula for the radial size of inner
rings as a function of galaxy type and bar classification.  For SBb
galaxies, this formula predicts inner rings to have radii 0.303
times the $B_{25}$ isophotal radius, where the latter is very similar
to the $R_{24}$ isophotal radius used in the present study.  The
$R$-band outer peaks are centred on 0.38 times the $R_{24}$ isophotal
radius, $\sim$25\% larger than is expected for inner rings; and the
\Ha\ outer peaks are at somewhat larger radii still, 0.4 - 0.5 times
this isophotal radius.  This implies that the SF we are seeing is not
directly associated with the bar ends, but is triggered in the spiral
arms lying just beyond this radius.  In this context, it should be
noted that NGC~4548, one of the SBb galaxies in the present study, is
the prototype for the `bracket type' of barred galaxies
\citep{buta02}, where short spiral arms lie just outside the radius of the bar
ends, and `overshoot' the bar, rather than starting where the bar
terminates.  In this galaxy at least, the SF causing the outer peak is
clearly located in these arm segments.

\subsection{Bars and star formation}

The impact of the presence of a bar on the total SFR in a galaxy has been
studied by many authors \citep [e.g.,] [] {hawa86,dres88,puxl88,pomp90,
isob92,ryde94,tomi96,huan96,mart97a,ague99,shet02,rous01,verl07}, 
but no consensus has been reached as to whether
the presence of bars causes a global enhancement of the SFR.

One of the most important characteristics of bars is that the
non-axisymmetric mass distribution in their host galaxy can lead to
the outward transport of angular momentum, and thus to the inward
transport of gas \citep [see, e.g., the review of] [for the
  theoretical view] {shlo90}. The resulting enhanced central
concentration of gas in barred galaxies as compared to non-barred
galaxies has indeed been observed \citep [e.g.,] [] {saka99,shet05}, 
although \citet{komu08} note that varying Hubble
type, indicative of the effect of the bulge, is more important for
central gas concentration than the presence of a bar.

The central gas concentration caused by bars may lead to an enhanced
SFR rate in bars, quite possibly in the form of circumnuclear rings of
SF which can occur between the inner Lindblad resonances set up by the
bar.  Indeed, almost all of these nuclear rings occur in barred
galaxies, and in the few that do not the influence of a past
interaction can often be deduced \citep [e.g.,] [] {knap05}. Whether
nuclear starburst and AGN activity is, statistically, induced or
facilitated by the presence of a bar is not a settled question,
  with theoretical studies supporting such a link
  \citep{shlo89,wada95,wada04,shlo00}, whilst observations have hinted
  both in favour of \citep{knap00,lain02,hunt04} and against
  \citep{mulc97,mart03,laur04,duma07} this connection.  A review by
\citet{knap04} concludes that the evidence favours a slight effect
connecting bars to both these types of activity, but the link is
statistical and not direct, and subject to important caveats.

What has been studied perhaps less in the literature is what we
present here: an analysis of the effect on bars on the radial
distribution of current and past SF in disk galaxies. In addition, all
past studies have considered radial profiles in surface brightness or
equivalent, rather than the kind of flux profiles we present here, and
which show much more clearly the effects of different components on
the radial profiles. The classical study of observed bar properties,
including radial surface brightness profiles, is that by
\citet{elme85}, although before that other papers, such as those by
\citet{deva72} and \citet{elme80} had considered the Large Magellanic
Cloud (LMC) and other late-type galaxies that resemble it. A common
finding in all these, and subsequent, papers is that important areas
of massive SF appear near the ends of the bar.


\section{Conclusions}
\label{sec:conc}

The spatial distributions of \Ha\ and $R$-band light have been
investigated for a sample of 313 nearby S0a - Im galaxies.  The
initial analysis employed three concentration indices, all of which
showed declining central concentration of light in both bandpasses as
a function of increasing $T$-type; somewhat lower concentration in
\Ha\ than $R$-band for most $T$-types; and a moderate decrease in the
central concentration of luminosity for barred cf. unbarred galaxies
for $T=$ 0 - 5 early to intermediate spiral types. Mean normalised
profiles were then introduced as a more detailed probe of light
distributions in the two bandpasses.  These profiles show
a smooth sequence in properties as a function of galaxy type, with a
hump containing most of the flux shifting to larger values of the
normalised radius as $T$ increases.  The profiles at a given $T$-type
were found to show a remarkably small scatter from galaxy to galaxy;
this was investigated at $R$ for the two spiral types with the largest
number of individual profiles, and profile shape was found to show no
significant dependence on galaxy size, luminosity or distance.

The central regions of the \Ha\ profiles of most galaxy types were
found to show a dip relative to the $R$-band profiles, which we
interpret as the signature of the older stellar populations in bulges
and bars, cf. disks.  Surprisingly, this central dip is not present in
the mean light profiles of the earliest types studied here ($T=$ 0 -
2), probably due to a combination of nuclear SF rings and
non-SF-related ENER emission in the central regions of many of these
galaxies.  The mean \Ha\ and $R$-band profiles of the latest types
(particularly $T=$ 10 Im types) show excellent agreement in overall
shape.  There is no evidence for outer truncation of the mean \Ha\ profiles 
relative to the $R$-band profiles for any type.

The mean \Ha\ profile for SAb, SABb and SBb ($T=$ 3) galaxies showed a
central spike, and an outer peak at $\sim$0.45 $r_{24}$.  Both were
found to be characteristic features of barred $T=$ 3 galaxies, 
occurring in almost all of the individual galaxies of this type, and
were present at a lower level in the $R$-band mean $T=$3 profile, and
in the mean profiles for galaxies of $T$-types 4 and 5.  The outer
peaks are at radii similar to or somewhat larger than those expected
for inner rings, and they constitute at least
20\% of the total SF activity in the SBb galaxies.  The central spikes
are resolved in most cases, implying that they are not purely powered
by AGN activity, and contribute $\sim$10\% of the total \Ha\ flux.

\begin{acknowledgements}
The Jacobus Kapteyn Telescope was operated on the island of La Palma
by the Isaac Newton Group in the Spanish Observatorio del Roque de los
Muchachos of the Instituto de Astrof\'isica de Canarias. This research
has made use of the NASA/IPAC Extragalactic Database (NED) which is
operated by the Jet Propulsion Laboratory, California Institute of
Technology, under contract with the National Aeronautics and Space
Administration.  
\end{acknowledgements}

\bibliographystyle{bibtex/aa}
\bibliography{refs}
      
\end{document}